\documentclass{aa}
\usepackage{graphicx}
\usepackage{bm}
\usepackage{natbib}
\usepackage{subcaption}
\usepackage{footmisc}

\usepackage[breaklinks=true]{hyperref}
\hypersetup{colorlinks=true,linkcolor=blue,citecolor=blue,filecolor=blue,urlcolor=blue}
\usepackage{txfonts}

\DeclareUnicodeCharacter{2212}{-}
\DeclareUnicodeCharacter{2212}{-}
\DeclareUnicodeCharacter{0308}{-}
\DeclareUnicodeCharacter{0302}{-}

\makeatletter
\renewcommand*\aa@pageof{, page \thepage{} of \pageref*{LastPage}}
\makeatother

\begin{document} 

\title{Galaxy And Mass Assembly (GAMA): Environment-dependent galaxy stellar mass functions in the low-redshift Universe}

\author{A.~Sbaffoni
  \inst{1,2}
  \and
  J.~Liske \inst{1,2}
  \and
  S.P.~Driver \inst{3}
  \and
  A.S.G.~Robotham \inst{3}
  \and
  E.N.~Taylor \inst{4}
}

\institute{Hamburger Sternwarte, Universität Hamburg, Gojenbergsweg 112,
  21029 Hamburg, Germany
  \and
  Cluster of Excellence Quantum Universe, Universität Hamburg,
  Luruper Chaussee 149, 22761 Hamburg, Germany
  \and
  International Centre for Radio Astronomy Research (ICRAR),
  University of Western Australia, Crawley, WA 6009, Australia
  \and
  Centre for Astrophysics and Supercomputing, Swinburne University
  of Technology, Hawthorn, VIC 3122, Australia
}

\titlerunning{GAMA: Environment-dependent GSMFs}
\authorrunning{Sbaffoni A. et al.}

\abstract{From a carefully selected sample of $52\,089$ galaxies and
  $10\,429$ groups, we investigate the variation of the low-redshift
  galaxy stellar mass function (GSMF) in the equatorial Galaxy And
  Mass Assembly (GAMA) dataset as a function of four different
  environmental properties. We find that: (i)~The GSMF is not strongly
  affected by distance to the nearest filament but rather by group
  membership. (ii)~More massive halos tend to host more massive
  galaxies and exhibit a steeper decline with stellar mass in the
  number of intermediate-mass galaxies. This result is robust against
  the choice of dynamical and luminosity-based group halo mass
  estimates. (iii)~The GSMF of group galaxies does not depend on the
  position within a filament, but for groups outside of filaments, the
  characteristic mass of the GSMF is lower. Finally, our global GSMF
  is well described by a double Schechter function with the following
  parameters: $\log [M^{\star} / (M_{\odot} \, h_{70}^{-2})] = 10.76
  \pm 0.01$, $\Phi_1^{\star} = (3.75 \pm 0.09) \times
  10^{-3}$~Mpc$^{-3}$~$h_{70}^3$, $\alpha_{1} = -0.86 \pm 0.03$,
  $\Phi_2^{\star} = (0.13 \pm 0.05) \times
  10^{-3}$~Mpc$^{-3}$~$h_{70}^3$, and $\alpha_{2} = -1.71 \pm
  0.06$. This result is consistent with previous GAMA studies in terms
  of $M^{\star}$, although we find lower values for both $\alpha_{1}$
  and $\alpha_{2}$.}
      

\maketitle


\section{Introduction}

According to the $\Lambda$ cold dark matter ($\Lambda$CDM)
cosmological paradigm, structures in the present Universe arose from
hierarchical clustering, whereby smaller dark matter halos (DMHs) grew
by gravitational attraction first and progressively merged into larger
halos \citep{press+74,blumenthal+84,davis+85}. Baryons gathered in the
centres of these halos, and a small fraction of them cooled and
subsequently condensed into stars to form galaxies
\citep{white+78,white+91}. However, DMHs -- and hence the galaxies
residing within them -- are not isolated. They are part of a
large-scale structure \citep[LSS;][]{kaiser+84,efstathiou+88}, also
known as the `cosmic web' \citep{joeveer+78,bond+96b}. Under the
effect of gravity across cosmic time, the cosmic web emerged from the
anisotropic collapse of the initial fluctuations of the matter density
field \citep{zeldovich+70}. This web consists of sizeable, nearly
empty voids surrounded by sheet-like walls formed by filaments
intersecting at the locations of clusters of galaxies. These nodes
represent the densest regions of the LSS, containing a large fraction
of the dark matter mass \citep{bond+96a,pogosyan+96}.

The galaxy stellar mass function
\citep[GSMF;][]{bell+03,baldry+08,baldry+12,wright+17,driver+22},
which gives the number density of galaxies per unit mass, is one of
the most fundamental measurements in extragalactic astronomy,
containing valuable information about the physical processes of galaxy
formation and evolution. GSMFs were first measured in the optical and
later in the near-infrared (NIR), where the light better traces the
low-mass stellar populations that dominate the stellar mass
reservoir. In fact, the stellar masses of galaxies may be derived from
optical and/or NIR photometry \citep{larson+78,jablonka+92,bell+01},
where, generally speaking, a broader wavelength coverage results in a
higher precision, or from spectroscopy
\citep{kauffmann+03,panter+04,gallazzi+05}. Multi-wavelength
photometric and spectroscopic surveys such as the Sloan Digital Sky
Survey (SDSS) and the Galaxy And Mass Assembly (GAMA) survey are thus
able to estimate stellar masses with a precision of $\sim$$0.2$ dex
\citep{taylor+11}, and the most significant measurements of the GSMF
obtained from SDSS \citep{bell+03,baldry+08} and GAMA
\citep{baldry+12,wright+17}, in which the GSMF was probed to a stellar
mass limit of $10^{8}$ $M_{\odot}$, are in reasonable agreement.

While the integral of the GSMF defines the density of baryonic mass
currently bound in stars, and thus the global star formation (SF)
efficiency, its shape (a power-law with an exponential cut-off at high
masses) is related to the evolutionary processes governing
SF. Explaining the shape of the GSMF (relative to that of the robustly
predicted halo mass function) is one of the major aims of galaxy
formation theory. The shape of the GSMF is generally thought to be
related to two different feedback mechanisms that are responsible for
suppressing SF. At low halo masses, supernova feedback creating
galactic winds plays a crucial role in regulating SF
\citep{larson+74,dekel+86,pillepich+18,scharre+24}, while
\cite{oppenheimer+10} claimed that the re-accretion of these winds is
essential in shaping the GSMF. At higher masses, initially infalling
gas is heated by a virial shock \citep{dekel+06} and subsequently
prevented from cooling by feedback from active galactic nuclei,
thereby suppressing SF
\citep{keres+05,croton+06,pillepich+18,scharre+24}. Both of these
feedback processes have a mass dependence; that is, the efficiency
with which they suppress SF depends on halo mass. Additionally,
mergers will also modify the GSMF by moving galaxies from lower to
higher masses, which may even be the dominant process shaping the GSMF
at masses greater than $10^{10.8}$~$M_{\odot}$ \citep{robotham+14}.
The mass dependencies of all of these processes shape the GSMF, and
since these are in turn expected to depend on the environment, the
GSMF should also be expected to vary as a function of environment.

In galaxy evolution theory, galaxy properties (including their stellar
mass) are often assumed to only depend on halo mass and not on any
other properties of the environment \citep[][and references
  therein]{moster+10}. However, several recent studies have failed to
confirm a dependence of the GSMF on halo mass. For example,
\cite{calvi+13} found that the GSMFs of the general field and of
clusters are essentially indistinguishable. A similar result was
obtained by \cite{vulcani+13} at intermediate
redshifts. \cite{guglielmo+18} also failed to find a difference
between the field and cluster GSMFs or a dependence of the cluster
GSMF on X-ray luminosity (a proxy for the halo mass). In contrast,
\cite{balogh+01} found a statistically significant difference between
the general field and the cluster GSMFs, whereas \cite{yang+09} showed
that the mass distribution of central as well as of satellite galaxies
appears to be a function of halo mass.

Clearly, a consensus on the environmental dependence of the GSMF has
not yet been reached, even at low redshift. The main aim of the
present study is to help clarify this issue by providing a meticulous
characterisation of the GSMF as a function of different environmental
measures using GAMA, which is arguably the survey that best probes the
environments of galaxies over a large range of scales at low redshift.

This paper is organised as follows. In Sect.~\ref{2}, we first
describe the precise GAMA data products that have been used for the
present analysis. In Sect.~\ref{3}, we explain in detail the
complexities of our target selection. In Sects.~\ref{4} and \ref{5},
we describe four different environmental properties and our precise
method of deriving the GSMF, repectively. In Sect.~\ref{6}, we present
our results separately for each environmental property. In
Sect.~\ref{7}, we compare our results to previous GAMA and SDSS
studies on the GSMF as well as to other studies that also investigated
the dependence of the GSMF on the environment. Finally, in
Sect.~\ref{8}, we draw our conclusions. Throughout this paper, we
assume a `737' cosmology with $(H_{0}, \Omega_{\rm M}, \Omega_{\rm
  \Lambda}) = (70, 0.3, 0.7)$, which corresponds to the same
cosmological model used in most GAMA studies on the GSMF. For all
physical quantities that depend on $H_{0}$, we include this dependency
using $h_{70} = H_{0}/(70$~km~s$^{-1}$~Mpc$^{-1})$.

\begin{figure}
\centering
\includegraphics[width=0.45\textwidth]{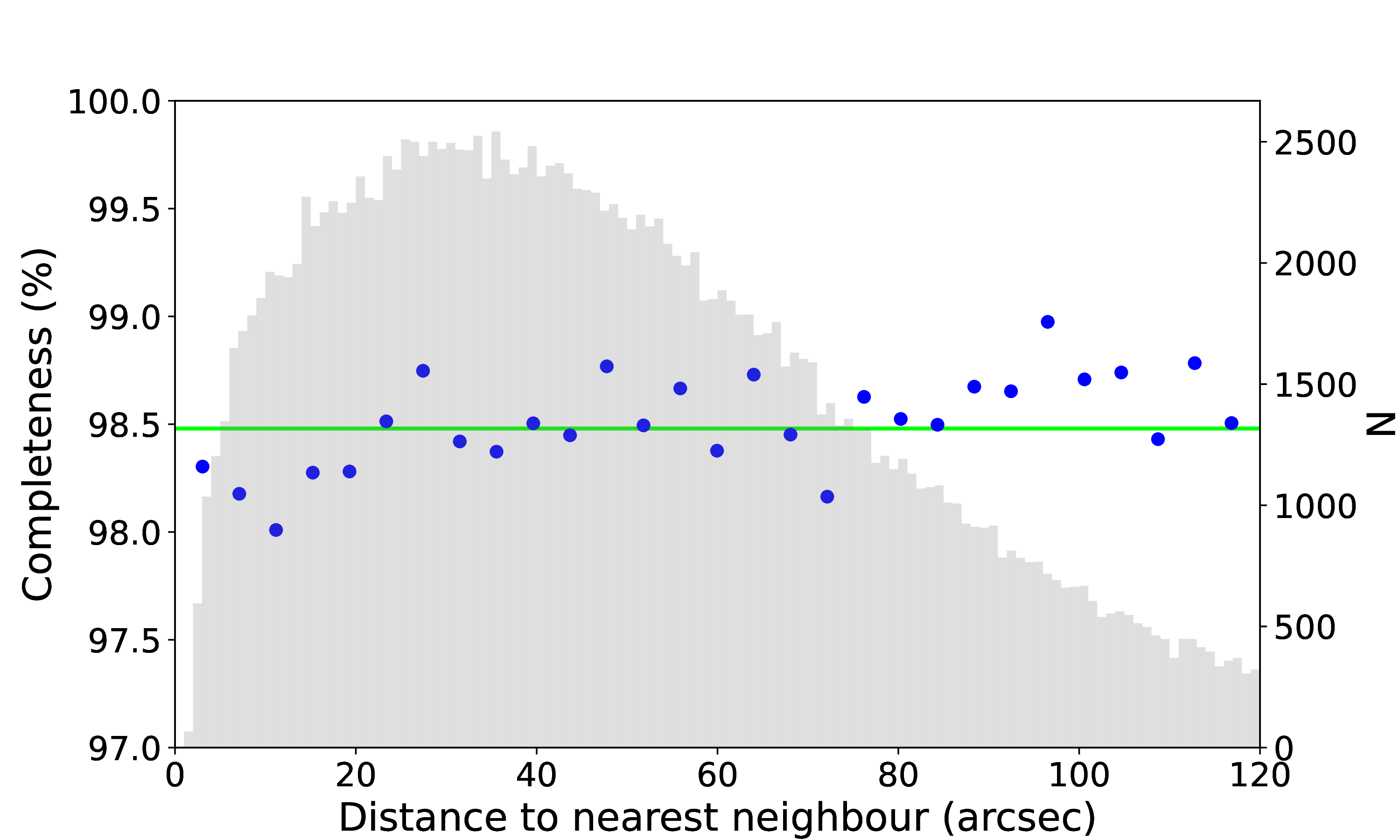}
\caption{Redshift completeness of the three equatorial GAMA regions
  G09, G12, and G15 as a function of the distance to the nearest
  neighbour among main survey targets (blue dots). As in
  \citet{liske+15}, the horizontal green line and the grey shaded
  histogram show the overall average redshift completeness and the
  distribution of all nearest neighbour distances, respectively.}
    \label{fig:zcompl}
\end{figure}

\section{Data}\label{2}
Our data are part of the
GAMA\footnote{\url{https://www.gama-survey.org/}}~II survey
\citep{liske+15}. GAMA is a large, low-redshift spectroscopic survey
covering $\sim$$238\,000$ galaxies down to $r<19.8$~mag over
$\sim$$286$~deg$^{2}$ of sky, split into 5 survey regions, 3 of which
are equatorial regions measuring $12 \times 5$~deg$^{2} =
60$~deg$^{2}$ each, and out to a redshift of approximately $0.6$. The
observations were completed in 2014, using the AAOmega spectrograph
\citep{saunders+04,smith+04,sharp+06} on the $3.9$-m Anglo-Australian
Telescope. The survey strategy and spectroscopic data reduction are
described in detail in
\cite{baldry+10,robotham+10a,driver+09,driver+11,hopkins+13,baldry+14,liske+15,baldry+18,driver+22}. In
addition, the GAMA team collected imaging data for the same survey
regions from a number of independent surveys in more than 20 bands,
with wavelengths between $1$~nm and $1$~m. Details of these imaging
surveys and the photometry derived from them are given in
\citet{liske+15}, \citet{driver+16,driver+22}, and references
therein. The combined spectroscopic and multi-wavelength photometric
data at the depth, imaging resolution, area and spectroscopic
completeness of GAMA provide a uniquely comprehensive survey of the
low redshift galaxy population, capturing a wide range of scales
relevant to galaxies.

In this study, we only consider the three equatorial GAMA survey
regions G09, G12, and G15, for which the overall redshift completeness
is $\sim$$98.5\%$ down to the magnitude limit of $r = 19.8$~mag. One
of the unique features of GAMA is that this exceptionally high average
redshift completeness is maintained even in the densest environments
such as pairs, groups, and clusters of galaxies. To illustrate this
point, we show in Fig.~\ref{fig:zcompl} the redshift completeness of
GAMA as a function of the distance to the nearest neighbour among main
survey targets \citep[see also][]{liske+15}. Thanks to the large
number of visits to each patch of sky during the spectroscopic
campaign, the redshift completeness stays roughly constant with
nearest neighbour separation. The only residual effect is a tiny
reduction of the completeness by $\sim$$0.5$\% at nearest neighbour
distances of $\sim$$10$ and $\sim$$70$~arcsec.

In the rest of this section, we describe the various GAMA data
products that we make use of in our work. These include stellar masses
(Sect.~\ref{2.1}), the group catalogue (Sect.~\ref{2.2}) and the
filament catalogue (Sect.~\ref{2.3}).

\subsection{Stellar masses}\label{2.1}
For the stellar mass measurements and uncertainties we make use of the
table \textit{StellarMassesLambdarv24} \citep{taylor+11}, which
provides stellar masses, restframe photometry, and other ancillary
stellar population parameters from stellar population fits to
multi-band spectral energy distributions for all objects across the
five GAMA survey regions. The values in this catalogue were derived
using a `737' cosmology and distances that were computed using the
flow-corrected redshifts derived from the \citet{tonry+00} flow model
for very low redshifts, and then tapering to a CMB-centric frame for
$z>0.03$, as described by \citet{baldry+12}. The broadband photometry
that was modelled to derive the stellar masses (in the restframe
wavelength range of $0.3$ -- $1.1$~$\mu$m) is the 21-band matched
aperture photometry presented by \citet[and published in the table
  \textit{LambdarCatv01}]{wright+16}. We corrected the masses derived
from this aperture photometry for the light lying outside of the
apertures as described by \citet{taylor+11}. However, following the
approach of \citet{vazquez+20}, we only applied this correction if the
correction factor is in the range $0.8$ -- $10$; otherwise, no
correction is applied. We note that we repeated our analyses with
different limits for the correction factor and found that our results
do not change qualitatively.

We point out that since the inception of the present work, the GAMA
collaboration has updated both their preferred multi-band photometry
[now derived using the code \textsc{ProFound} \citep{robotham+18} and
  published by \cite{bellstedt+20}] as well as their preferred method
of deriving stellar masses [now using the code \textsc{ProSpect}
  \citep{robotham+20}]. As shown by \cite{robotham+20}, the joint net
effect of these changes is a global shift of the stellar masses by
$+0.1$~dex (with a scatter of $0.11$~dex), approximately independent
of stellar mass. This shift is of course irrelevant when performing
purely internal comparisons of the GSMF as a function of different
environmental properties (Sect.~\ref{6}) but should be kept in mind
when comparing our GSMF to literature values (Sect.~\ref{7}).

\subsection{Group catalogue}\label{2.2}
 The GAMA galaxy group catalogue (G$^{3}$C)\footnote{We note for
 completeness that the cosmological parameters used in the
 construction of the G$^3$C differed somewhat from the model used
 here: $\Omega_{\rm M} = 0.25$, $\Omega_{\rm \Lambda} = 0.75$, $H_{0}
 = 100$ km s$^{-1}$ Mpc$^{-1}$, corresponding to the cosmology of the
 Millennium $\Lambda$CDM N-body simulations that were used to
 construct the mock GAMA light cones.} is one of the major data
 products of the GAMA project \citep{robotham+11}. The G$^{3}$C was
 constructed by employing a modified Friends-of-Friends (FoF) grouping
 algorithm that considers galaxies to be in a group if they are
 `close' both along the line of sight and when projected on to the
 sky; this successfully accounts for redshift space distortions caused
 by the peculiar velocities of galaxies in groups. This FoF algorithm
 was first run on mock GAMA light cones in order to tune the grouping
 parameters and to test the quality of the grouping, before applying
 it to the real GAMA data.

In particular, we make use of the following tables:
(i)~\textit{G3CGalv08}\footnote{Newer versions of this catalogue are
available, but to be more consistent with the filament catalogue (see
Sect.~\ref{2.3}) we use this slightly older version. In any case, the
differences between the version used here and the latest version are
irrelevant for this paper.\label{refnote}}, which lists the $184\,081$
galaxies on which the FoF algorithm was run. The exact selection of
these galaxies is given by: $nQ \geq 2$, SURVEY\_CLASS $\geq 4$, and
$0.003 < z_{\rm CMB} < 0.6$ \citep{baldry+12}, as taken from the
tables \textit{TilingCatv46} and \textit{DistancesFramesv14}.
(ii)~\textit{G3CFoFGroupv09}\footref{refnote}, which lists the
properties of the $23\,654$ groups comprising 2 or more members that
were identified among the galaxy sample above and which contain
$\sim$$40\%$ of these galaxies.

\subsection{Filament catalogue}\label{2.3}
The filament catalogue (FC)\footnote{The cosmological parameters used
in the construction of this catalogue were the same as those used in
the construction of the G$^{3}$C.} identifies the LSS in the three
equatorial survey regions \citep{alpaslan+14}. Using a volume-limited
subsample of the G$^{3}$C with a redshift cut of $z \leq 0.213$ and an
absolute magnitude cut of $M_r = -19.77 + 5 \log\,h_{100}$, and
discarding all groups with fewer than two members remaining after
these cuts, \cite{alpaslan+14} used the remaining groups as nodes to
generate a minimal spanning tree, thus identifying $643$ individual
filaments spanned by $5152$ groups. The $45\,468$ galaxies of their
subsample were then classified as belonging to filaments, tendrils, or
voids. However, since we do not want to restrict our analysis to this
volume-limited subsample, we do not make use of these classifications
here. Instead, we only use the group-defined filaments and make use of
our own environmental classification as described in Sect.~\ref{4.1}.

The filaments identified in this catalogue are composed of an average
of eight groups and span up to $100 \, h_{100}^{−1}$~Mpc. In
particular, groups in filaments are connected through straight lines
called links. All groups that are in the same set of unbroken links
are considered to be part of the same filament. Additional
substructures are defined within a filament. The backbone refers to
the combination of the two paths with the most links that begin from
the edges of the filament and meet at its centre, defined as the group
which is furthest away from all edges. In other words, the backbone is
the longest path that travels from one end of a filament to the other
through its most central group. All other paths that are connected to
the backbone are referred to as branches, and their order determines
how close they are to the backbone. A branch of order $n$ always
connects to a branch of order $n-1$, which connects to a branch of
order $n-2$, and so on, where the backbone is considered a branch of
order $1$.

\section{Sample selection}\label{3}
In this section we describe the selection defining our parent sample
of galaxies. The final selection, which results from the application
of the redshift-dependent stellar mass limit, is described in
Sect.~\ref{smcl}.

The spectroscopic GAMA observations targeted objects with
dust-corrected Petrosian SDSS Data Release 7 (DR7) $r$-band magnitudes
of $r < 19.8$~mag \citep{baldry+10}. The reference table
\textit{TilingCatv46} defines the class of each survey target
(SURVEY\_CLASS) as well as various redshift quality parameters ($nQ$
and $nQ2\_{\rm FLAG}$, which records the result of tests of $nQ = 2$
redshifts against independent measurements). Our initial sample
selection was defined as follows:

\begin{enumerate}
    \item[(i)] Survey regions G09, G12, and G15;
    \item[(ii)] $r<19.8$ mag;
    \item[(iii)] SURVEY\_CLASS $\geq 4$ to select only main survey
      targets, excluding additional filler targets from the sample;
    \item[(iv)] $nQ \geq 3$ or ($nQ = 2$ and $nQ2\_{\rm FLAG} \geq 1$)
      to select targets with reliable redshifts.
\end{enumerate}

Next, because we make use of the filaments identified by the FC, we
applied the same upper redshift cut of $z = 0.213$. Because the
various GAMA data products above used slightly different versions of
the redshift data (i.e.\ \textit{TilingCatv46}), small inconsistencies
arose between these data products. As a result, our sample does not
include one FC group because it was left with no members after our
selection criteria were applied. Furthermore, $1088$ galaxies
catalogued in the G$^{3}$C with $nQ = 2$ are not included in our
sample because of our additional requirement of $nQ2_{\rm FLAG} \geq
1$ (see above). We note that the redshift cut applied above is based
on different measurements: For ungrouped galaxies, we considered the
CMB frame redshift $z_{\rm CMB}$ coming from the table
\textit{DistanceFramev14}; for a grouped galaxy, we just took the
median redshift $z_{\rm FOF}$ of the galaxy's group coming from
\textit{G3CFoFGroupv09}. Therefore, either all or none of the galaxies
in a group are considered as being part of our sample.

\begin{figure}
\centering
\includegraphics[width=0.53\textwidth]{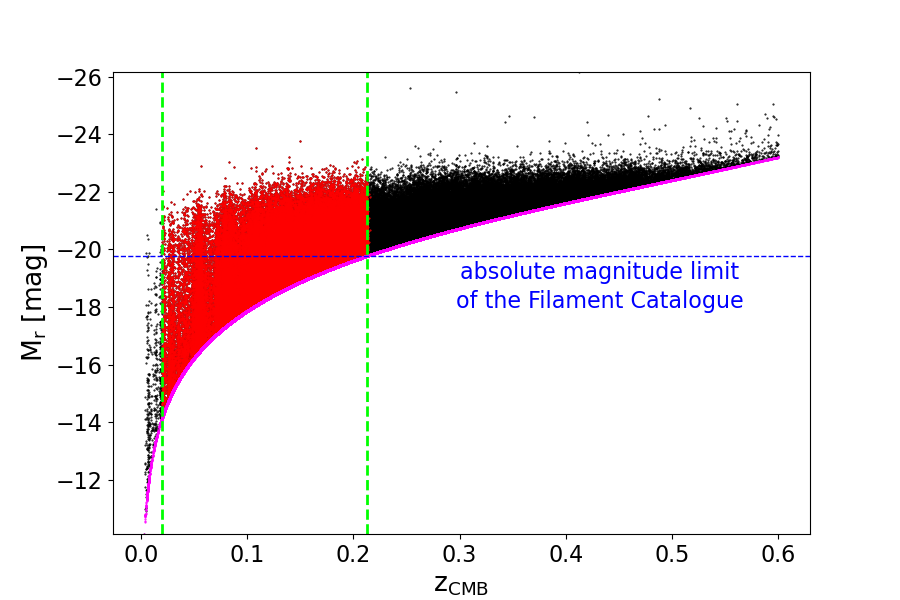}
\caption{Distribution of the absolute $r$-band magnitude as a function
  of redshift for all the galaxies in \textit{G3CGalv08} (black
  points). The magenta curve represents our selection function and
  gives the faintest possible galaxy that is visible in GAMA at each
  redshift, given our apparent magnitude limit of $r = 19.8$~mag. Red
  points show our parent galaxy sample after our initial selection
  criteria were applied. The two dashed green lines determine our
  redshift limits (see Sect.~\ref{lowz} for a discussion of the lower
  redshift cut). The dashed blue line shows the absolute magnitude cut
  applied by \citet{alpaslan+14}.}
\label{fig:sampleselection}
\end{figure}

In Fig.~\ref{fig:sampleselection} we show the distribution of the
absolute $r$-band magnitude as a function of redshift for all the
galaxies in \textit{G3CGalv08} (black dots), with the red dots
highlighting our parent galaxy sample after our initial selection
criteria were applied. To determine the faintest possible galaxy that
is visible in GAMA at each redshift (shown as a magenta curve in
Fig.~\ref{fig:sampleselection}), we first calculated the distance
modulus $DM$ as a function of redshift using the cosmological
luminosity distance $D_L$ (in units of $h_{70}^{-1}$~Mpc):
\begin{equation}
    DM = 5\log (D_L) + 25,
\end{equation}
with $D_L = (1 + z) R_0 S_k(r)$. Here, $R_0 S_k(r)$ refers to the
radial comoving distance. We then calculated the absolute magnitude of
the faintest possible galaxy that can be seen within the GAMA survey
given our apparent magnitude limit of $r=19.8$~mag and using the
$k$-correction from \cite{robotham+11} as
\begin{equation}
    M_r(z) = 19.8 − DM(z) − k(z).
\end{equation}
This equation describes our initial selection function, that is, a
function that precisely bounds our parent sample data. Unlike
\cite{alpaslan+14}, we decided not to apply an absolute magnitude cut
here in order to be able to probe the GSMFs more deeply. Our parent
galaxy sample now contains a total of $89\,451$ galaxies across G09,
G12, and G15 and $11\,820$ groups, $5151$ of which are part of a
filament.

We note that some black dots in Fig.~\ref{fig:sampleselection} are
actually still present below the upper redshift cut: these galaxies
have $nQ = 2$ and $nQ2\_{\rm FLAG} = 0$. Also, some red dots lie above
the same cut. These $69$ galaxies have higher redshifts than $0.213$
but belong to a group with a median redshift below this value.
According to our selection process, the reverse is also possible but
less clear from the plot. Fifty-two galaxies are not included in our
final sample, even though they meet all the other requirements,
because the group they belong to has a median redshift higher than
$0.213$.

In the rest of this section we now describe two small additional
restrictions that have to be applied to our sample.

\begin{table*}
\caption{Our parent sample selection process.}
\begin{center}
\begin{tabular}{c|c|c|c}
\hline
Selection step & GAMA table & No of galaxies & No of groups (in filaments) \\
\hline
\hline 
$r<19.8$ mag & \textit{TilingCatv46} & $184\,081$ & $23\,654$ ($5152$) \\
SURVEY$_{\rm CLASS} \geq 4$ & \textit{TilingCatv46} & & \\
$nQ \geq3$ or ($nQ = 2$ and $nQ2_{\rm FLAG} \geq 1$) & \textit{TilingCatv46} & $182\,993$ & $23\,652$ ($5151$) \\
\hline
$z_{\rm CMB} \leq 0.213$ & \textit{DistancesFramesv14} & & \\
$z_{\rm FOF} \leq 0.213$ & \textit{G3CFoFGroupv09} & $89\,451$ & $11\,820$ ($5151$) \\
\hline
Stellar mass availability & \textit{StellarMassesLambdarv24} & $89\,406$ & $11\,820$ ($5151$) \\
PPP $\geq 0.01$ & \textit{StellarMassesLambdarv24} & $88\,581$ & $11\,789$ ($5142$)\\
\hline
$z_{\rm CMB} \geq0.02$ & \textit{DistancesFramesv14} & $88\,093$ & $11\,725$ ($5142$) \\
\hline
\end{tabular}
\tablefoot{For each step in the selection, we list the selection
  criterion, the GAMA table used in this step, and the numbers of
  remaining galaxies and groups after the selection was applied.}
\end{center}
\label{tab:ss}
\end{table*}

\subsection{Additional sample selection}
\subsubsection{Stellar mass availability}\label{3.1.1}
A small mismatch of $45$ galaxies between our spectroscopic and the
available stellar mass sample described in Sect.~\ref{2.1}, due to
different selection processes, reduces our sample to $89\,406$
galaxies. Furthermore, the stellar mass catalogue provides the
Posterior Predictive P-value (PPP) as a goodness of fit statistic,
which can be roughly thought of as a Bayesian analogue to the
frequentist reduced $\chi^2$. The PPP is a useful tool for checking
whether or to what extent the model is consistent with the data. Given
the observed data and the posterior probability density function on
the properties of the model, the PPP quantifies the fraction of future
observations that would be predicted to be as discrepant as the
observed data. Here, we impose a lower limit of PPP $\geq 0.01$ to
discard those galaxies where the data invalidate even the `most
likely' model at $99\%$ confidence. After the PPP cut, our galaxy
sample consists of $88\,581$ objects.

\begin{figure}
\centering
\includegraphics[width=0.45\textwidth]{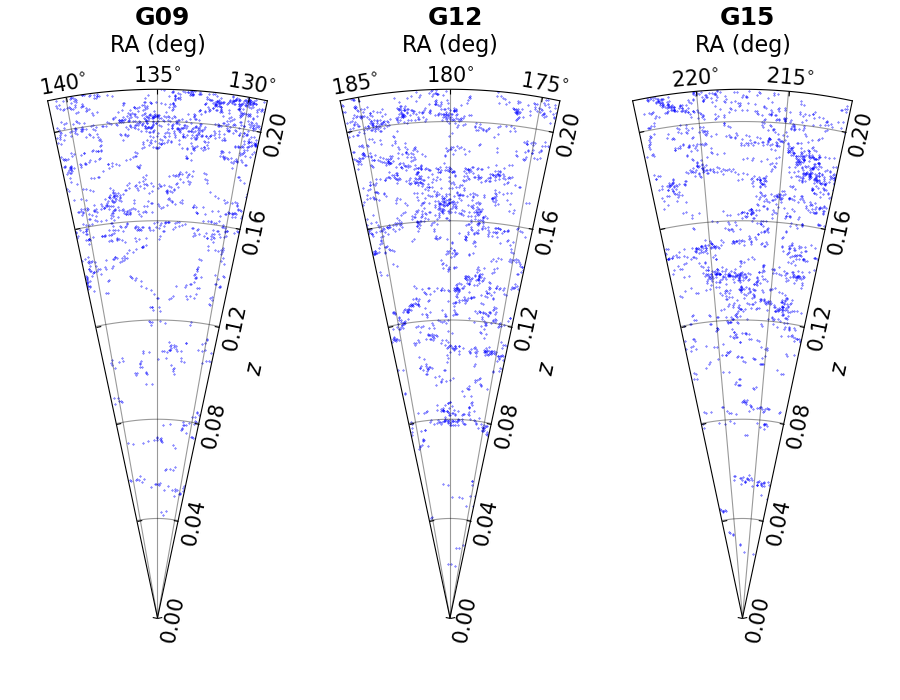}
\caption{Light cones of the three GAMA equatorial survey regions
  showing our FC group sample out to $z = 0.213$ (blue dots). All
  three cones span the full $5^{\circ}$ declination range.}
\label{fig:lightcones1}
\end{figure}

\subsubsection{Lower redshift limit}\label{lowz}
The three panels of Fig.~\ref{fig:lightcones1} show the distribution
of our groups in the filaments across the three GAMA equatorial
fields. As one can see, no filament is found at redshifts $z <
0.02$. Below this redshift, the volume is simply not large enough to
accommodate a typical filament, and the environment of a galaxy cannot
be reliably determined. We thus imposed a lower redshift limit of $z =
0.02$ on our sample (Fig.~\ref{fig:sampleselection}, left dashed green
line), in addition to the upper limit of $z = 0.213$ already applied
above. Our final parent sample thus consists of a total of $88\,093$
galaxies and $11\,725$ groups, $5142$ of which are part of a filament.

We summarise the complete selection process of the parent sample in
Table~\ref{tab:ss}. The final step of the selection process, that is,
the application of the redshift-dependent stellar mass limit, is
described in Sect.~\ref{smcl}.

\section{Environmental properties}\label{4}
In this section, we present four different environmental properties we
make use of in this work. These are the orthogonal distance to
the nearest filament $D_{\perp,\rm min}$ (Sect.~\ref{4.1}), group
membership (Sect.~\ref{4.2}), group halo mass $M_{\rm halo}$
(Sect.~\ref{4.3}), and a combination of the group's branch order (BO)
and the group's number of connecting links ($N_{\rm links}$;
Sect.~\ref{4.4}).

\subsection{Orthogonal distance to nearest filament}\label{4.1}
\begin{figure}
\centering
\includegraphics[width=0.5\textwidth]{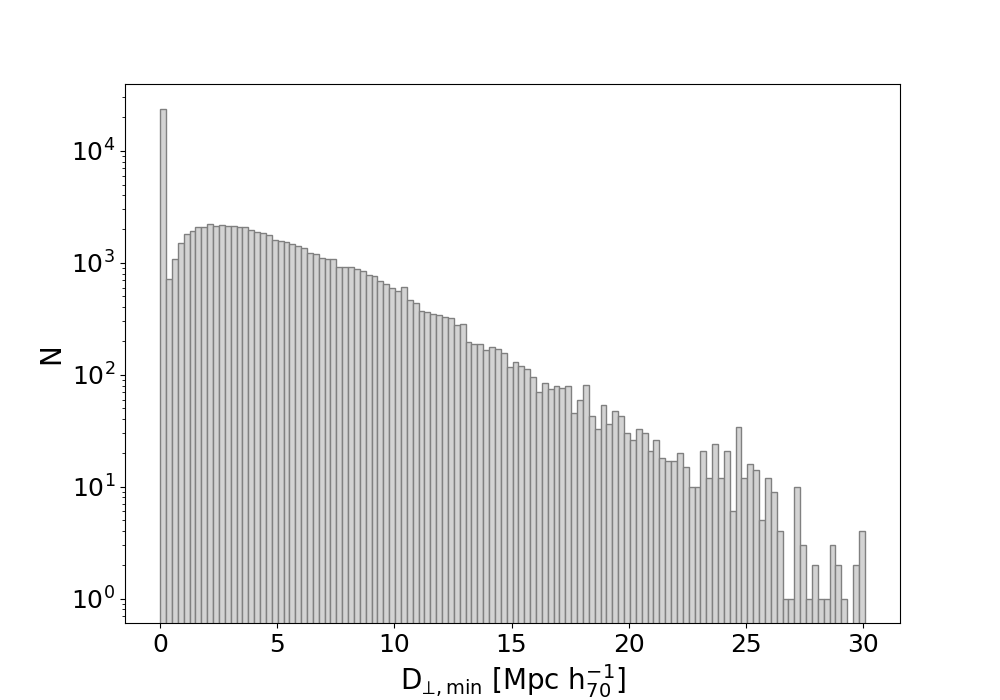}
\caption{Distribution of the orthogonal distance of a galaxy to its
  nearest filament. The galaxies in subsample G1 (galaxies in groups
  that are part of a filament) are responsible for the bump at $0$~Mpc
  since they are assigned a value of $D_{\perp,\rm min}=0$~Mpc by
  definition.}
\label{fig:distr}
\end{figure}

As discussed in Sect.~\ref{2.3}, from \cite{alpaslan+14} we only use
the filaments identified in the FC but we do not use their galaxy
environmental classification. Therefore, we provided our own
classification of whether a galaxy is part of a filament or a void,
which is simply based on the distance of a galaxy to the nearest
filament. In particular, for each galaxy we consider the orthogonal
distance to its nearest filament, $D_{\perp,\rm min}$, as our first
environmental measure. Our sample consists of $88\,093$ galaxies,
$39\,341$ of which are assigned to a group and $48\,752$ of which are
ungrouped (U subsample hereafter); of the grouped galaxies, $23\,657$
galaxies belong to groups in filaments (G1 subsample hereafter) and
the remaining $15\,684$ galaxies to groups which are not part of a
filament (G2 subsample hereafter). All galaxies in G1 were assigned
$D_{\perp,\rm min}=0$~Mpc by definition. For galaxies in U and G2, we
first calculated their 3D Cartesian coordinates and then determined
the values of $D_{\perp,\rm min}$. We note that to diminish the
effects of redshift space distortions in this step of the process, the
Cartesian coordinates of galaxies in G2 belonging to the same group
were derived using the RA and Dec values of the iterative central
galaxy as well as the median redshift of the group as stated in
Sect.~\ref{3}; in this way, all galaxies in a given group ended up
with the same Cartesian coordinates. In Fig.~\ref{fig:distr} we show
the distribution of $D_{\perp,\rm min}$ for our galaxy sample. As
expected, this distribution peaks at $D_{\perp,\rm min} \approx
2$--$3~h_{70}^{-1}$~Mpc and then decreases exponentially. The galaxies
in G1 are responsible for the bump at $D_{\perp,\rm min}=0$~Mpc.

\subsection{Group membership}\label{4.2}
A galaxy's membership of a group or not, as defined by the G$^{3}$C in
Sect.~\ref{2.2}, represents our second environmental property. As
described above, this categorical classification comprises $39\,341$
grouped galaxies (G1 $+$ G2 subsamples) and $48\,752$ ungrouped
galaxies (U subsample).

\subsection{Group halo mass}\label{4.3}
As in \cite{vazquez+20}, we make use of two different methods for
estimating a group's halo mass $M_{\rm halo}$, which represents our
third environmental measure. The first method provides a dynamical
mass estimate, $M_{\rm dyn}$, based on the virial theorem and the
galaxy dynamics within each group; the second provides a
luminosity-based mass estimate, $M_{\rm lum}$, using the group's total
$r$-band luminosity and the weak-lensing calibrated $M$-$L$ relation
of \citet[Eq. 37]{viola+15}. As discussed in \cite{robotham+11}, the
total group $r$-band luminosity is not simply the sum of the
luminosities of all galaxies belonging to that group, but is instead
corrected for the fraction of light in galaxies below the survey
magnitude limit using the global GAMA luminosity function. Since most
of the luminosity is contributed by galaxies around $L^\star$, and
since most GAMA groups are sampled well beyond $L^\star$ due to the
survey's depth of $r < 19.8$~mag, these corrections typically amount
to only a few percent.

The G$^{3}$C provides the dynamical group halo masses, both using a
constant calibration factor of $A = 10.0$ required to get a median
unbiased mass estimate (column \texttt{MassA}), as well as using a
calibration factor that is a function of group multiplicity and
redshift (column \texttt{MassAfunc}). Also listed are the group
$r$-band luminosities integrated down to an absolute magnitude of $M_r
- 5 \log h_{100} = -14$, again both considering a constant calibration
factor of $B = 1.04$ for the median unbiased luminosity estimate
(\texttt{LumB}), and its functional form (\texttt{LumBfunc}). See
Sects.~4.3 and 4.4 of \cite{robotham+11} for detailed explanations of
these different calibration factors.

In Fig.~\ref{fig:corr1} we show the halo mass distribution of our
group sample using both the dynamical estimate (black dashed line) and
the luminosity-based one (solid line). Our groups mainly reside in
halos of $10^{12}$--$10^{14}~M_{\odot}~h_{70}^{-1}$. Moreover, the
dynamical mass estimates have a broader distribution towards lower
masses compared to the luminosity-based ones, as already noted by
\cite{han+15} and \cite{vazquez+20}. We also show as blue and red
dashed lines the dynamical halo mass distributions of the groups that
define the filaments and those that lie outside of the
filaments. Although there is an almost complete overlap, these
distributions are nevertheless quite different, with the groups in
filaments clearly being more massive. In fact, the group sample is
dominated by the filament groups at halo masses $M_{\rm dyn} \ga
10^{13}~M_\odot~h_{70}^{-1}$, and by the groups outside of filaments
below that value.

\subsection{Group branch order and number of connecting links}\label{4.4}
In the FC, $656$ filaments spanned by $5152$ groups were identified
across the three equatorial GAMA regions. As explained in
Sect.~\ref{2.3}, a filament is defined as a collection of
branches. Each branch (and therefore each group within a branch as
well as each galaxy within those groups) can be assigned a branch
order (BO). In particular, groups with BO $=1$ define the backbone;
groups with BO $=2$ belong to a second-order branch (connected to the
backbone); and so on. The backbone and all the other branches of a
filament make up its morphology, with the backbone representing the
most central route through the filament, and the other branches being
paths that emerge from the backbone.

Apart from the order of its branch, the position of a group within its
filament can be further characterised by its total number of links to
neighbouring groups, $N_{\rm links}$. A group at the end of its branch
has $N_{\rm links} = 1$, a group in the middle of a branch has $N_{\rm
  links} = 2$, while a group belonging to multiple branches has
$N_{\rm links} > 2$. To obtain a comprehensive picture of how much a
group is embedded in a denser (or less dense) part of a filament, we
consider both its BO and its $N_{\rm links}$, the combination of which
represent our final environmental property.

\section{Method}\label{5}
In this section we describe the methods we used in our work: the
Modified Maximum Likelihood estimator for the construction of the
GSMFs (Sect.~\ref{5.1}), the stellar mass completeness limit for the
derivation of the selection function (Sect.~\ref{smcl}) and the
random sample generation for the application of volume corrections to
the GSMFs (Sect.~\ref{5.3}).

\subsection{Modified maximum likelihood estimator}\label{5.1}
Massive galaxies are much rarer compared to low-mass ones. In a fixed
cosmic volume, the number of galaxies per unit mass decreases with
mass according to a power law until reaching a specific cut-off,
beyond which the number density falls off exponentially. To
characterise this behaviour, it is helpful to define the GSMF
$\phi(M)$ as the density of galaxies per unit volume and per unit
stellar mass $M$. Specifically, in a given cosmic volume $dV$, the
expected number of galaxies within the interval $[M, M + dM]$ is given
by $dN = \phi(M) dV dM$. Analytical parametric functions intended to
match observed GSMFs can be expressed in the form
$\phi(M|\bm{\theta})$, where $\bm{\theta}$ denotes a vector of $P$
scalar model parameters. The Schechter function \citep{schechter+76}
represents the most well-known model accurately capturing the
truncated power-law behaviour:
\begin{equation}\label{schechter}
   \phi(M)dM = e^{-\frac{M}{M^{\star}}} \phi^{\star} \biggl(\frac{M}{M^{\star}}\biggr)^\alpha \frac{dM}{M^{\star}}.
\end{equation}
Here, $\phi^{\star}$ is the normalisation factor, $M^{\star}$ is the
mass at the normalisation point (i.e.\ near the exponential break),
and $\alpha$ is the faint-end slope parameter. However, the shape of
the GSMF is not always well represented by a single Schechter function
due to an often observed steepening below $10^{10}$~$M_{\odot}$,
giving rise to a double Schechter function \citep{baldry+08}:
\begin{equation}\label{doubleschechter}
    \phi(M)dM = e^{-\frac{M}{M^{\star}}} \Biggl( \phi_{1}^{\star} \biggl( \frac{M}{M^{\star}} \biggr)^{\alpha_{1}} + \phi_{2}^{\star} \biggl( \frac{M}{M^{\star}} \biggr)^{\alpha_{2}} \Biggr) \frac{dM}{M^{\star}} ,
\end{equation}
where $\phi_{1}^{\star}$, $\phi_{2}^{\star}$ and $\alpha_{1}$,
$\alpha_{2}$ describe the normalisation and slope parameters,
respectively, for the two components. Without loss of generality, we
can always choose $\alpha_{1} > \alpha_{2}$ such that the second term
in Eq. \ref{doubleschechter} dominates at lower masses
\citep{baldry+12}.

\begin{figure}
\centering
\includegraphics[width=0.5\textwidth]{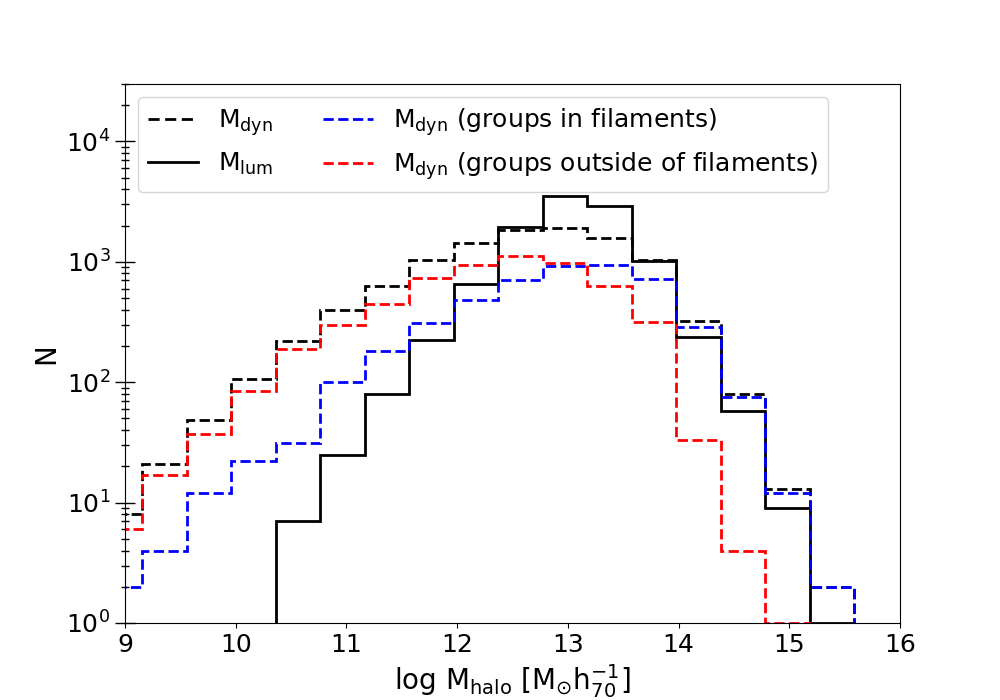}
\caption{Black solid and dashed lines show the distributions of the
  luminosity-based and dynamical halo mass estimates for our group
  sample, respectively. The blue and red dashed lines show the
  distributions of the dynamical halo mass estimates for those groups
  that are part of a filament and those that are not, respectively.}
\label{fig:corr1}
\end{figure}

\begin{figure*}
\centering
\includegraphics[width=0.49\textwidth]{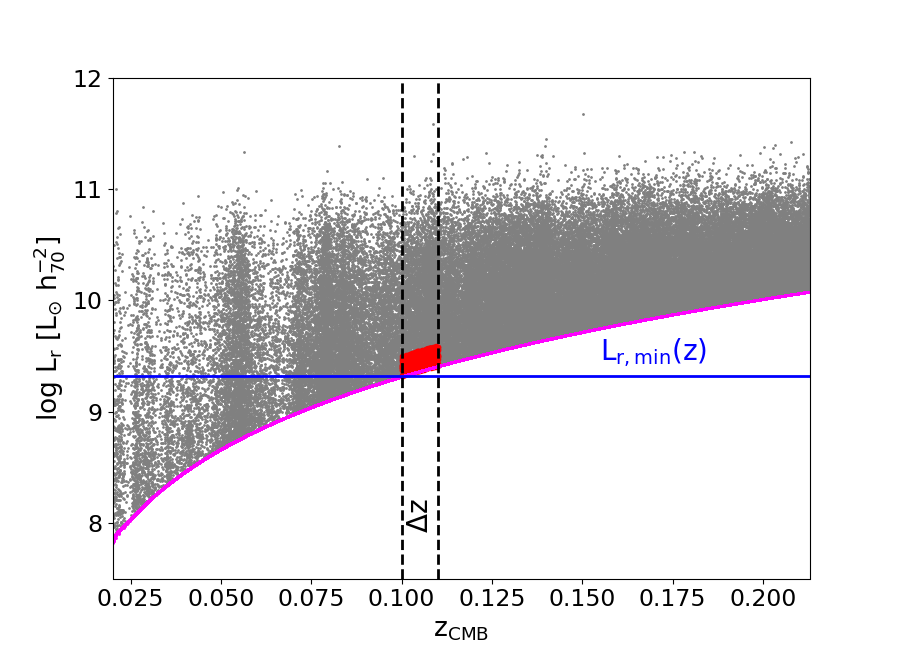}\quad\includegraphics[width=0.49\textwidth]{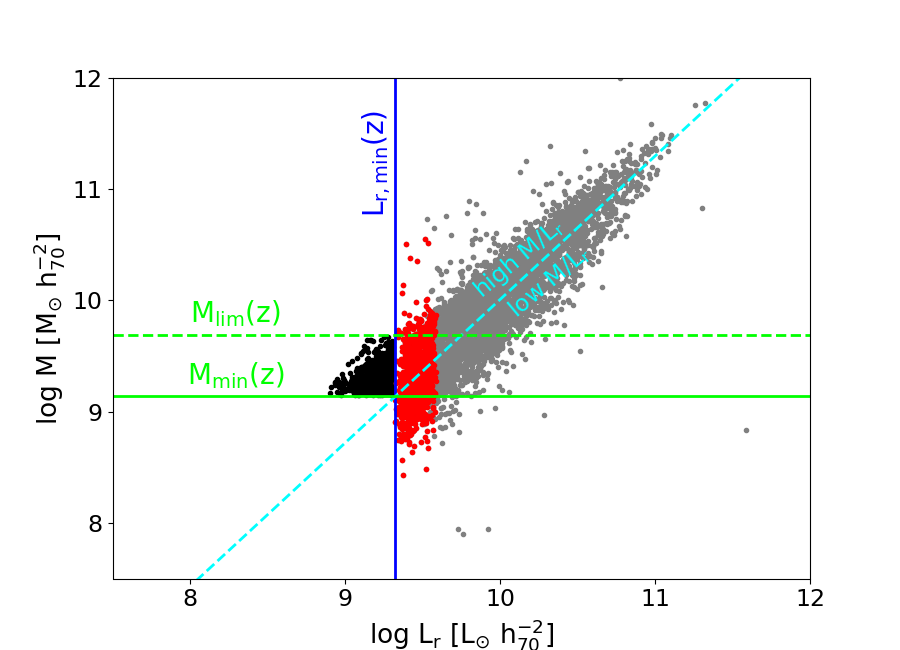}
\caption{Left-hand panel: Distribution of the total $r$-band
  luminosity as a function of redshift for our parent galaxy sample
  (grey dots). The magenta curve represents our selection function and
  gives the lowest luminosity of a galaxy for which a redshift could
  have been collected in GAMA, given our apparent magnitude limit of
  $r = 19.8$~mag. The two dashed black lines define a narrow redshift
  range $\Delta z$ for illustrative purposes. The solid blue line and
  the red box correspond to the luminosity limit and the $20\%$
  faintest galaxies associated with this redshift range,
  respectively. Right-hand panel: Stellar mass versus $r$-band
  luminosity for the galaxy subsample in the narrow redshift range,
  $\Delta z$, defined in the left-hand panel. The dashed cyan line
  gives the average correlation between the two galactic
  properties. Galaxies lying above this line have a higher than
  average $M/L_r$ and vice versa. The luminosity limit from the
  left-hand panel is also shown here in blue. The intersection of this
  limit with the average correlation defines the corresponding mass
  limit at this redshift shown as a solid green line. However, this
  mass limit would not in fact result in a complete sample as all
  galaxies to the left of the blue line, shown in black, would be
  missed. To be complete, it is necessary to move the mass limit up
  sufficiently such that (almost) no more galaxies are missed. The
  dashed green line thus gives the final stellar mass limit,
  representing the $95\%$ $M/L_r$ completeness limit of the $20\%$
  faintest galaxies at this redshift (see text).}
\label{fig:mcompleteness}
\end{figure*}

The most straightforward and intuitive technique for fitting a GSMF
model, as described in \cite{schmidt+68}, is to estimate the observed
space densities for different mass intervals of the data. This is
achieved by computing for each mass interval the ratio between the
number of detected galaxies and the maximum volume $V_{\rm max}$ in
which galaxies of that mass could have been observed. This procedure
is also known as $1/V_{\rm max}$ method. The model function
$\phi(M|\bm{\theta})$ is then fitted to these values. However, this
method has several drawbacks: the fitting process is influenced by the
division into arbitrary mass intervals; Poisson errors cannot be
assigned to non-detections (i.e.\ mass intervals with no galaxy); the
choice of $V_{\rm max}$ is sometimes uncertain due to complex
detection limits with source-dependent completeness; and systematic
errors can be introduced by the cosmic LSS. Just observing more
galaxies will not solve most of these limitations, which indeed remain
a pertinent issue for modern spectroscopic redshift surveys.

In this study, we hence employ the modified maximum likelihood (MML)
estimation, which was comprehensively documented by
\cite{obreschkow+18}. This method bypasses the need for data binning
and operates within a Bayesian framework tailored for fitting
distribution functions (e.g.\ GSMFs) to complex multi-dimensional
datasets. The MML framework meticulously takes into account the
observational measurement errors for individual objects, incorporates
complex observational selection functions, and provides the option to
internally correct for the underlying LSS detected within the survey
volume. The core of the MML approach consists of a fit-and-debias
procedure, an iterative fitting algorithm that iteratively solves a
standard maximum likelihood estimation, revising the data by
accounting for the previous fit and observational uncertainties. The
MML framework is accessible via \texttt{DFTOOLS}
\citep{obreschkow+18}, an open-source software package for the
\texttt{R} statistical programming language. \texttt{DFTOOLS} gives
the most likely solution and full co-variance matrix of the relevant
model parameters in order to derive volume-corrected binned mass
functions. In our analysis, we fit a double Schechter function
(Eq.~\ref{doubleschechter}), which has been demonstrated to
effectively address the notable upturn observed at intermediate
stellar masses \citep{baldry+08}.

\subsection{Stellar mass completeness limit}\label{smcl}
One of the most crucial steps in properly estimating the number
density $\phi (M)$ consists in the derivation of $M_{\rm lim}(z)$,
which represents the stellar mass limit above which our sample is
complete at a given redshift, and which depends both on redshift and
the stellar mass-to-light $M/L$ distribution of the sample, given our
flux-limited sample.

The determination of the stellar mass completeness limit is
particularly challenging for a flux-limited sample like ours, since a
sharp limit in luminosity does not correspond to a sharp limit in
stellar mass, as shown in Fig.~\ref{fig:mcompleteness}. Taking a
narrow redshift range (dashed black lines, left-hand panel) and simply
using as a mass limit (solid green line, right-hand panel) the mass
that corresponds to the luminosity limit and the average $M/L$
(respectively, solid blue and dashed cyan line, right-hand panel)
would result in an incomplete sample, since all the massive yet
lower-luminous sources falling below the absolute magnitude cutoff are
already missing from the sample. Therefore, it is necessary to move
the mass limit sufficiently upwards (dashed green line, right-hand
panel) to reduce the incompleteness to an acceptable level.

Different strategies for the estimation of $M_{\rm lim}(z)$ can be
found in the literature. For example, the approach presented in
\cite{dickinson+03}, \cite{fontana+06}, \cite{perezgonzalez+08} and
others rely on a single stellar population (SSP) paradigm; the minimum
stellar mass of a galaxy with magnitude equal to the magnitude limit
$r_{\rm lim}$ can be estimated by scaling the flux of a synthetised
spectrum of a passively evolving SSP formed at high redshift. The
technique introduced by \cite{marchesini+09} considers deeper survey
data, where the flux and mass of each object are scaled to match
$r_{\rm lim}$; in this context, the most massive sources have the
lowest $M/L$ and can therefore be excluded by simply applying $r_{\rm
  lim}$ as a cut. \cite{quadri+12} and \cite{tomczak+14} use a
slightly modified version of this approach, by considering the objects
above the flux completeness and scaling their masses and fluxes down
to $r_{\rm lim}$; the upper limit of those scaled-mass values defines
their mass completeness limit as a function of redshift.

\begin{figure}
\centering
\includegraphics[width=0.53\textwidth]{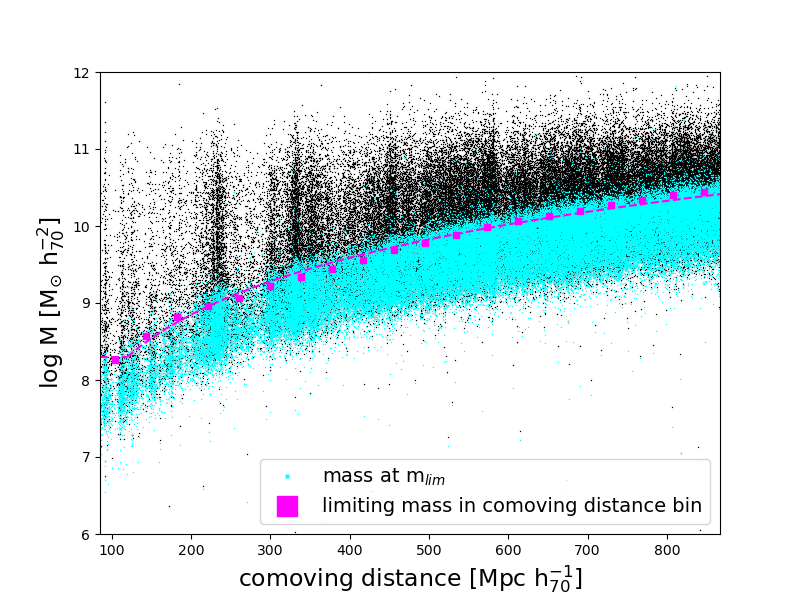}
\caption{Determination of the stellar mass completeness limit as a
  function of comoving distance for our galaxy sample (black
  dots). Due to the range in $M/L_r$, constraining the mass
  completeness limit of a flux limited sample is not straightforward.
  By keeping the $M/L_r$ and the redshift of each source constant, we
  determined the stellar mass that each object would have if its flux
  was equal to the flux limit. These limiting mass values are shown in
  cyan, with higher dots at a given redshift having larger $M/L_r$
  values. We then divided our sample into bins of comoving distance,
  sorted the sources in each bin according to their luminosity,
  selected the faintest $20\%$, and determined the stellar mass below
  which $95\%$ of these faint objects lie \citep{pozzetti+10}. We
  repeated this procedure for each bin and thus obtained the limiting
  mass values that are shown as magenta squares. The mass completeness
  function $M_{\rm lim}(d)$, shown as the magenta dashed line, was then
  estimated by fitting an exponential curve to these limiting mass
  value in each comoving distance bin.}
\label{fig:mlim}
\end{figure}

In this work, we followed the approach presented by \cite{pozzetti+10}
and illustrated in Fig.~\ref{fig:mlim}. For each galaxy in our
sample, we first considered $M_{{\rm lim},i}$, that is, the mass that
source $i$ would have if it had the same $M/L_r$ but a magnitude equal
to our spectroscopic magnitude limit of $r_{\rm lim} = 19.8$~mag. In
Fig.~\ref{fig:mlim} we show the resulting distribution in $M_{{\rm
    lim},i}$ which reflects that of the $M_i/L_{r,i}$ at each redshift
in our sample. In other words, $M_{{\rm lim},i}$ is $M/L_r$
dependent. Under the assumption of a constant $M/L_r$, $M_{\rm lim,i}$
is simply given by
\begin{equation}\label{mlim}
    \log (M_{{\rm lim},i}) = \log (M_i) + 0.4 (r_i - r_{\rm lim}).
\end{equation}
Next, we divided our sample into 20 bins of comoving distance $d$ and
then, in each bin, sorted all sources by magnitude, selecting the
$20\%$ faintest galaxies (shown in red in
Fig.~\ref{fig:mcompleteness}). For each bin, we defined $M_{\rm lim}$
as the upper envelope of the $M_{\rm lim,i}$ distribution, below which
$95\%$ of these faint objects lie. The function $M_{\rm lim}(d)$ was
then determined by fitting an exponential curve to the estimated mass
limits in each bin defined as
\begin{equation}
  \log [M_{\rm lim} / (M_\odot\,h_{70}^{-2})] (d) = a \log [d / ({\rm
      Mpc}\,h_{70}^{-1})] + b ,
\end{equation}
with $a$ and $b$ being the free parameters. The fit resulted in $a =
1.0668$ and $b = 3.1956$. This $M_{\rm lim}(d)$ therefore corresponds
to the $95\%$ completeness limit of the $M/L_r$ distribution of the
$20\%$ lowest-luminosity galaxies at each redshift and represents the
selection function for our sample (shown as the magenta dashed line in
Fig.~\ref{fig:mlim}).

Unavoidably, however, the (small) remaining incompleteness of the
sample thus selected is not random. Instead, it depends on luminosity
and $M/L_r$, with faint, high-$M/L_r$ galaxies most affected. Since
any subsample of galaxies that we might choose to select below may
have an $L_r$-$M/L_r$ distribution that is slightly different from
that of the full sample (at a given redshift), the incompleteness of
this subsample will also be slightly different from that of the full
sample. This could be avoided by re-determining $M_{\rm lim}$
individually for each subsample. However, this would result in the
union of a complete set of subsamples not necessarily being identical
to the full sample, and these subsamples covering slightly different
volumes, which may in turn result in slightly different LSS
corrections. In practice, however, we found that the two methods
produce essentially the same results for all of our subsamples. For
simplicity, we hence use the same selection function for all
subsamples.

Finally, we amended our selection function at the lowest redshifts by
requiring $\log [M / (M_\odot \, h_{70}^{-1})] > 8.3$
(cf.\ Fig.~\ref{fig:mlim}) because we found our mass function
measurements below this value to be unreliable for many of the
subsamples that we investigate in Sect.~\ref{6} below.

\begin{table}
\caption{Fraction of the total survey volume occupied by filaments and
  voids (i.e.\ everything outside the filaments) as a function of
  the assumed filament radius.}
\begin{center}
\begin{tabular}{c|c|c}
\hline
Filament radius & Fractional volume &  Fractional volume\\
  & of filaments & of voids \\
($h^{-1}_{70}$ Mpc)  & (\%)  & (\%)\\
\hline
\hline
3 & 13 & 87\\
4 & 23 & 77\\
5 & 34 & 66\\
6 & 45 & 55\\
7 & 56 & 44\\
\hline
\end{tabular}
\end{center}
\label{tab:distances}
\end{table}

At this point, the sample was restricted to the galaxies that lie
above this curve. This resulted in our final sample of $52\,089$
galaxies ($59\%$ of the parent sample defined in Sect.~\ref{3}) and
$10\,429$ groups. We note that when referring to any subsamples below,
in particular the samples G1, G2 and U defined in Sect.~\ref{4.1}
above, these samples should be understood to have had the selection
function shown in Fig.~\ref{fig:mlim} applied to them.

To measure a given sample's GSMF, we input the selected galaxies’
comoving distances, stellar masses, stellar mass errors, selection
function, and desired functional form to fit (i.e.\ double Schechter)
into the \texttt{DFTOOLS} routine \texttt{DFFIT}. We note that the
code computes both the functional fit as well as the full covariance
matrix for the fitted parameters.

\begin{table*}
\caption{Best-fit double Schechter function parameters for the mass
  functions in voids (left) and filaments (right), when considering
  our entire galaxy sample.}
\begin{center}
\begin{tabular}{c|c|c||c|||c||c|c|c}
\hline
$\log M^{\star}$ & $\alpha_{1}$ & $\alpha_{2}$ & Void & Filament & $\log M^{\star}$ & $\alpha_{1}$ & $\alpha_{2}$\\
$(M_{\odot} h_{70}^{-2})$ &  & & sample & sample & $(M_{\odot} h_{70}^{-2})$ &  &\\ 
\hline
\hline 
10.52 $\pm$ 0.01 & $-$0.49 $\pm$ 0.07 & $-$1.51 $\pm$ 0.05 & $D_{\perp,\rm min}>3$ & $D_{\perp,\rm min}\leq3$ & 10.82 $\pm$ 0.01 & $-$0.83 $\pm$ 0.03 & $-$1.75 $\pm$ 0.11 \\
10.50 $\pm$ 0.02 & $-$0.43 $\pm$ 0.08 & $-$1.51 $\pm$ 0.05 & 
$D_{\perp,\rm min}>4$ & $D_{\perp,\rm min}\leq4$ & 10.81 $\pm$ 0.01 & $-$0.86 $\pm$ 0.02 & $-$1.77 $\pm$ 0.11\\
10.51 $\pm$ 0.02 & $-$0.49 $\pm$ 0.08 & $-$1.54 $\pm$ 0.06 & 
$D_{\perp,\rm min}>5$ & $D_{\perp,\rm min}\leq5$ & 10.80 $\pm$ 0.01 & $-$0.85 $\pm$ 0.02 & $-$1.75 $\pm$ 0.09\\
10.50 $\pm$ 0.02 & $-$0.47 $\pm$ 0.09 & $-$1.54 $\pm$ 0.06 & 
$D_{\perp,\rm min}>6$ & $D_{\perp,\rm min}\leq6$ & 10.79 $\pm$ 0.01 & $-$0.86 $\pm$ 0.02 & $-$1.75 $\pm$ 0.09\\
10.49 $\pm$ 0.02 & $-$0.38 $\pm$ 0.11 & $-$1.50 $\pm$ 0.06 &
$D_{\perp,\rm min}>7$ & $D_{\perp,\rm min}\leq7$ & 10.78 $\pm$ 0.01 & $-$0.87 $\pm$ 0.02 & $-$1.78 $\pm$ 0.09\\ 
\hline
\end{tabular}
\tablefoot{The $D_{\perp,\rm min}$ limits are given in units of
  Mpc~$h_{70}^{-1}$.}
\end{center}
\label{tab:dmin1}
\end{table*}

\begin{table*}
\caption{Same as Table~\ref{tab:dmin1} but now discarding all the
  grouped galaxies in filaments from our filament samples.}
\begin{center}
\begin{tabular}{c|c|c||c|||c||c|c|c}
\hline
$\log M^{\star}$ & $\alpha_{1}$ & $\alpha_{2}$ & Void & Filament & $\log M^{\star}$ & $\alpha_{1}$ & $\alpha_{2}$\\
$(M_{\odot} h_{70}^{-2})$ &  & & sample & sample & $(M_{\odot} h_{70}^{-2})$ &  &\\ 
\hline
\hline 
10.52 $\pm$ 0.01 & $-$0.49 $\pm$ 0.07 & $-$1.51 $\pm$ 0.05 & $D_{\perp,\rm min}>3$ & $D_{\perp,\rm min}\leq3$ & 10.49 $\pm$ 0.02 & $-$0.34 $\pm$ 0.11 & $-$1.45 $\pm$ 0.07 \\
10.50 $\pm$ 0.02 & $-$0.43 $\pm$ 0.08 & $-$1.51 $\pm$ 0.05 & 
$D_{\perp,\rm min}>4$ & $D_{\perp,\rm min}\leq4$ & 10.52 $\pm$ 0.02 & $-$0.47 $\pm$ 0.09 & $-$1.46 $\pm$ 0.06\\
10.51 $\pm$ 0.02 & $-$0.49 $\pm$ 0.08 & $-$1.54 $\pm$ 0.06 & 
$D_{\perp,\rm min}>5$ & $D_{\perp,\rm min}\leq5$ & 10.51 $\pm$ 0.02 & $-$0.41 $\pm$ 0.08 & $-$1.45 $\pm$ 0.05\\
10.50 $\pm$ 0.02 & $-$0.47 $\pm$ 0.09 & $-$1.54 $\pm$ 0.06 & 
$D_{\perp,\rm min}>6$ & $D_{\perp,\rm min}\leq6$ & 10.51 $\pm$ 0.01 & $-$0.44 $\pm$ 0.07 & $-$1.47 $\pm$ 0.05\\
10.49 $\pm$ 0.02 & $-$0.38 $\pm$ 0.11 & $-$1.50 $\pm$ 0.06 &
$D_{\perp,\rm min}>7$ & $D_{\perp,\rm min}\leq7$ & 10.52 $\pm$ 0.01 & $-$0.48 $\pm$ 0.07 & $-$1.50 $\pm$ 0.05\\ 
\hline
\end{tabular}
\end{center}
\label{tab:dmin2}
\end{table*}

\subsection{Random sample generation}\label{5.3}
The general mass function of filament galaxies just considers all the
galaxies in filaments, using the total survey volume, and describes
the overall average density of galaxies living in filaments. In
contrast, the 'conditional' GSMF (cGSMF hereafter) represents the
local density of galaxies in filaments, given that they are a member
of a filament. To see how the local density of galaxies changes as a
function of environment, we need to know which fraction of the total
survey volume is occupied by filaments (assuming a certain radius) and
voids. To this end, we distribute a random sample of points in the
total survey volume and then ask which fraction of this sample lies
within various environments. The density of this random sample is
determined by the requirement to have the volume of a sphere with the
minimum filament radius considered in our analysis
(i.e.\ $3$~Mpc~$h_{70}^{-1}$), probed by at least $30$ random
points. This leads us to a total number of $\sim$$3$ billion random
points spread across the three equatorial regions.

In Table~\ref{tab:distances} we provide the fractional volumes of both
filaments and voids as a function of filament radius. We note that
\cite{alpaslan+14} used a radius of $4.13$~Mpc~$h_{100}^{-1}$ to
divide galaxies into filament and void galaxies in the FC.

\begin{figure*}
\captionsetup[subfigure]{labelformat=empty}
    \centering
    \begin{subfigure}[b]{0.47\textwidth}
        \centering
        \includegraphics[width=\textwidth]{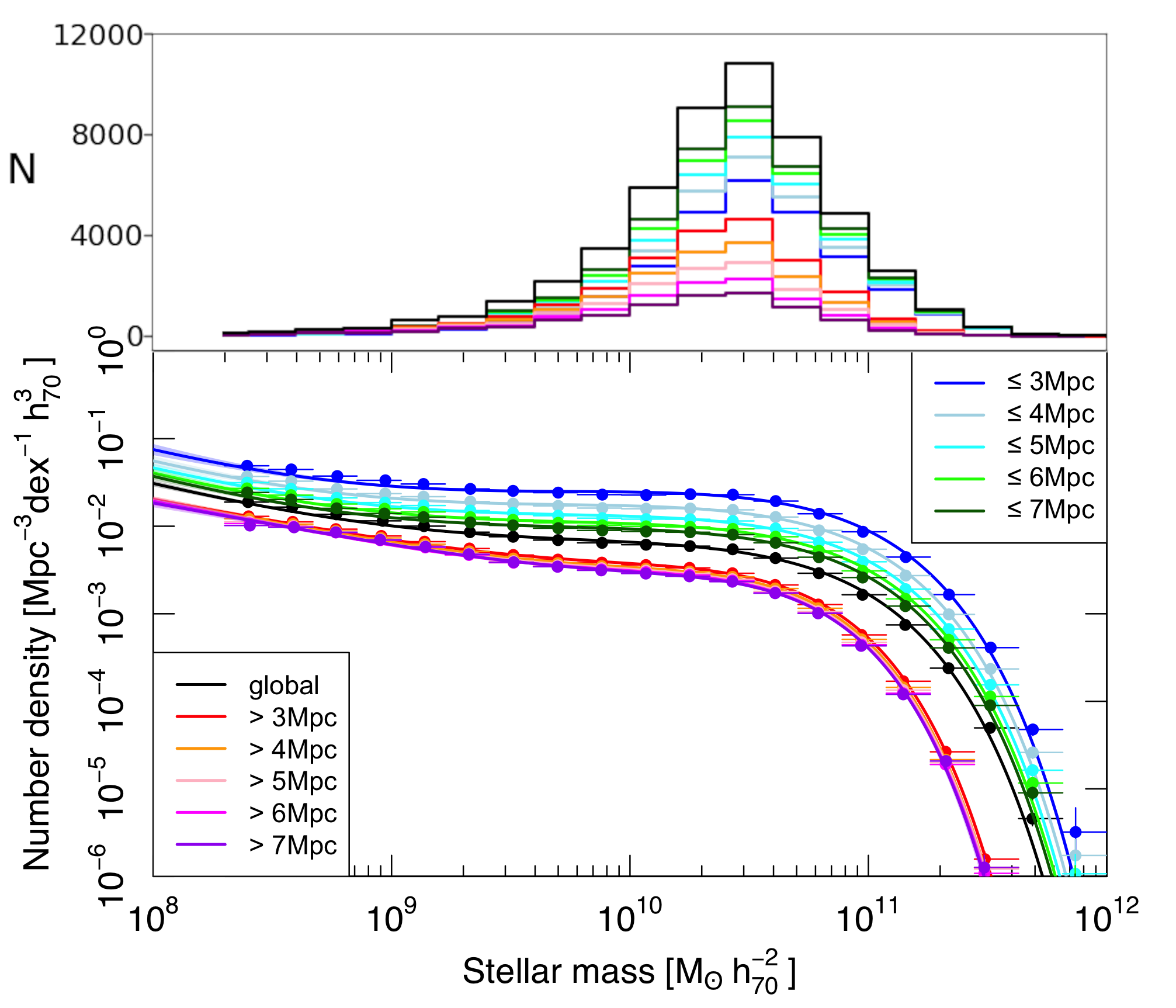}
        \caption{(a) All galaxies}
        \label{fig:group1}
    \end{subfigure}
    \begin{subfigure}[b]{0.47\textwidth}
        \centering
        \includegraphics[width=\textwidth]{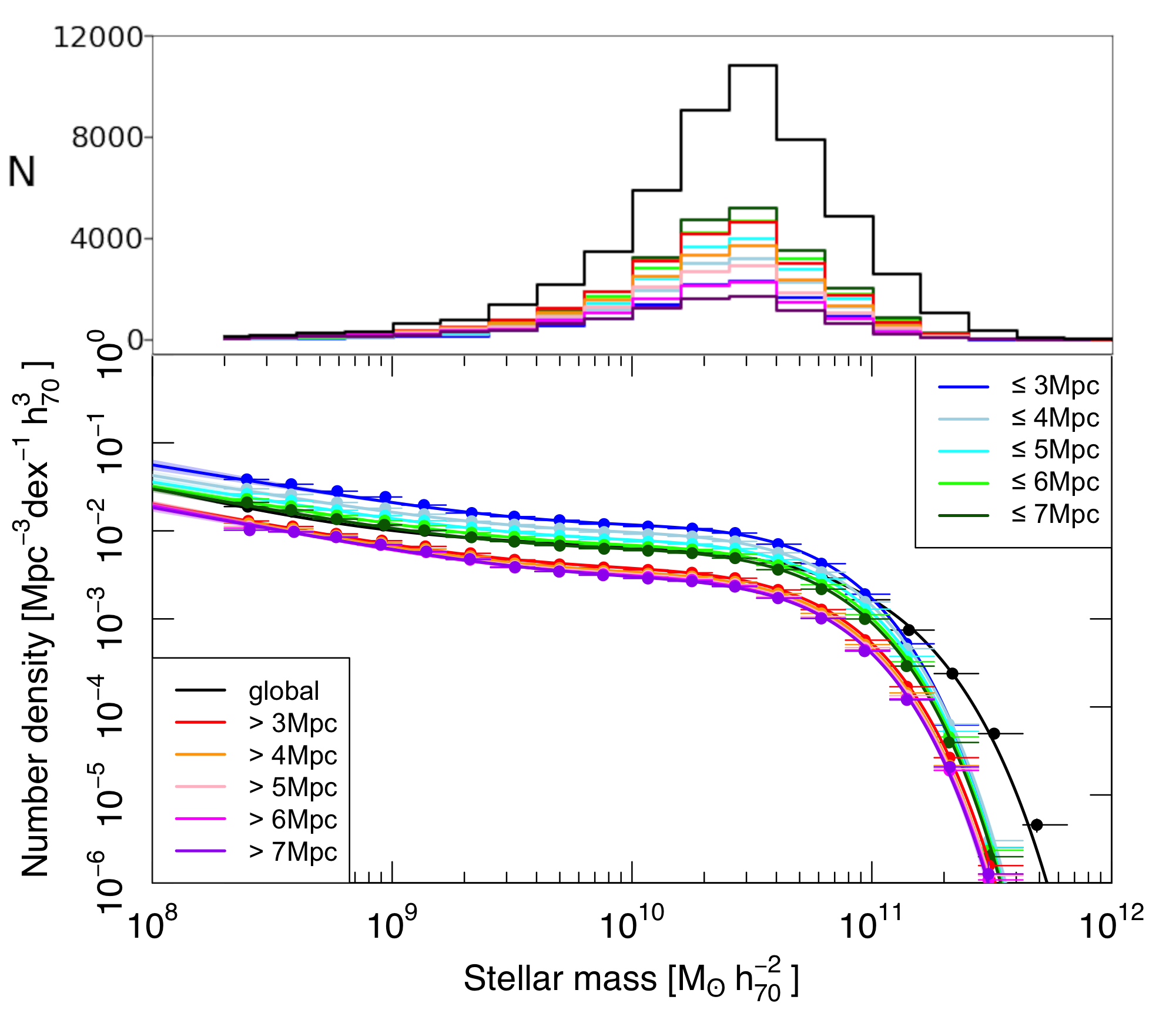}
        \caption{(b) Excluding grouped galaxies in filaments}
        \label{fig:group3}
    \end{subfigure}    
    \caption{Lower panels: cGSMFs in voids and filaments colour-coded
      by filament radius (as indicated in the legend). In panel (a),
      we use our entire galaxy sample, while in panel (b) we discard all
      the grouped galaxies from our filament samples. Our global GSMF
      is also shown in black. Upper panels: Raw number of galaxies as
      a function of stellar mass in each sample, as indicated.}
    \label{fig:group_uno}
\end{figure*}

\begin{figure*}
\captionsetup[subfigure]{labelformat=empty}
    \centering
    \begin{subfigure}[b]{0.44\textwidth}
        \centering
        \includegraphics[width=\textwidth]{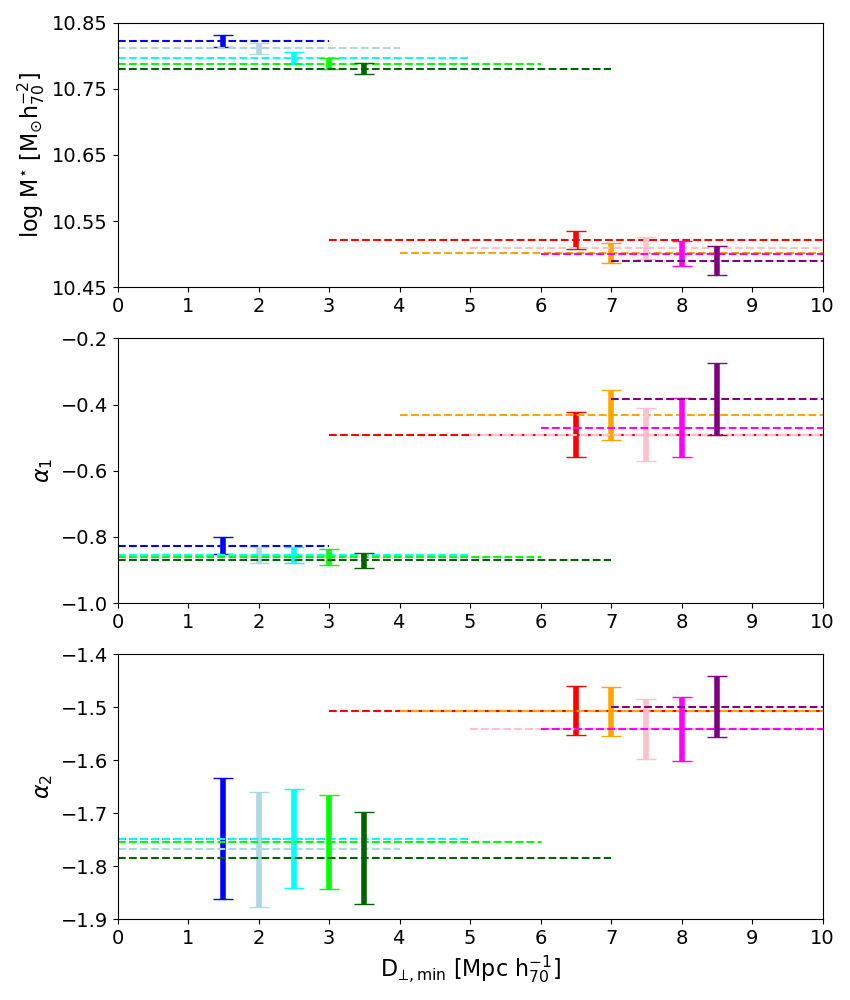}
        \caption{(a) All galaxies}
        \label{fig:group2}
    \end{subfigure}
    \begin{subfigure}[b]{0.44\textwidth}
        \centering
        \includegraphics[width=\textwidth]{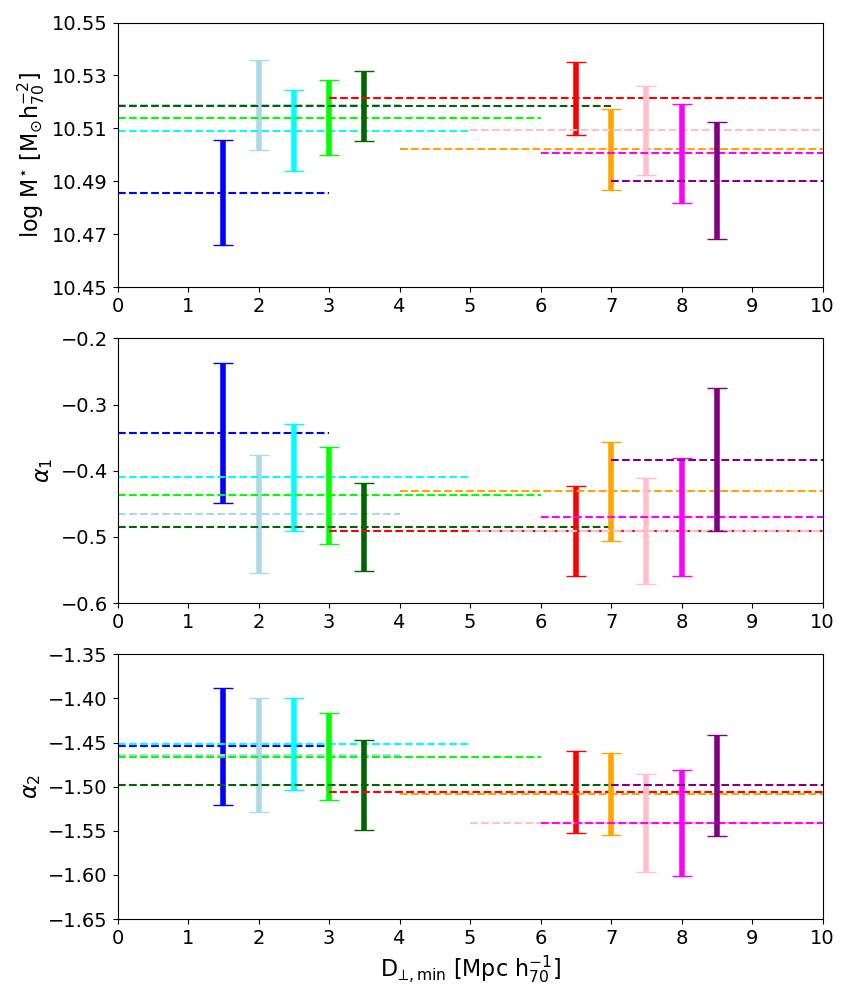}
        \caption{(b) Excluding grouped galaxies in filaments}
        \label{fig:group4}
    \end{subfigure}    
    \caption{Best-fit double Schechter function parameters of the
      cGSMFs shown in Fig.~\ref{fig:group_uno} using the same
      colour-coding by filament radius. The differences between
      filaments and voids found for $M^{\star}$, $\alpha_{1}$, and
      $\alpha_{2}$ when using the entire galaxy sample in panel (a)
      vanish after removing all grouped galaxies from the filament
      samples in panel (b). We note the different scaling of the
      y-axes in (a) and (b).}
    \label{fig:group_due}
\end{figure*}

\section{Results}\label{6}
In this section we present our results on the variation of the cGSMF
as a function of each environmental property described in
Sect.~\ref{4}: orthogonal distance to the nearest filament
$D_{\perp,\rm min}$ (Sect.~\ref{6.1}), group membership
(Sect.~\ref{6.2}), group halo mass $M_{\rm halo}$ (in particular, its
dynamical estimate $M_{\rm dyn}$ in Sect.~\ref{dyn} and its
luminosity-based one $M_{\rm lum}$ in Sect.~\ref{lum}), and the
combination of group branch order BO and group number of connecting
links $N_{\rm links}$ (Sect.~\ref{6.4}).

\subsection{How the GSMF differs in filaments and voids}\label{6.1}
We now investigate the cGSMF in filaments and in voids. As discussed
in Sect.~\ref{2.3}, we use the definition of the quasi one-dimensional
filaments identified in the equatorial GAMA regions by
\cite{alpaslan+14}, but we do not use their classification of galaxies
as belonging to filaments or voids. Instead, we prefer to implement
our own classification in order to be able to investigate the effect
of assuming different filament radii. Since we do not know a priori
how thick a typical filament should be, we used different filament
radii, as listed in Table~\ref{tab:distances}. Having calculated, for
each galaxy in our sample, its orthogonal distance to the nearest
filament, $D_{\perp,\rm min}$, we consider all galaxies with
$D_{\perp,\rm min}$ smaller than the filament radius under
consideration as being part of a filament, while all others are
considered void galaxies.

In Fig.~\ref{fig:group_uno} (a) we show the cGSMFs of void and
filament galaxies as a function of the filament radius. Their best-fit
double Schechter function parameters are tabulated in
Table~\ref{tab:dmin1} and shown in Fig.~\ref{fig:group_due} (a). We
note that in these and similar tables and figures throughout this
paper, we only present our results regarding $M^{\star}$, $\alpha_{1}$
and $\alpha_{2}$ since we are more interested in any change of the
shape of the mass function and less in its normalisation. From
Figs.~\ref{fig:group_uno} (a) and \ref{fig:group_due} (a) we see that
the shape of the void GSMF is essentially unaffected by the filament
radius. On the other hand, the filament GSMF does show some small
variation in $M^{\star}$. However, by far the most striking feature of
these figures is the very significant difference in the shapes of the
filament and void GSMFs, which is manifested in all three of the
Schechter function parameters shown.

\begin{figure}
\centering
\includegraphics[width=0.50\textwidth]{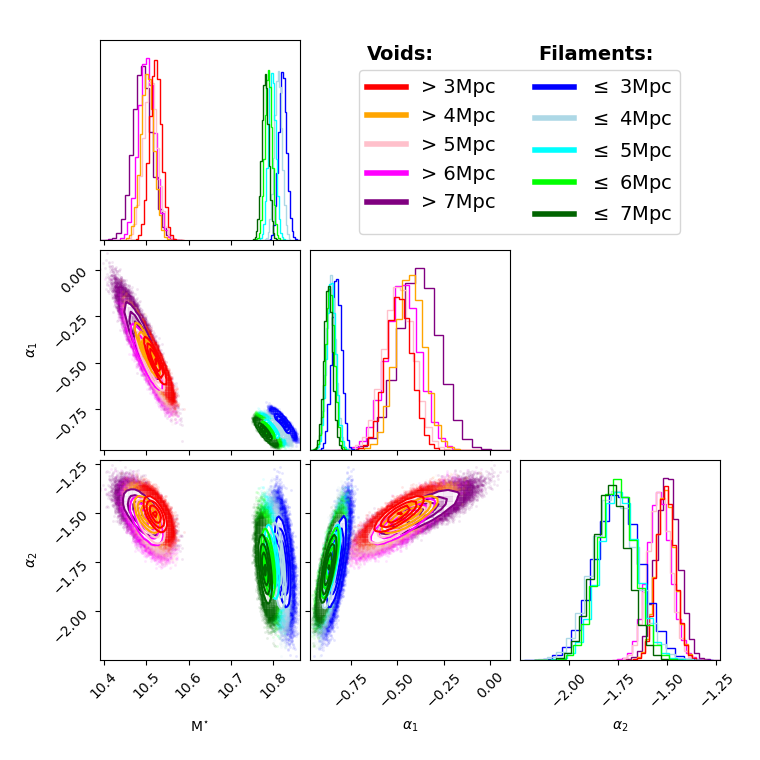}
\caption{Likelihood contours at the $1$$\sigma$, $2$$\sigma$, and
  $3$$\sigma$ levels of the cGSMFs in voids and filaments colour-coded
  by filament radius as indicated in the legend. For each
  distribution, $10^{5}$ random samples were generated.}
\label{fig:corner}
\end{figure}

\begin{figure}
\centering
\includegraphics[width=0.47\textwidth]{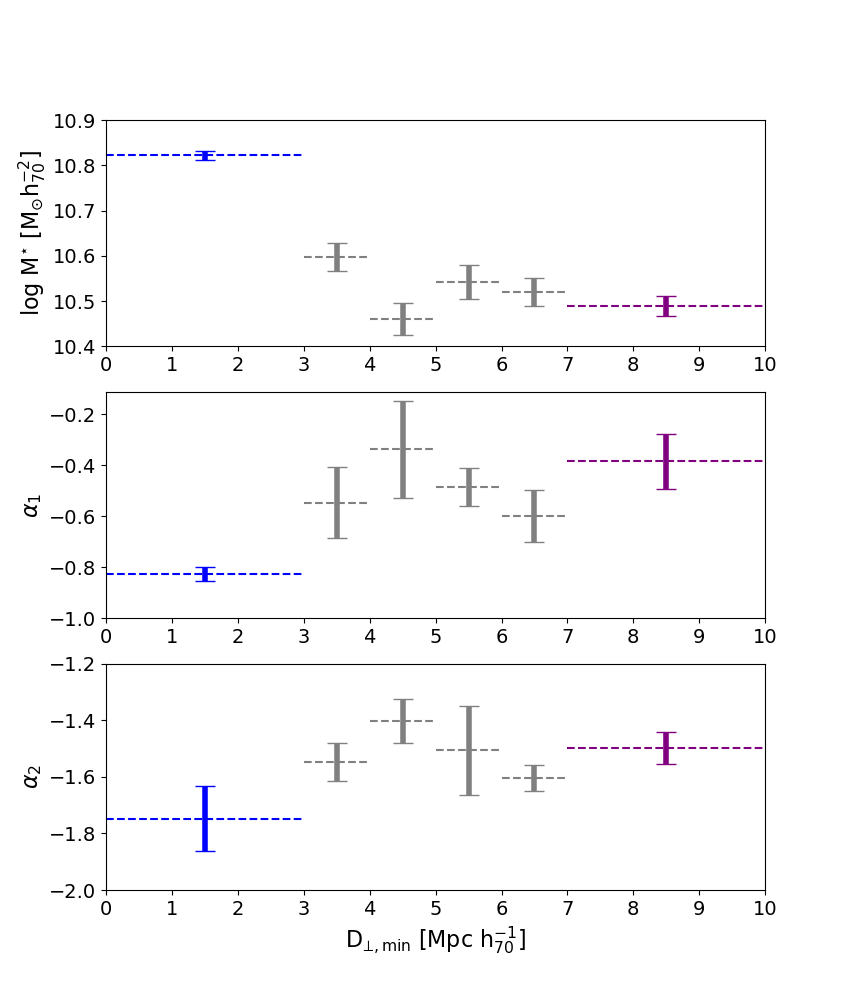}
\caption{Best-fit double Schechter function parameters for the
  differential cGSMFs in the range $3$--$7$~Mpc~$h_{70}^{-1}$ (shown
  in grey). The results for $D_{\perp,\rm min} \leq 3$ and
  $D_{\perp,\rm min} > 7$~Mpc~$h_{70}^{-1}$ are shown in blue
  and purple, respectively.}
\label{fig:group5}
\end{figure}

\begin{figure}
\centering
\includegraphics[width=0.48\textwidth]{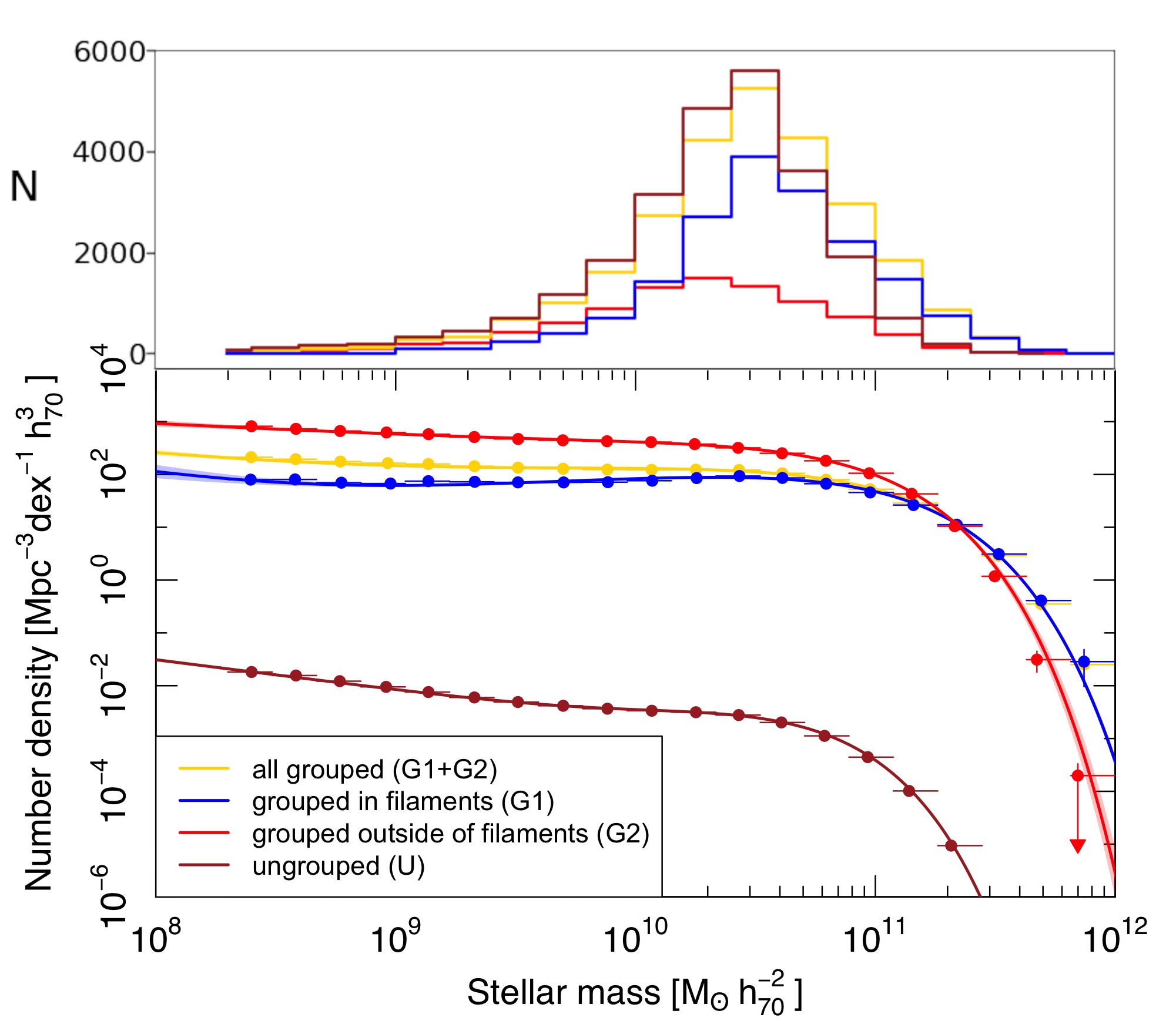}
\caption{Lower panel: cGSMFs of grouped and ungrouped galaxies, as
  indicated in the legend. The grouped galaxies have further been
  subdivided into those that are part of a filament and those that are
  not. We note that our total group sample occupies a volume of just
  $\sim$$2.4 \times 10^{2}$~Mpc$^{3}~h_{70}^{-3}$. This explains why
  the cGSMFs of grouped and ungrouped galaxies differ by $\sim$$4$
  orders of magnitude. Upper panel: Raw number of galaxies as a
  function of stellar mass in each sample, as indicated.}
\label{fig:nuovo}
\end{figure}

In Fig.~\ref{fig:corner} we show the $1$$\sigma$, $2$$\sigma$ and
$3$$\sigma$ likelihood contours of the cGSMFs in voids and
filaments. Clearly, the differences in the double Schechter function
parameters $M^{\star}$, $\alpha_{1}$ and $\alpha_{2}$ between voids
and filaments, already seen in Fig.~\ref{fig:group_due} (a), remain
significant when taking into account the correlations among the
parameters.

So far, we have only considered the cumulative (w.r.t.\ filament
radius) mass functions. In Fig.~\ref{fig:group5} we now show the
Schechter function parameters of the cGSMFs in differential bins of
$D_{\perp,\rm min}$. This clearly demonstrates that the difference
between the shapes of the void and filament cGSMFs is entirely driven
by the innermost $0$--$3$~Mpc~$h_{70}^{-1}$ bin. In fact, as we
subsequently demonstrate, it is entirely driven by the galaxies
belonging to the groups that define the filaments.

In Sect.~\ref{4.1} we defined the G1 subsample as all those grouped
galaxies for which the groups are part of a filament. These galaxies
were assigned $D_{\perp,\rm min}=0$ by definition. We now discard the
G1 subsample from our filament samples. The new cGSMFs as a function
of the filament radius are shown in Fig.~\ref{fig:group_uno} (b), and
their best-fit double Schechter function parameters are tabulated in
Table~\ref{tab:dmin2} and shown in Fig.~\ref{fig:group_due} (b). With
the removal of the grouped galaxies from our filament samples, the
differences between the shapes of the void and the filament cGSMFs
have disappeared almost entirely, as shown in Fig.~\ref{fig:group_uno}
(b). In particular, the new filament cGSMFs now have systematically
lower $M^{\star}$ and shallower $\alpha_{1}$ and $\alpha_{2}$ values,
almost indistinguishable from those of the void mass functions. In
other words, the galaxies in G1 contribute significantly to higher
masses and steeper slopes, leading us to conclude that the shape of
the GSMF is not strongly affected by the larger-scale environment
(i.e.\ voids versus filaments), but rather by group membership.

\subsection{How the GSMF differs for grouped and ungrouped galaxies}\label{6.2}
We now demonstrate the effect of group membership on the mass function
explicitly by directly comparing the cGSMFs of grouped and ungrouped
galaxies.

The volume correction for the group galaxy sample was calculated as
follows. First, each group was assigned the median \texttt{Rad50} of
the groups with the same number of members, where \texttt{Rad50} is
the group radius defined by the $50^{\rm th}$ percentile group member
(taken from the G$^{3}$C). The group's volume was then calculated
assuming a spherical shape, and the final volume correction was then
derived by summing up the volumes of all groups. When we divide our
sample into bins of a group property, as we do in Sects.~\ref{6.3} and
\ref{6.4}, each volume correction is derived by summing up the volumes
of those groups with the property within that bin.

\begin{table}
\centering
\caption{Best-fit double Schechter function parameters for the mass
  functions of grouped and ungrouped galaxies.}
\begin{center}
\noindent\begin{tabular}{c||c|c|c}
\hline
Galaxy & $\log M^{\star}$ & $\alpha_{1}$ & $\alpha_{2}$\\
sample & $(M_{\odot} \, h_{70}^{-2})$ &  &\\ 
\hline
\hline 
G1$+$G2  & $10.87 \pm 0.01$ & $-0.81 \pm 0.04$ & $-1.49 \pm 0.13$\\
G1       & $10.86 \pm 0.01$ & $-0.69 \pm 0.03$ & $-1.71 \pm 0.21$\\
G2       & $10.70 \pm 0.04$ & $-0.57 \pm 0.25$ & $-1.24 \pm 0.11$\\
U        & $10.41 \pm 0.01$ & $-0.22 \pm 0.06$ & $-1.57 \pm 0.03$\\ 
\hline
\end{tabular}
\end{center}
\label{tab:nuovo}
\end{table}

\begin{table*}
\centering
\caption{Best-fit double Schechter function parameters of the cGSMFs
  of grouped galaxies for different subsamples, as indicated.}
\begin{center}
\noindent\begin{tabular}{c|c||c|c|c}
\hline
Galaxy sample & Halo mass estimator & $\log M^{\star}$ & $\alpha_{1}$ & $\alpha_{2}$\\
 &  & $(M_{\odot} \, h_{70}^{-2})$ &  &\\ 
\hline
\hline 
$\log (\frac{M_{\rm dyn}}{M_{\odot} h^{-1}_{70}}$) $\leq 12.5$ & \texttt{MassA}  & 10.47 $\pm$ 0.02 & 0.14 $\pm$ 0.13 & $-$1.25 $\pm$ 0.06\\
$\log (\frac{M_{\rm dyn}}{M_{\odot} h^{-1}_{70}}$) $\leq 12.5$ & \texttt{MassAfunc}  & 10.47 $\pm$ 0.03 & 0.12 $\pm$ 0.16 & $-$1.24 $\pm$ 0.07\\ 
$12.5 < \log (\frac{M_{\rm dyn}}{M_{\odot} h^{-1}_{70}}$) $\leq 13$ & \texttt{MassA}  & 10.65 $\pm$ 0.03 & $-$0.36 $\pm$ 0.10 & $-$1.49 $\pm$ 0.12\\ 
$12.5 < \log (\frac{M_{\rm dyn}}{M_{\odot} h^{-1}_{70}}$) $\leq 13$ & \texttt{MassAfunc}  & 10.55 $\pm$ 0.03 & $-$0.12 $\pm$ 0.14 & $-$1.35 $\pm$ 0.09\\ 
$13 < \log (\frac{M_{\rm dyn}}{M_{\odot} h^{-1}_{70}}$) $\leq 13.5$ & \texttt{MassA}  & 10.78 $\pm$ 0.03 & $-$0.40 $\pm$ 0.14 & $-$1.34 $\pm$ 0.12\\ 
$13 < \log (\frac{M_{\rm dyn}}{M_{\odot} h^{-1}_{70}}$) $\leq 13.5$ & \texttt{MassAfunc}  & 10.75 $\pm$ 0.02 & $-$0.49 $\pm$ 0.09 & $-$1.50 $\pm$ 0.14\\
$\log (\frac{M_{\rm dyn}}{M_{\odot} h^{-1}_{70}}) > 13.5$ &  \texttt{MassA} & 11.05 $\pm$ 0.01 & $-$0.99 $\pm$ 0.04 & $-$0.99 $\pm$ 0.08\\  
$\log (\frac{M_{\rm dyn}}{M_{\odot} h^{-1}_{70}}) > 13.5$ &  \texttt{MassAfunc} & 11.03 $\pm$ 0.01 & $-$0.98 $\pm$ 0.06 & $-$0.98 $\pm$ 0.09\\ 
\hline
\end{tabular}
\end{center}
\label{tab:mdyn1}
\end{table*}

\begin{table*}
\caption{Best-fit double Schechter function parameters of the cGSMFs
  of grouped galaxies for different subsamples, as indicated.}
\begin{center}
\begin{tabular}{c|c||c|c|c}
\hline
Galaxy sample & Halo mass estimator & $\log M^{\star}$ & $\alpha_{1}$ & $\alpha_{2}$\\
$N_{\rm FOF} > 4$ &  & $(M_{\odot} \, h_{70}^{-2})$ &  &\\ 
\hline
\hline 
$\log (\frac{M_{\rm lum}}{M_{\odot} h^{-1}_{70}}$) $\leq$ 13.75 & \texttt{LumB}  & 10.74 $\pm$ 0.04 & $-$0.33 $\pm$ 0.22 & $-$1.23 $\pm$ 0.11\\
$\log (\frac{M_{\rm dyn}}{M_{\odot} h^{-1}_{70}}$) $\leq$ 13.75 & \texttt{MassA}  & 10.88 $\pm$ 0.03 & $-$0.67 $\pm$ 0.12 & $-$1.40 $\pm$ 0.18\\
$\log (\frac{M_{\rm lum}}{M_{\odot} h^{-1}_{70}}$) $\leq$ 13.75 & \texttt{LumBfunc}  & 10.88 $\pm$ 0.03 & $-$0.70 $\pm$ 0.12 & $-$1.48 $\pm$ 0.20\\
$\log (\frac{M_{\rm dyn}}{M_{\odot} h^{-1}_{70}}$) $\leq$ 13.75 & \texttt{MassAfunc}  & 10.87 $\pm$ 0.03 & $-$0.69 $\pm$ 0.11 & $-$1.47 $\pm$ 0.19\\
13.75 < $\log (\frac{M_{\rm dyn}}{M_{\odot} h^{-1}_{70}}$) $\leq$ 14.25 & \texttt{LumB} & 11.12 $\pm$ 0.02 & $-$1.01 $\pm$ 0.07 & $-$1.01 $\pm$ 0.10\\
13.75 < $\log (\frac{M_{\rm dyn}}{M_{\odot} h^{-1}_{70}}$) $\leq$ 14.25 & \texttt{MassA} & 11.10 $\pm$ 0.02 & $-$1.04 $\pm$ 0.17 & $-$1.04 $\pm$ 0.06\\
13.75 < $\log (\frac{M_{\rm dyn}}{M_{\odot} h^{-1}_{70}}$) $\leq$ 14.25 & \texttt{LumBfunc}  & 11.10 $\pm$ 0.02 & $-$1.00 $\pm$ 0.02 & $-$1.00 $\pm$ 0.04\\
13.75 < $\log (\frac{M_{\rm dyn}}{M_{\odot} h^{-1}_{70}}$) $\leq$ 14.25 & \texttt{MassAfunc}  & 11.08 $\pm$ 0.02 & $-$1.01 $\pm$ 0.18 & $-$1.01 $\pm$ 0.06\\
$\log (\frac{M_{\rm dyn}}{M_{\odot} h^{-1}_{70}}$) > 14.25 & \texttt{LumB} & 11.07 $\pm$ 0.03 & $-$1.02 $\pm$ 0.09 & $-$1.02 $\pm$ 0.08\\
$\log (\frac{M_{\rm dyn}}{M_{\odot} h^{-1}_{70}}$) > 14.25 & \texttt{MassA}  & 11.11 $\pm$ 0.02 & $-$1.08 $\pm$ 0.07 & $-$1.08 $\pm$ 0.06\\
$\log (\frac{M_{\rm dyn}}{M_{\odot} h^{-1}_{70}}$) > 14.25 & \texttt{LumBfunc}  & 11.06 $\pm$ 0.02 & $-$1.04 $\pm$ 0.13 & $-$1.04 $\pm$ 0.05\\
$\log (\frac{M_{\rm dyn}}{M_{\odot} h^{-1}_{70}}$) > 14.25 & \texttt{MassAfunc}  & 11.13 $\pm$ 0.02 & $-$1.10 $\pm$ 0.02 & $-$1.10 $\pm$ 0.05\\
\hline
\end{tabular}
\end{center}
\label{tab:vm1}
\end{table*}

Our resulting cGSMFs are shown in Fig.~\ref{fig:nuovo}, and the
best-fit double Schechter function parameters are tabulated in
Table~\ref{tab:nuovo} (cf.\ also Fig.~\ref{fig:dsmdyn}). Clearly, the
mass functions of the grouped (in yellow) and ungrouped galaxies (in
brown) differ substantially: the characteristic mass $M^\star$ of the
grouped cGSMF is larger, its intermediate-mass slope $\alpha_1$ is
steeper, while its low-mass slope $\alpha_2$ is essentially the
same. As we shall see in the next subsection, these differences are
likely due to the ungrouped galaxies being hosted by less massive
halos compared to the grouped galaxies.

Splitting the group galaxy sample into those that are part of a
filament (subsample G1) and those that are not (subsample G2), we also
find significant differences. In particular, the cGSMF of the grouped
filament galaxies has a larger characteristic mass than that of the
grouped galaxies outside of filaments. Again anticipating the results
of the next subsection, we attribute this difference to the larger
halo masses of the filament groups compared to the groups outside of
filaments, as shown in Fig.~\ref{fig:corr1}.

\begin{figure*}
\captionsetup[subfigure]{labelformat=empty}
    \centering
    \begin{subfigure}[b]{0.49\textwidth}
        \centering
        \includegraphics[width=\textwidth]{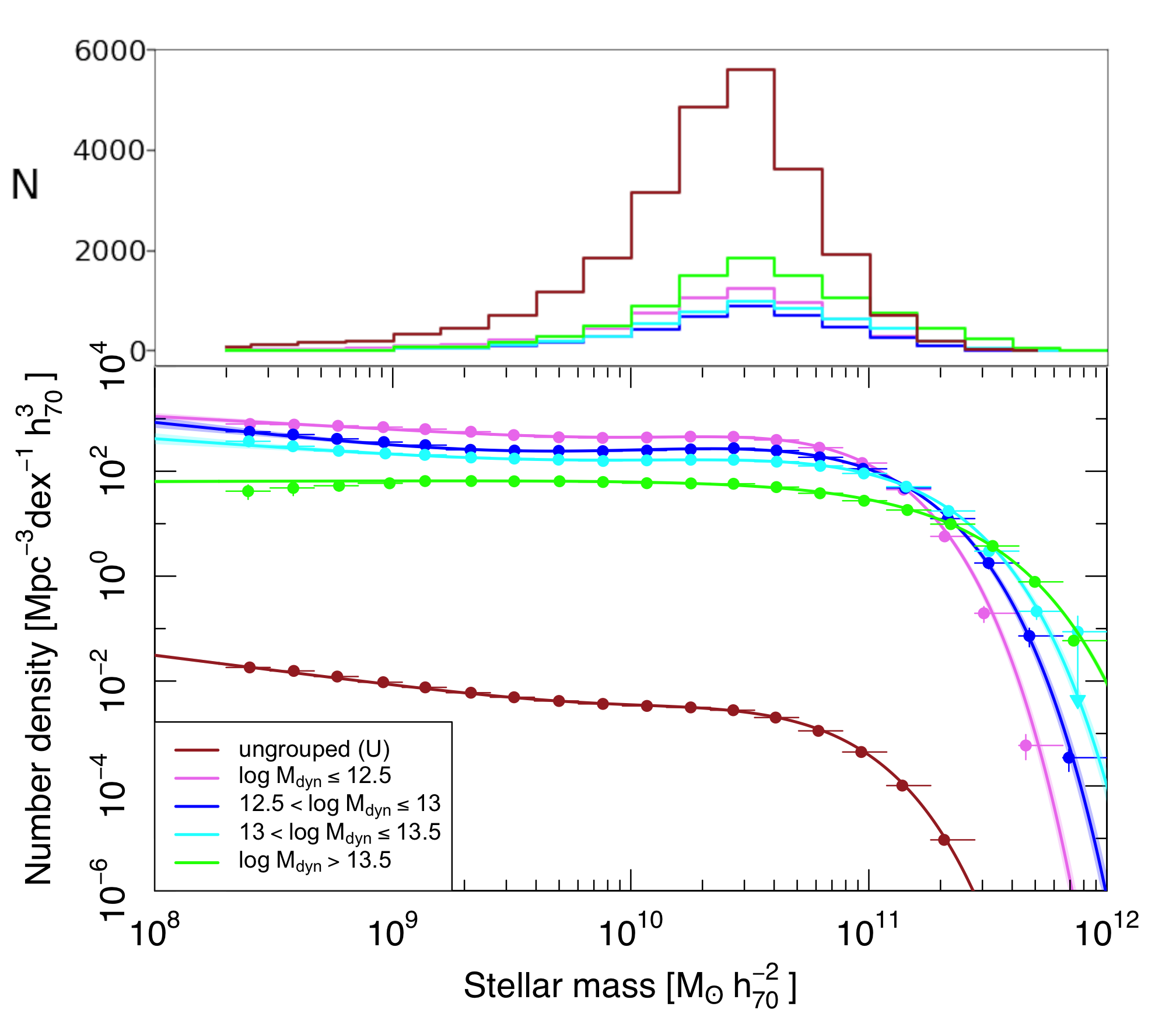}
        \caption{(a) \texttt{MassA} estimator}
        \label{fig:cGSMF2a}
    \end{subfigure}
    \begin{subfigure}[b]{0.49\textwidth}
        \centering
        \includegraphics[width=\textwidth]{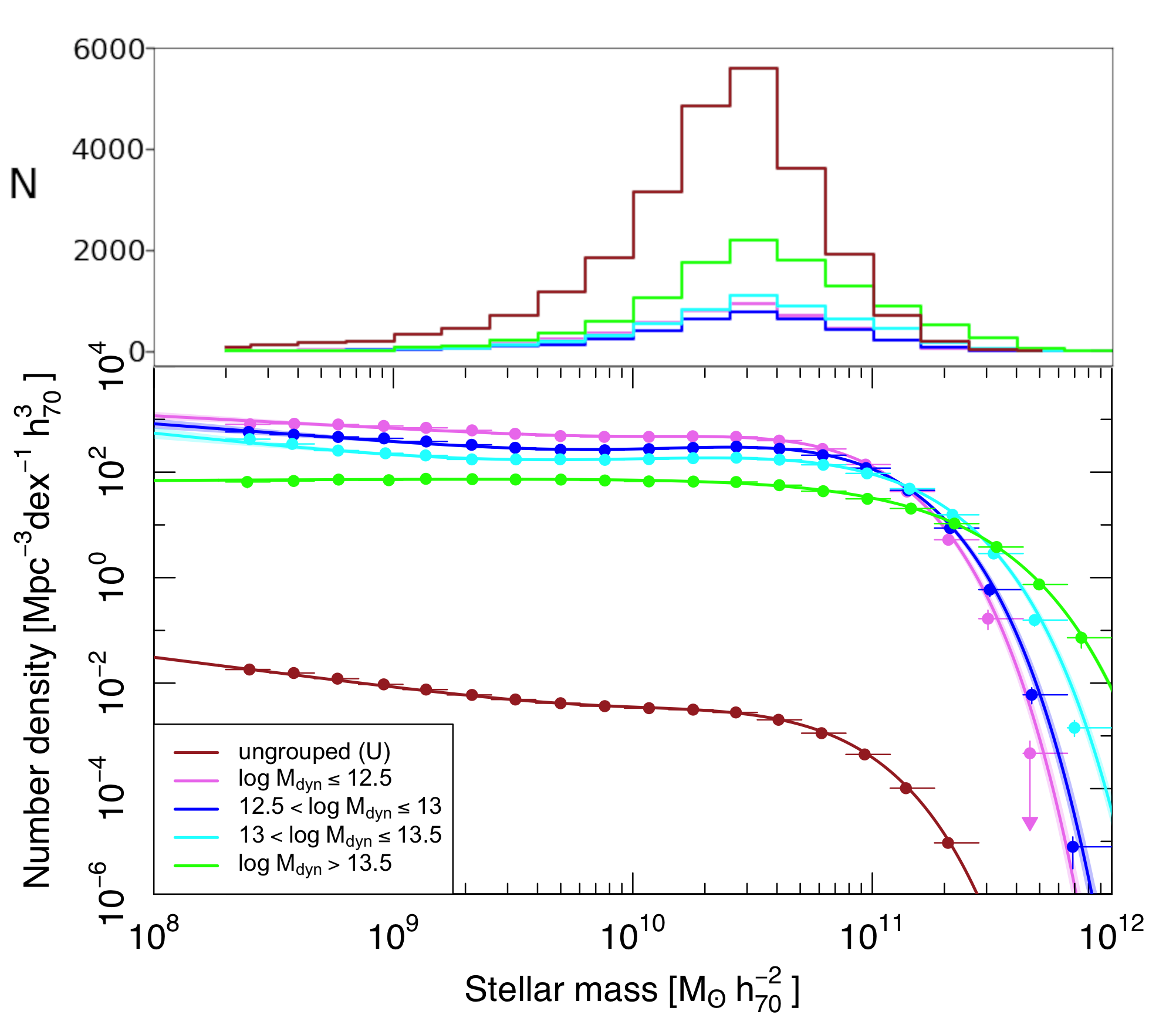}
        \caption{(b) \texttt{MassAfunc} estimator}
        \label{fig:cGSMF2b}
    \end{subfigure}    
    \caption{Lower panels: cGSMFs of the group galaxy subsample
      colour-coded by $M_{\rm dyn}$ as indicated in the legend. In
      panel (a), we use the \texttt{MassA} calibration factor, while
      in panel (b) we use the \texttt{MassAfunc} factor. For
      comparison, the cGSMF of the ungrouped galaxies is also shown in
      brown. Upper panels: Raw number of galaxies as a function of
      stellar mass in each sample, as indicated.}
    \label{fig:bigmdyn}
\end{figure*}

\begin{figure}[!h]
\centering
\vspace{-0.6cm}
\includegraphics[width=0.49\textwidth]{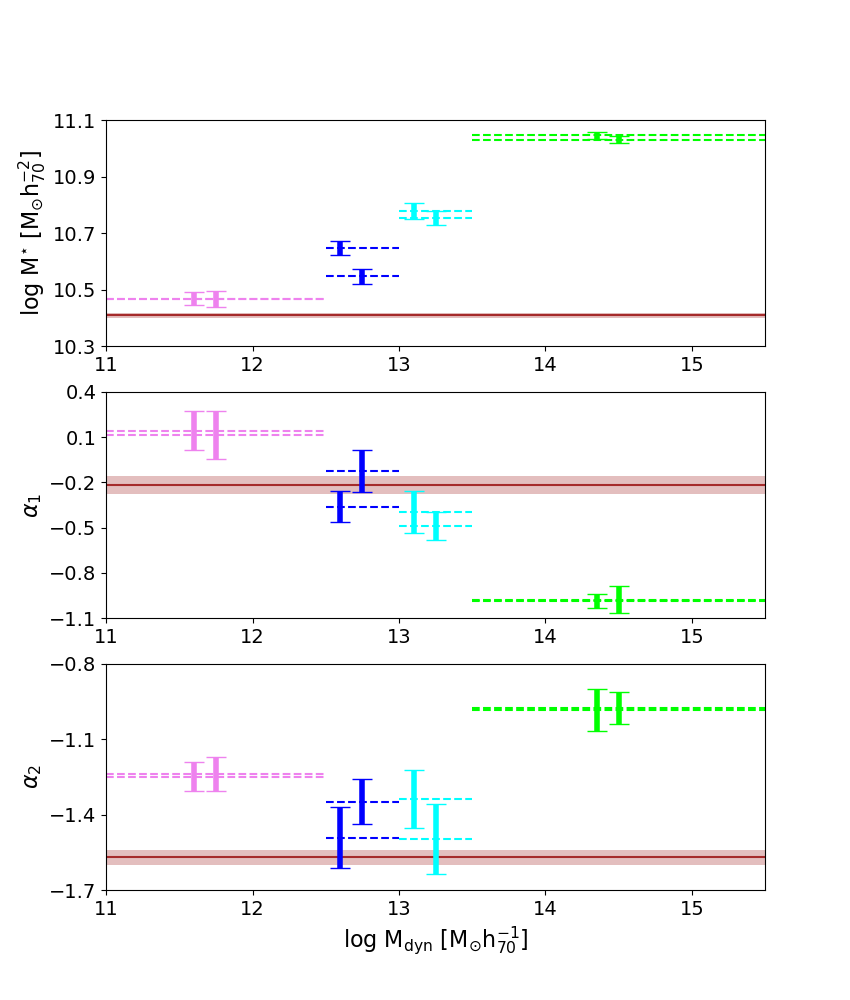}
\caption{Best-fit double Schechter function parameters of the cGSMFs
  shown in Fig.~\ref{fig:bigmdyn} using the same colour-coding by
  $M_{\rm dyn}$. For clarity, the vertical error bars corresponding to
  the \texttt{MassA} estimator have been slightly offset to the
  left. The brown horizontal bands show the results for the ungrouped
  galaxy sample.}
\label{fig:dsmdyn}
\end{figure}

We also note the large difference in the normalisation of the grouped
and ungrouped cGSMFs by $\sim$$4$ orders of magnitude. This is of
course due to the cGSMFs effectively measuring the typical local
density in groups and of ungrouped galaxies. While the raw numbers of
grouped and ungrouped galaxies in our sample are actually quite
similar (cf.\ Sect.~\ref{4.2} and upper panel of
Fig.~\ref{fig:nuovo}), the volumes occupied by them are vastly
different, generating the large difference in density. We hasten to
point out, though, that our method of determining the volume occupied
by the groups is only very approximate.

Having established the group environment as an important factor in
shaping the GSMF, we now turn to investigating the GSMF as a function
of group properties.

\subsection{GSMF dependence on group halo mass}\label{6.3}
In this section, we study the dependence of the GSMF on group halo
mass $M_{\rm halo}$. As described in Sect.~\ref{4.3}, we have two
different estimates of $M_{\rm halo}$: the dynamical mass estimate
$M_{\rm dyn}$ and the group $r$-band luminosity-based mass estimate
$M_{\rm lum}$. We now describe our GSMF measurements using these two
estimates in turn.

\subsubsection{GSMF dependence on dynamical group halo mass}\label{dyn}
To study the dependence of the GSMF on dynamical group halo mass
$M_{\rm dyn}$, we are forced to discard $879$/$10\,429$ ($8.4\%$) of
our groups for which the G$^3$C does not report any $M_{\rm dyn}$
values because the measured velocity dispersion of these groups is
smaller than its error. These are overwhelmingly groups with $N_{\rm
  FOF} = 2$. Our total sample now consists of $9550$ groups containing
$25\,567$ galaxies.

Next, we bin the galaxies into four different bins in $\log [M_{\rm
    dyn} / (M_{\odot} \, h^{-1}_{70})]$ according to the mass of the
group that they belong to: $\leq 12.5$, $12.5$--$13$, $13$--$13.5$,
and $> 13.5$ (cf.\ Fig.~\ref{fig:corr1}).

As explained in Sect.~\ref{4.3}, the G$^{3}$C provides the dynamical
group halo mass using two different calibration factors, which we
refer to as \texttt{MassA} and \texttt{MassAfunc}, respectively.

Our resulting cGSMFs, colour-coded by $M_{\rm dyn}$, are shown in
Fig.~\ref{fig:bigmdyn} (a) for the \texttt{MassA} calibration factor
and in (b) for the \texttt{MassAfunc} factor, respectively. The
best-fit double Schechter function parameters are tabulated in
Table~\ref{tab:mdyn1} and shown in Fig.~\ref{fig:dsmdyn}. We first
point out that the characteristic mass $M^{\star}$ clearly increases
with $M_{\rm dyn}$. In other words, more massive halos tend to host
more massive galaxies, a result to be expected in a hierarchical
structure formation paradigm, where larger halos form by the merging
of smaller ones, accumulating more mass and forming larger central
galaxies. Furthermore, the intermediate-mass slope $\alpha_{1}$
steepens with $M_{\rm dyn}$, that is, a more rapid decline with
stellar mass in the number of intermediate-mass galaxies is observed
in more massive halos. In contrast, there is no evidence for any trend
in the low-mass slope $\alpha_{2}$, except possibly for an increase in
the highest halo mass bin, where the cGSMF is nonetheless best
represented by a single Schechter function (with $\alpha_2 =
\alpha_1$). Reassuringly, these results are largely robust against the
choice of calibration factor.

We also find that the results of the previous subsection are broadly
consistent with these trends in that the differences between the
shapes of the cGSMFs of grouped galaxies in filaments and those
outside of filaments can be explained almost entirely by the fact that
the halo masses of groups in filaments are systematically higher than
those of groups outside of filaments (cf.\ Fig.~\ref{fig:corr1}). The
only exception is the very steep low-mass slope of the grouped
galaxies in filaments (cf.\ Table~\ref{tab:nuovo}).

To investigate whether the shape of the cGSMF of the ungrouped
galaxies also fits this picture we show in Fig.~\ref{fig:dsmdyn} (as
brown lines) the best-fit double Schechter function parameters of the
ungrouped galaxies derived in the previous subsection. Of course, for
these galaxies we lack an estimate of their halo masses but it seems
reasonable to assume that the halo mass distribution of these galaxies
will be skewed towards similar or even lower masses than that of the
group halos in our lowest halo mass bin. Under this assumption, at
least the $M^{\star}$ value of the ungrouped galaxies is entirely
consistent with the $M^{\star}$-halo mass trend of grouped galaxies.

Overall, we thus conclude that there is clear evidence of a dependence
of $M^{\star}$ and $\alpha_1$ on halo mass, while there is no clear
trend for $\alpha_2$.

\begin{figure*}
\vspace{-0.4cm}
\captionsetup[subfigure]{labelformat=empty}
    \centering
    \begin{subfigure}[b]{0.41\textwidth}
        \centering
        \includegraphics[width=\textwidth]{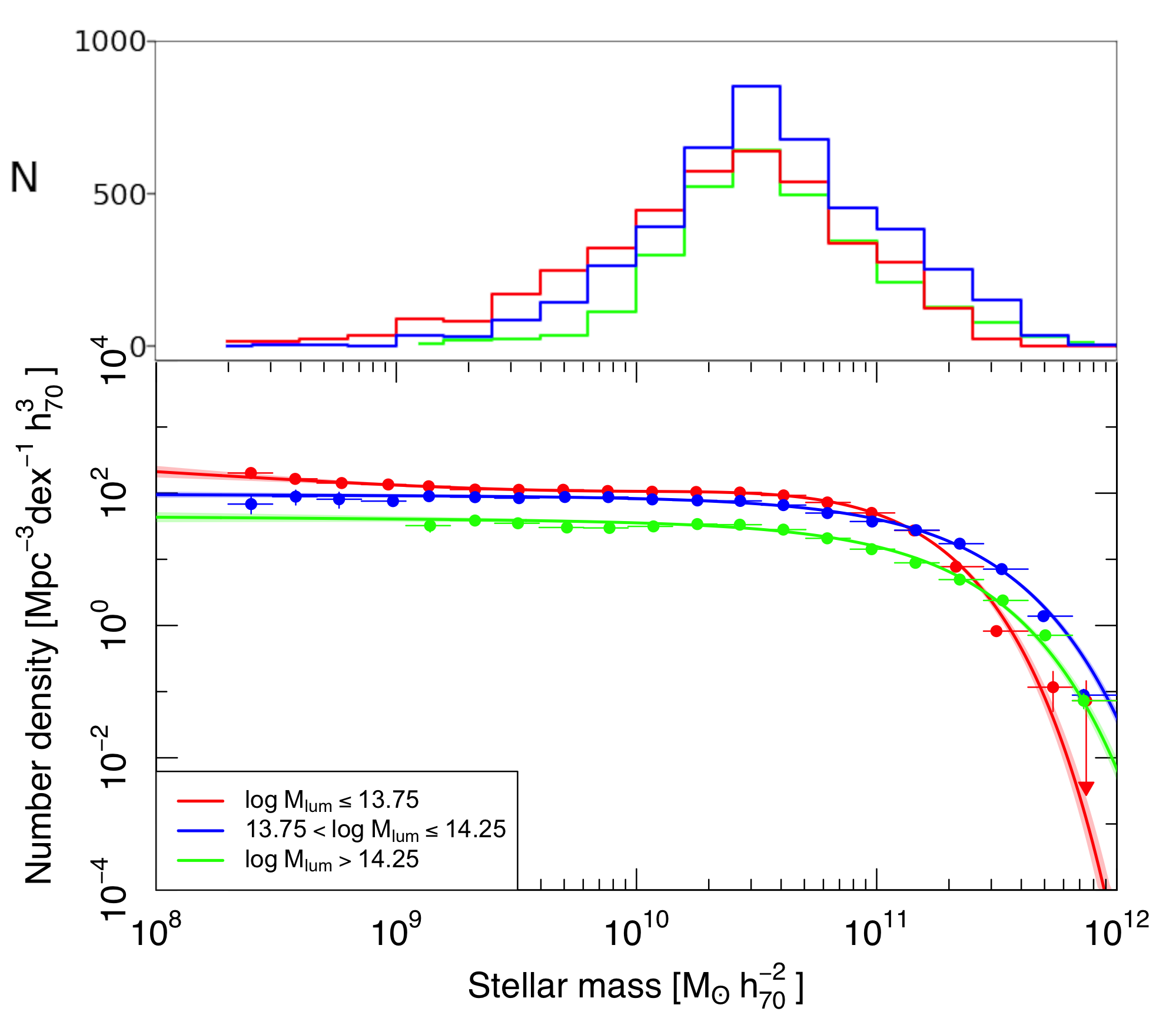}
        \caption{(a) \texttt{LumB} estimator}
        \label{fig:cGSMF3a}
    \end{subfigure}
    \hspace{0.5cm}
    \begin{subfigure}[b]{0.41\textwidth}
        \centering
        \includegraphics[width=\textwidth]{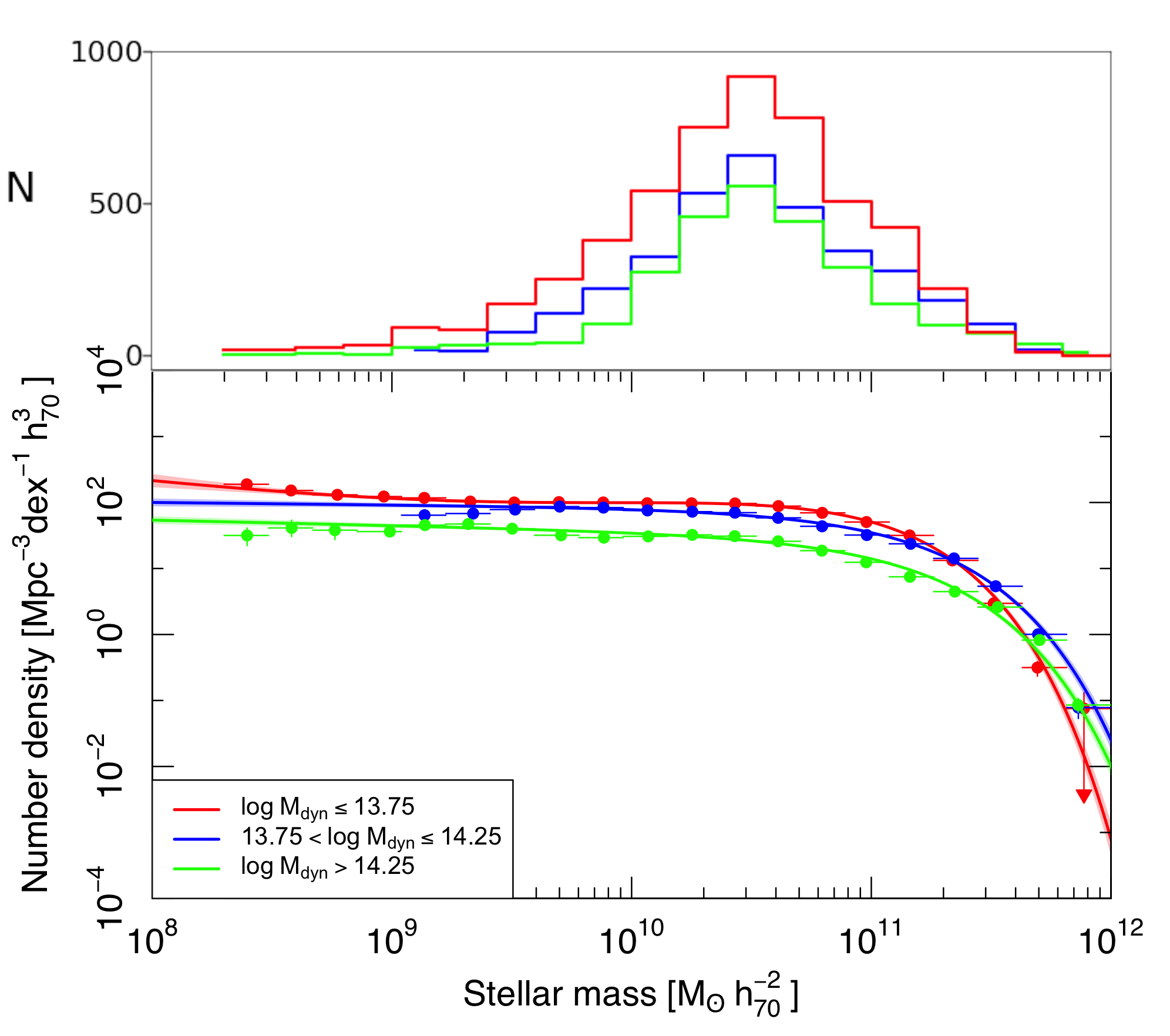}
        \caption{(c) \texttt{MassA} estimator}
        \label{fig:cGSMF3b}
    \end{subfigure}
    \begin{subfigure}[b]{0.405\textwidth}
        \centering
        \includegraphics[width=\textwidth]{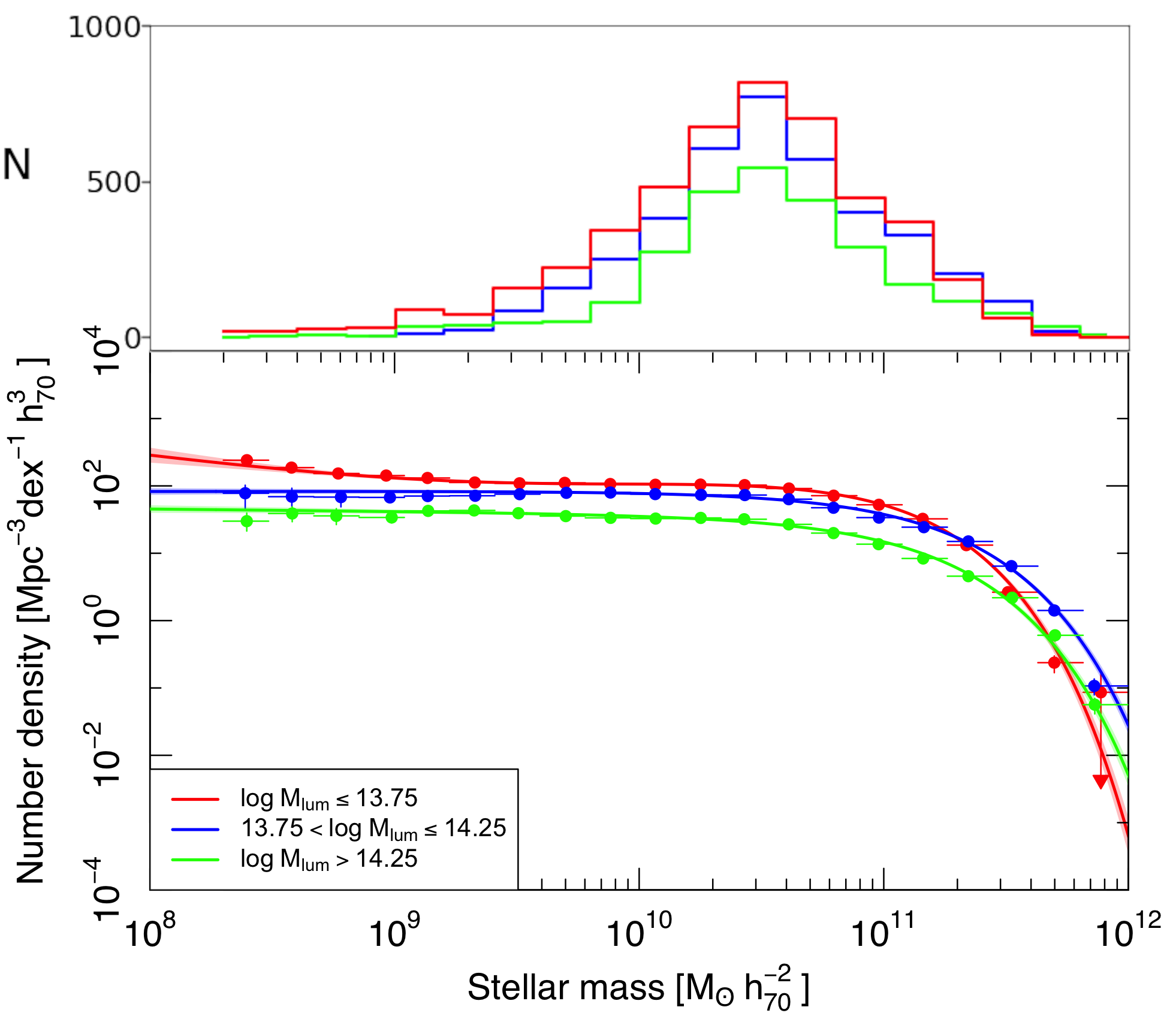}
        \caption{(b) \texttt{LumBfunc} estimator}
        \label{fig:cGSMF4a}
    \end{subfigure}
    \hspace{0.5cm}
    \begin{subfigure}[b]{0.41\textwidth}
        \centering
        \includegraphics[width=\textwidth]{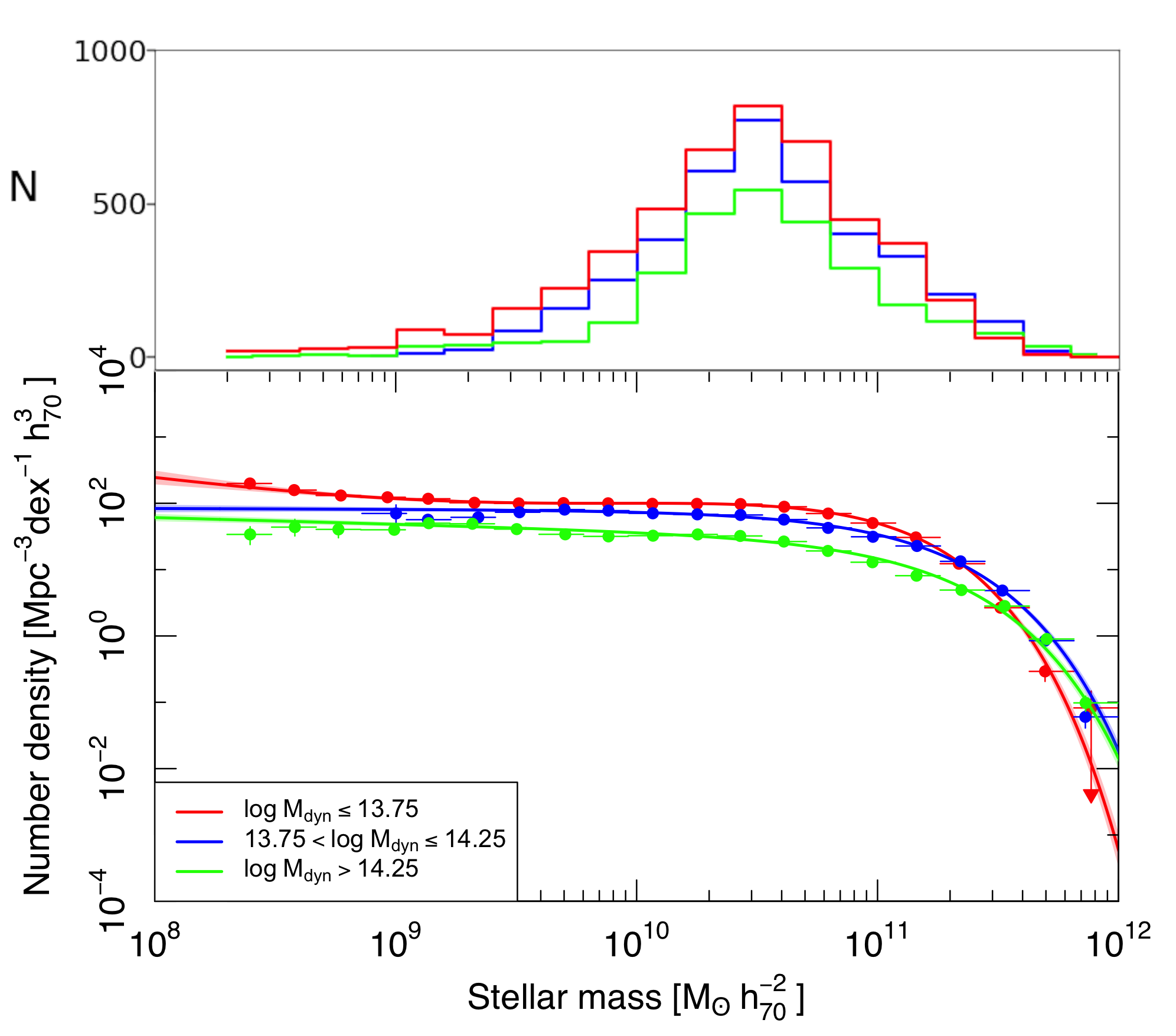}
        \caption{(d) \texttt{MassAfunc} estimator}
        \label{fig:cGSMF4b}
    \end{subfigure} 
    \caption{Left column, lower panels: cGSMFs of our high-fidelity
      group galaxy subsample colour-coded by $M_{\rm lum}$, as
      indicated in the legend. In panel (a) we use the \texttt{LumB}
      calibration factor, while in panel (b) we use the
      \texttt{LumBfunc} factor. Right column, lower panels: Same as
      the left but using the dynamical halo mas estimates with the
      \texttt{MassA} calibration factor in panel (c) and the
      \texttt{MassAfunc} factor in panel (d). Upper panels: Raw number
      of galaxies as a function of stellar mass in each sample, as
      indicated.}
    \label{fig:bigmlum1}
\end{figure*}

\begin{figure*}
\vspace{-0.4cm}
\centering
\quad\quad\includegraphics[width=0.34\textwidth]{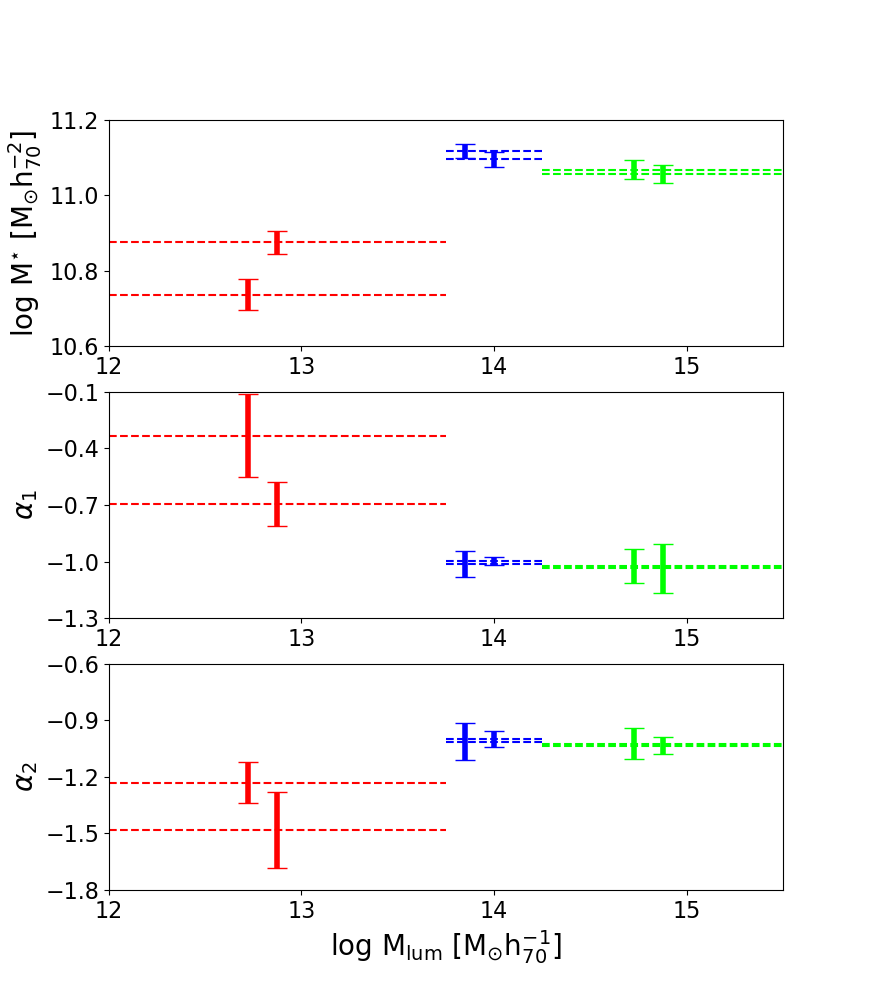}\quad\includegraphics[width=0.34\textwidth]{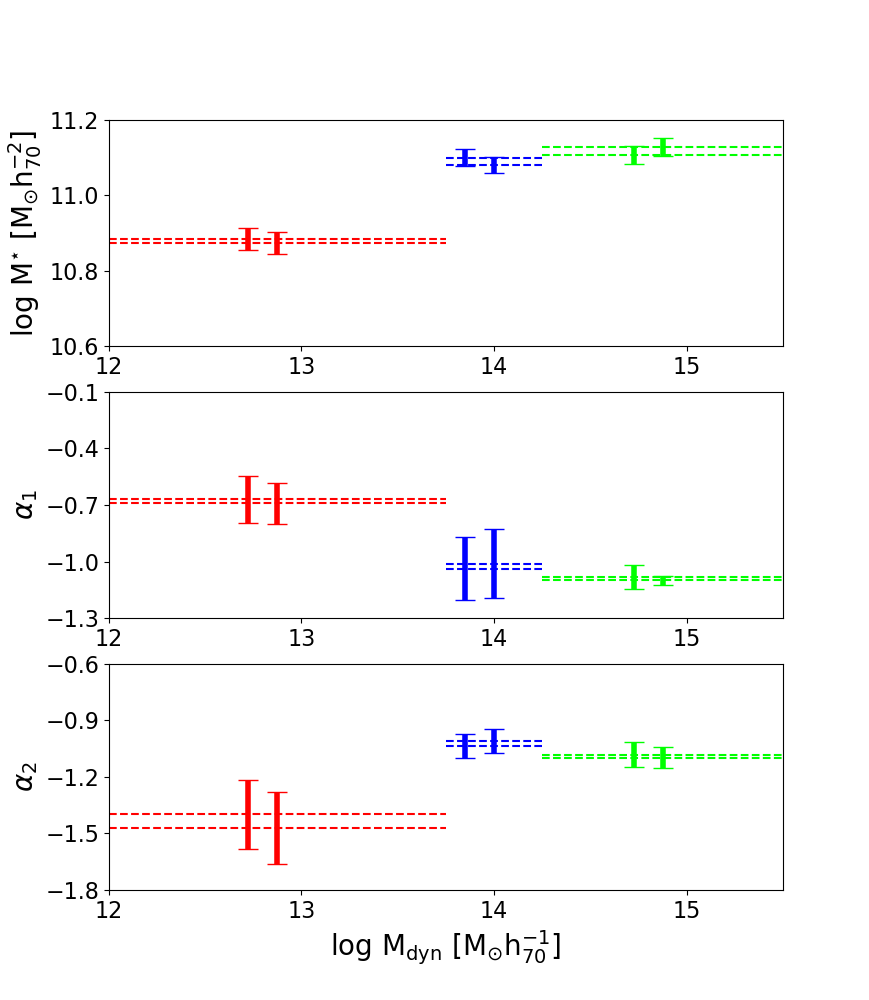}
\caption{Best-fit double Schechter function parameters of the cGSMFs
  shown in Fig.~\ref{fig:bigmlum1} using the same colour-coding by
  $M_{\rm lum}$ (left) or $M_{\rm dyn}$ (right). For clarity, the
  vertical error bars corresponding to the \texttt{LumB} (left) and
  \texttt{MassA} (right) estimators have been slightly offset to the
  left.}
\label{fig:dsmlum}
\end{figure*}

\subsubsection{GSMF dependence on luminosity-based group halo mass}\label{lum}
To study the dependence of the GSMF on luminosity-based group halo
mass $M_{\rm lum}$, we follow the work of \cite{vazquez+20} and only
consider high-fidelity groups with multiplicity $N_{\rm FOF} > 4$ and
$M_{\rm lum} > 10^{12}~M_{\odot}~h_{70}^{-1}$, as this is the sample
for which the luminosity-halo mass scaling relation of \cite{viola+15}
is well calibrated. We are thus forced to discard $8917$/$10\,429$
($85.5\%$) of our groups, leaving us with $1512$ groups containing
$11\,417$ galaxies.

We now bin the remaining galaxies into three different bins in $\log
[M_{\rm lum} / (M_{\odot} \, h_{70}^{-1})]$ according to the mass of
the group that they belong to: $\leq 13.75$, $13.75$--$14.25$, and $>
14.25$ (cf.\ Fig.~\ref{fig:corr1}).

As explained in Sect.~\ref{4.3}, the G$^{3}$C provides the total group
luminosity that we use to derive the luminosity-based halo mass using
two different calibration factors, which we refer to as
(\texttt{LumB}) and (\texttt{LumBfunc}), respectively.

Our resulting cGSMFs, colour-coded by $M_{\rm lum}$, are shown in
Fig.~\ref{fig:bigmlum1} (a) for the \texttt{LumB} calibration factor
and in (b) for the \texttt{LumBfunc} factor, respectively. For a
direct comparison, panels (c) and (d) show the corresponding cGSMFs
for the same mass bins using the dynamical halo mass estimator
(i.e.\ using $M_{\rm dyn}$ for the bin assignment, and using the same
$N_{\rm FOF} > 4$ selection). The best-fit double Schechter function
parameters are tabulated in Table~\ref{tab:vm1} and shown in
Fig.~\ref{fig:dsmlum}, where the left panels show the results for
$M_{\rm lum}$ and the right panels those for $M_{\rm dyn}$. We see
that all four mass estimators provide results that are largely
consistent with each other for all three Schechter function
parameters.  As in the previous subsection, we also find here that the
cGSMF is best represented by a single Schechter function at $M_{\rm
  halo} > 10^{13.5}~M_\odot~h_{70}^{-1}$. However, with the cGSMFs of
the intermediate and high halo mass bins being essentially identical,
it is not possible to confidently confirm the trends with halo mass
observed in the previous subsection with this more limited (albeit
higher fidelity) sample. On the other hand, the results are not
inconsistent with these trends either.

\subsection{GSMF dependence on group branch order and number of connecting links}\label{6.4}
In Sect.~\ref{6.1} we analysed the change (or lack thereof) of the
mass function when moving perpendicularly to the filaments. We now
investigate the dependence of the GSMF on the position within the
filamentary structure.

\begin{table}
\caption{Selection of $2 \times 3 = 6$ different subsamples from the
  FC group galaxy sample (G1) based on the group branch order (BO) and
  number of connecting links ($N_{\rm links}$).}
\begin{center}
\begin{tabular}{c||c|c}
\hline
& Backbone & Outskirts \\
\hline
\hline 
edge        & BO $= 1$ \& $N_{\rm links} = 1$ & BO $> 1$ \& $N_{\rm links} = 1$ \\
intermediate& BO $= 1$ \& $N_{\rm links} = 2$ & BO $> 1$ \& $N_{\rm links} = 2$ \\
centre      & BO $= 1$ \& $N_{\rm links} > 2$ & BO $> 1$ \& $N_{\rm links} > 2$ \\
\hline
\end{tabular}

\end{center}
\label{tab:subsamples}
\end{table}

\begin{figure}[!h]
\centering
\includegraphics[width=0.48\textwidth]{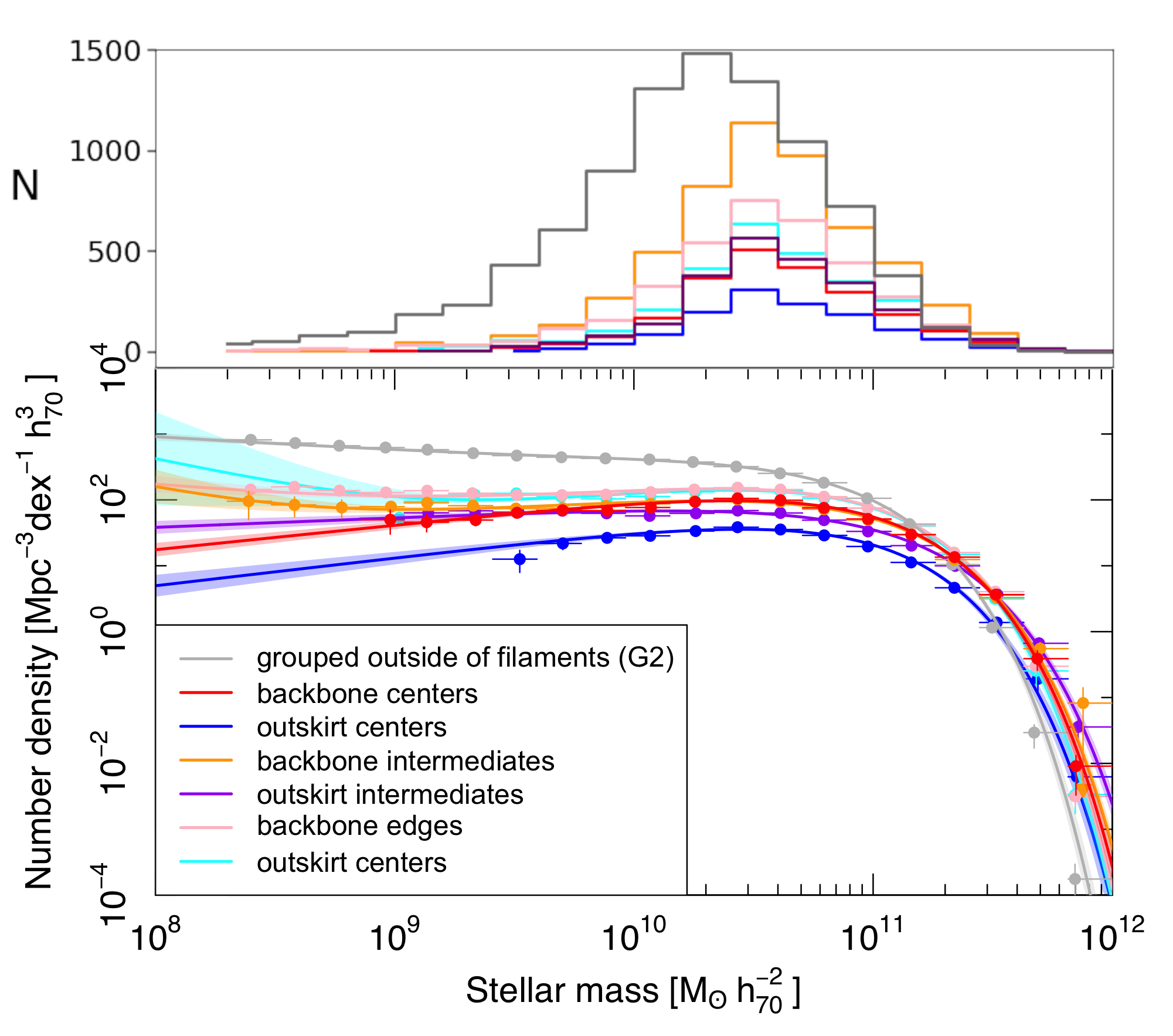}
\caption{Lower panel: cGSMFs of grouped galaxies in filaments (G1
  subsample), colour-coded by a combination of BO and $N_{\rm links}$
  as indicated in the legend. For comparison, the cGSMF of grouped
  galaxies outside of filaments (G2) is also shown in grey.  Upper
  panel: Raw number of galaxies as a function of stellar mass in each
  sample, as indicated.}
\label{fig:subsample1}
\end{figure}

\begin{figure}[!h]
\centering
\includegraphics[width=0.49\textwidth]{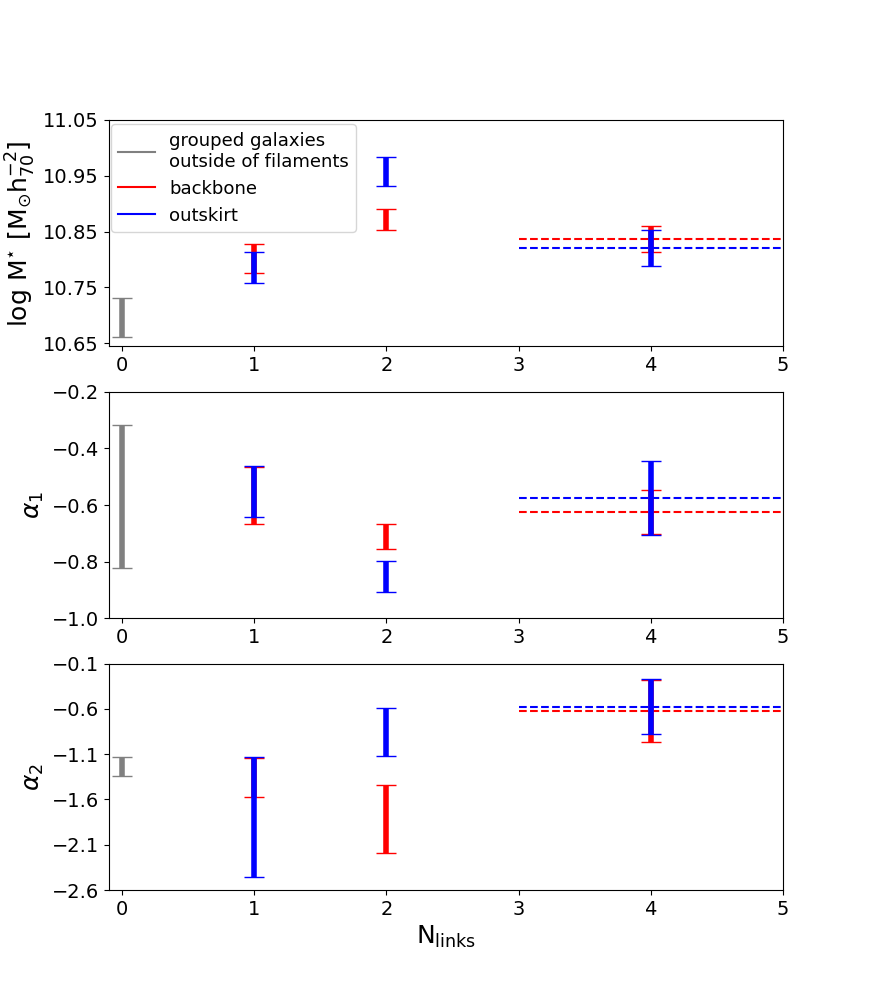}
\caption{Best-fit double Schechter function parameters of the cGSMFs
  shown in Fig.~\ref{fig:subsample1}, as indicated in the legend. The
  grouped galaxies outside of filaments (in grey) are somewhat
  arbitrarily shown at $N_{\rm links} = 0$.}
\label{fig:subsample2}
\end{figure}

From our groups-in-filaments galaxy subsample (G1) defined in
Sect.~\ref{4.1}, we select six different subsamples according to
specific combinations of group branch order BO and group number of
connecting links $N_{\rm links}$, as listed in
Table~\ref{tab:subsamples}. In particular, we distinguish between
galaxies in groups located in their filaments' backbone (BO $= 1$) and
in their outskirts (BO $> 1$), as well as between galaxies inhabiting
groups located at the edge ($N_{\rm links} = 1$), at an intermediate
position ($N_{\rm links} = 2$), and at the centre ($N_{\rm links} >
2$) of their filament. We note that the full range of both parameters
is $1$--$5$.

The resulting cGSMFs are shown in Fig.~\ref{fig:subsample1} as a
function of the combination of BO and $N_{\rm links}$. The best-fit
double Schechter function parameters are tabulated in
Table~\ref{tab:subsamples1} and shown in Fig.~\ref{fig:subsample2}. We
first note that, when moving from the backbone (in red) to the
outskirts (in blue), none of the three double Schechter parameters
seem to vary very significantly, regardless of the value of $N_{\rm
  links}$. Both for centres and edges, the backbone and outskirts are
in excellent agreement in all three Schechter parameters, while for
the intermediate regions some mildly significant differences appear,
with higher $M^{\star}$ and $\alpha_{2}$ values and a lower
$\alpha_{1}$ value in the outskirts compared to the backbone. However,
this may be affected by the fact that the cGSMF of the intermediate
regions in outskirts are best fit with a single Schechter function
(where $\alpha_{1} = \alpha_{2}$), which also applies to the central
regions of both the backbone and the outskirts
(cf.\ Table~\ref{tab:subsamples1}).

In addition, we did not find any significant trends as a function of
$N_{\rm links}$ for any of the Schechter function parameters for the
backbone or the outskirts sample, or their combination. This remains
true if we add, somewhat arbitrarily, the Schechter function
parameters of grouped galaxies outside of filaments at $N_{\rm links}
= 0$ (shown in grey in Fig.~\ref{fig:subsample2}). All we observed
here is that the $M^{\star}$ value of grouped galaxies outside of
filaments is significantly lower that that of grouped galaxies in
filaments, as we had already noted in Sect.~\ref{6.2}.

We thus conclude that there is no evidence for any differences in the
shape of the GSMF of grouped galaxies in filaments as a function of
position within the filamentary structure.

\section{Discussion}\label{7}
In this section we first compare our global GSMF with other GAMA and
SDSS DR7 GSMFs (Sect.~\ref{7.1}), and then discuss our results on the
dependence of the mass function on different environmental measures in
the context of other similar studies (Sect.~\ref{7.2}).

\begin{table*}
\caption{Best-fit double Schechter function parameters for the mass
  functions of grouped galaxies in filaments for different subsamples
  and for grouped galaxies outside of filaments, as indicated.}
\begin{center}
\begin{tabular}{l||c|c|c}
\hline
Galaxy sample & $\log M^{\star}$ & $\alpha_{1}$ & $\alpha_{2}$\\
 & $(M_{\odot} \, h_{70}^{-2})$ &  &\\ 
\hline
\hline 
BO $= 1$ \& $N_{\rm links} > 2$ (backbone centres)  & 10.84 $\pm$ 0.02 & $-$0.62 $\pm$ 0.08& $-$0.62 $\pm$ 0.34\\
BO $> 1$ \& $N_{\rm links} > 2$ (outskirts centres)  & 10.82 $\pm$ 0.03 & $-$0.57 $\pm$ 0.13& $-$0.57 $\pm$ 0.30\\
BO $= 1$ \& $N_{\rm links} = 2$ (backbone intermediates)  & 10.87 $\pm$ 0.02 & $-$0.71 $\pm$ 0.04& $-$1.82 $\pm$ 0.38\\
BO $> 1$ \& $N_{\rm links} = 2$ (outskirts intermediates) & 10.96 $\pm$ 0.03 & $-$0.85 $\pm$ 0.05& $-$0.85 $\pm$ 0.27\\
BO $= 1$ \& $N_{\rm links} = 1$  (backbone edges) & 10.80 $\pm$ 0.03 & $-$0.57 $\pm$ 0.10& $-$1.36 $\pm$ 0.21\\
BO $> 1$ \& $N_{\rm links} = 1$ (outskirts edges) & 10.79 $\pm$ 0.03 & $-$0.55 $\pm$ 0.09& $-$1.79 $\pm$ 0.66\\
Grouped galaxies outside of filaments (G2)  & 10.70 $\pm$ 0.04 & $-$0.57 $\pm$ 0.25 & $-$1.24 $\pm$ 0.11\\
\hline
\end{tabular}
\end{center}
\label{tab:subsamples1}
\end{table*}

\subsection{Global GAMA and SDSS DR7 GSMFs}\label{7.1}
Many studies have already been undertaken inside the GAMA project
concerning GSMFs. \cite{baldry+12} selected a flux-limited sample of
$5210$ galaxies from the earliest phase of the GAMA project, called
GAMA~I, covering an area of $143$~deg$^2$. This sample was complete to
$r = 19.4$~mag in the G09 and G15 regions and to $r = 19.8$~mag in
G12, and covered the redshift range $0.002 < z <
0.06$. \cite{kelvin+14} selected a local volume-limited subsample of
$2711$ galaxies, also from GAMA~I, to a lower stellar mass limit of $M
= 10^9~M_{\odot}$, covering $0.025 < z < 0.06$. Similarly,
\cite{alpaslan+15} defined a local volume-limited subsample of $7195$
galaxies to a lower stellar mass limit of $M = 10^{9.5}~M_{\odot}$, up
to $z = 0.1$. \cite{wright+17} expanded the analysis by
\cite{baldry+12} to the full GAMA~II dataset. Their flux-limited
sample, covering the same GAMA~II equatorial regions of $180$~deg$^2$
used in this study, was complete down to $r = 19.8$~mag and covered
the redshift range $0.002 < z < 0.1$. \cite{baldry+12},
\cite{kelvin+14} and \cite{wright+17} all computed their global GSMFs
using a density-corrected $1/V_{\rm max}$ method. Completing the set
of GSMF studied with GAMA, \cite{driver+22} most recently considered a
flux-limited sample of $13\,957$ galaxies from the latest version of
GAMA, called GAMA~III, adding the G23 region to the equatorial regions
for a total area of $250$~deg$^2$. The magnitude limit of this sample
was $r = 19.65$~mag, covering a redshift range of $0.0013 < z <
0.1$. They computed their GSMF using the same MML estimator as in the
present study. Finally, we include in our comparison the GSMF of
\cite{weigel+16}, which was measured from a sample of $\sim$$110\,000$
galaxies selected from the SDSS DR7, covering an area of
$7748$~deg$^2$, with a magnitude limit of $r = 17.77$~mag and a
redshift range of $0.02 < z < 0.06$. Their GSMF was computed using the
parametric maximum likelihood method, originally developed by
\cite{sandage+79}.

\begin{table*}
\caption{Best-fit double Schechter function parameters of the GSMFs
  shown in Fig.~\ref{fig:comp1}, as indicated.}
\begin{center}
\begin{tabular}{c|c|c|c|c|c}
\hline
Dataset & $\log M^{\star}$  & $\Phi_1^{\star}$  & $\alpha_1$ & $\Phi_2^{\star}$  & $\alpha_2$\\
 & $(M_{\odot} \, h_{70}^{-2})$ & $(10^{-3}$~Mpc$^{-3} \, h_{70}^3)$ & & $(10^{-3}$~Mpc$^{-3} \, h_{70}^3)$ & \\
\hline
\hline
\cite{baldry+12} & 10.66 $\pm$ 0.05 & 3.96 $\pm$ 0.34 & $-$0.35 $\pm$ 0.18 & 0.79 $\pm$ 0.23 & $-$1.47 $\pm$ 0.05\\
\cite{kelvin+14} & 10.64 $\pm$ 0.07 & 4.18 $\pm$ 1.52 & $-$0.43 $\pm$ 0.35 & 0.74 $\pm$ 1.13 & $-$1.50 $\pm$ 0.22\\
\cite{alpaslan+15} & 10.82 $\pm$ 0.02 & 2.00 $\pm$ 0.13 & $-$0.97 $\pm$ 0.02 & 2.00 $\pm$ 0.13 & $-$0.97 $\pm$ 0.02 \\
\cite{wright+17} & 10.78 $\pm$ 0.01 & 2.93 $\pm$ 0.40 & $-$0.62 $\pm$ 0.03 & 0.63 $\pm$ 0.10 & $-$1.50 $\pm$ 0.01\\ 
\cite{driver+22} & 10.75 $\pm$ 0.02 & 3.66 $\pm$ 0.14 & $-$0.47 $\pm$ 0.07 & 0.63 $\pm$ 0.09 & $-$1.53 $\pm$ 0.03\\ 
This work & 10.76 $\pm$ 0.01 & 3.75 $\pm$ 0.09 & $-$0.86 $\pm$ 0.03 & 0.13 $\pm$ 0.05 & $-$1.71 $\pm$ 0.06\\ 
\hline
\cite{weigel+16} & 10.79 $\pm$ 0.01 & 9.77 $\pm$ 6.30 & $-$0.79 $\pm$ 0.04 & 0.49 $\pm$ 0.23 & $-$1.69 $\pm$ 0.10\\ 
\hline
\end{tabular}
\tablefoot{We show our global GSMF, other GAMA GSMFs \citep{baldry+12, kelvin+14, alpaslan+15, wright+17, driver+22} and the SDSS DR7 GSMF derived by \cite{weigel+16}. The GSMF of \cite{alpaslan+15} was fit with a single Schechter function, which we represent as $\Phi_1^{\star} = \Phi_2^{\star} = \Phi^{\star}/2$ and $\alpha_1 = \alpha_2 = \alpha$ here. Furthermore, in contrast to all other studies listed here, \cite{alpaslan+15} used cosmological parameters of $(H_{0}, \Omega_{\rm M}, \Omega_{\rm \Lambda}) = (100, 0.25, 0.75)$. We have thus rescaled their values to $H_0 = 70$~km~s$^{-1}$~Mpc$^{-1}$.}
\end{center}
\label{tab:otherworks}
\end{table*}

\begin{figure}
\centering
\includegraphics[width=0.49\textwidth]{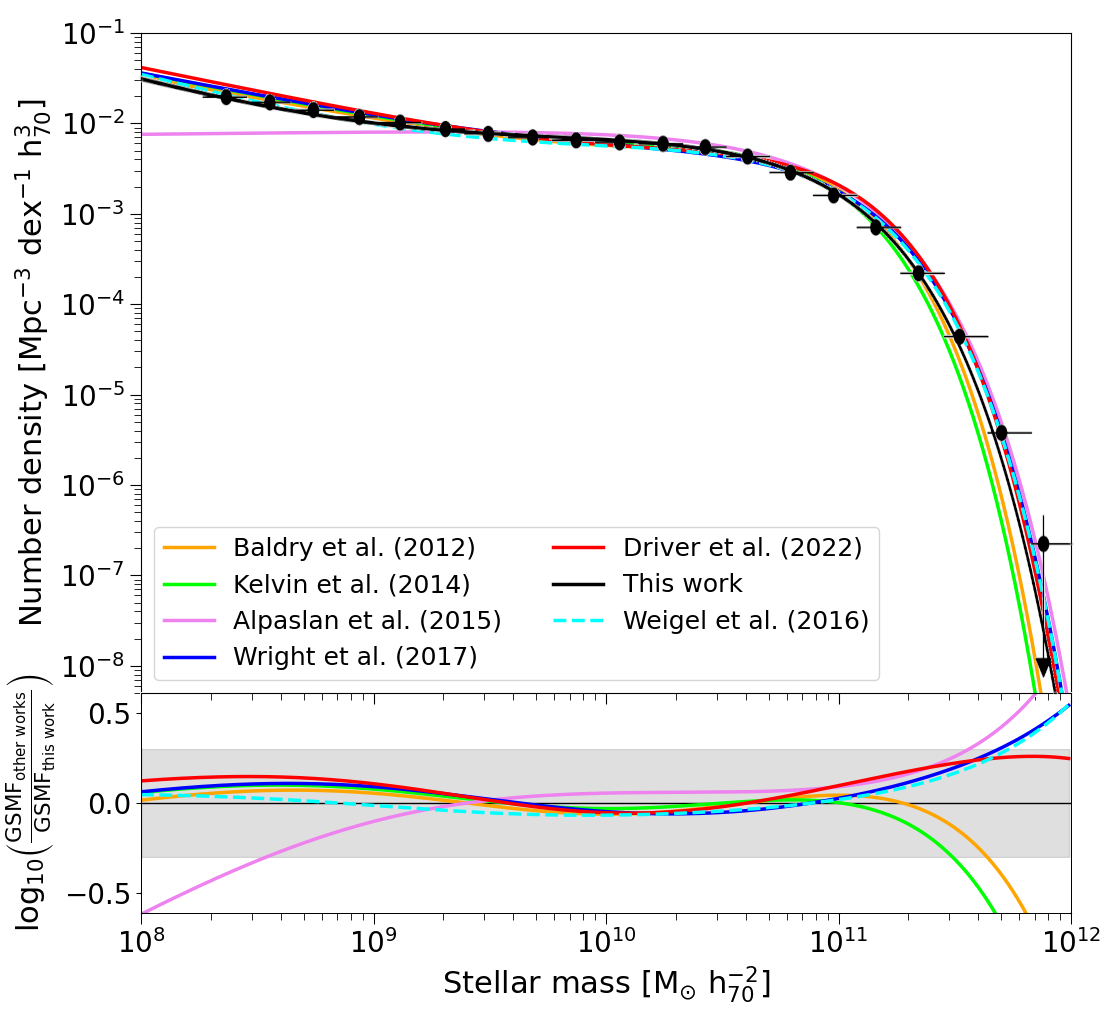}
\caption{Upper panel: Comparison of our global GSMF with other recent
  measurements derived from GAMA and SDSS data, as indicated. Lower
  panel: Logarithmic ratio of the literature GSMFs relative to
  ours. The grey shaded area indicates agreement within a factor of
  two.}
\label{fig:comp1}
\end{figure}

The global GSMF was fit with a double Schechter function in almost all
of theses studies. The only exception is \cite{alpaslan+15} who used a
single Schechter function. We list the best-fit Schechter parameters
in Table~\ref{tab:otherworks} and compare the corresponding functions
to our data and best-fit double Schechter function in
Fig.~\ref{fig:comp1}. Evidently, our global GSMF, shown in black, is
in very good agreement with all of the mass functions of the previous
studies described above, except that of \cite{alpaslan+15}. The
agreement is best with the GSMF of \cite{driver+22}, with the
differences being less than a factor of two over the entire range of
four orders of magnitude in stellar mass considered here (cf.\ lower
panel of Fig.~\ref{fig:comp1}). The overall excellent agreement
between these various mass functions is quite remarkable, given the
differences in photometry, stellar mass estimation techniques, and
GSMF measurement procedures.

Nevertheless, some differences remain. At high stellar masses, the
lower $M^\star$ values of the GAMA~I studies by \citep{baldry+12} and
\cite{kelvin+14} are likely due to the over-fragmentation of bright
galaxies relative to fainter and smaller systems in the original GAMA
aperture photometry, causing their fluxes to be underestimated and
thus leading to an underestimate of their stellar masses as well. This
issue was addressed in the photometry presented by \cite{wright+17},
which we have also used in this work. We note that our $M^\star$ value
compares well with that of \cite{driver+22} despite their use of
further improved photometry and the improved stellar mass estimation
technique of \cite{robotham+20}. At intermediate masses, where the
inter-GSMF agreement is best, our global GSMF confirms a plateau from
$\sim$$10^{9.5}$ to $\sim$$10^{10.5}~M_{\odot}~h_{70}^{-2}$, which was
already noticed by \cite{baldry+12} in the GAMA~I data. This
characteristic seems to emerge at lower redshifts ($z < 1$) and may be
caused by mergers \citep{robotham+14}. Moving towards lower masses,
the GSMF rises noticeably, clearly requiring a double Schechter
function for an adequate description, as demonstrated by the
disagreement with the single Schechter function GSMF of
\cite{alpaslan+15} at low masses. Despite the good agreement of the
various double Schechter functions in Fig.~\ref{fig:comp1} in the
intermediate- and low-mass regimes, their parameters in this regime
($\Phi^\star_1, \alpha_1, \Phi^\star_2, \alpha_2$) are actually
somewhat different. For example, our $\alpha_1$ and $\alpha_2$ values
are significantly steeper than those of \cite{wright+17} and \cite[but
  very similar to those of \citealp{weigel+16}]{driver+22}, whereas
our $\Phi^\star_2$ value is significantly lower. Apart from
correlations among these parameters and different lower mass limits
used, another reason for this variance is presumably the inability
even of a double Schechter function to accurately model the
intermediate-mass, low-mass, and transition regimes simultaneously.

\subsection{Environment-dependent GSMFs}\label{7.2}
As soon as the environment is a concern in describing any change in
the shape of the mass function, it is important to stress that many
definitions of environment are found in the literature
(e.g. \citealp{muldrew+12}). One approach relies on geometry,
according to which galaxies are classified into categories such as
voids (i.e.\ three-dimensional structures), sheets (two-dimensional),
filaments (one-dimensional), and clusters, groups or knots
(zero-dimensional). A second approach considers the distinction
between grouped and ungrouped galaxies, where the latter are often
referred to as field galaxies. A third approach classifies galaxies
based on halo mass. A fourth approach describes the environment
through measurements of the local density (averaged over some scale),
which can be parametrised in several ways and following different
techniques, for example by counting the number of neighbours of a
galaxy within a specific aperture or measuring the distance to the
$n^{\rm th}$ nearest neighbour. Although these different measures of
environment are generally correlated, they cannot be considered
equivalent.

Exploiting the capabilities of two surveys in the redshift range $0.03
\leq z \leq 0.11$, \cite{vulcani+12} investigated the dependence of
the low-$z$ GSMF on density, using a $5^{\rm th}$ nearest neighbour
approach. They found that density regulates the shape of the mass
function at any mass above their mass limit, claiming that lower
density regions host relatively a larger population of low-mass
galaxies than higher density regions. However, when examining clusters
in particular, using a $10^{\rm th}$ nearest neighbour approach, the
situation was found to be slightly different: In the highest density
regions, the slope of the GSMF flattens out only at very low masses,
suggestive of a sort of deficit of low-mass galaxies with respect to
intermediate-mass ones compared to lower density regions. Moreover,
they found that density determines not only the shape of the mass
function, but also the highest mass encountered. The highest density
regions host the most massive galaxies, which seem to be absent in the
lowest density regions. Using the same surveys, \cite{calvi+13}
studied the dependence of the GSMF as a function of halo mass, finding
that the mass functions of galaxies inside and outside clusters look
indistinguishable. When comparing grouped, binary and ungrouped
galaxies, they found that the GSMF of single galaxies was relatively
richer in low-mass galaxies (i.e.\ showed a steeper slope) than any
other subsample.

In another complementary study, making use of two surveys at
intermediate redshifts in the range $0.3 \leq z \leq 0.8$,
\cite{vulcani+13} also analysed the dependence of the GSMF as a
function of halo mass, identifying clusters, groups and ungrouped
galaxies lying in the field. Except for the brightest cluster
galaxies, they found that the GSMF is invariant, i.e.\ shows
comparable values for $M^{\star}$ and $\alpha$, from clusters to the
field. Comparing the virialised regions of clusters and their
outskirts, they also failed to detect any difference. In their view,
this result confirmed that halo mass does not alter the shape of the
mass distribution from clusters to ungrouped galaxies, since the
outskirts of clusters can be considered as transition regions between
the cluster virialised regions and the field. Their conclusions are
quite surprising, considering that galaxies in halos with different
masses are subject to different physical processes which can suppress
SF, resulting in different mass growth rates and timescales.
Nevertheless, at the redshifts considered, \cite{vulcani+13} showed
that most of the galaxy mass appears to have already been assembled,
and that environment-dependent processes have had no significant
influence on galaxy mass.

\cite{guglielmo+18} also investigated the dependence of the GSMF on
halo mass, identifying from the XXL survey a magnitude-complete groups
and clusters (G$\&$C) galaxy sample with redshifts $z \leq 0.6$,
spanning a wide range of X-ray luminosities. Similarly to the results
described above, they found no significant overall difference in the
shape of the GSMF between galaxies belonging to the G$\&$C sample and
those that do not, also when binning in X-ray luminosity.

In a more recent work, \citet{okane+24} investigated the dependence of
the GSMF on large-scale environment using a mass-complete sample from
the SDSS DR8 Main Galaxy Sample. When matching in local density, using
a $3^{\rm rd}$ nearest neighbour approach, as well as in stellar mass,
they found that the differences observed as a function of the position
in the cosmic web vanish, concluding that local density can fully
explain the environmental effect of filaments.

Our results are inconsistent with the conclusions drawn by
\cite{vulcani+13}, \cite{calvi+13} and \cite{guglielmo+18}. In
Sect.~\ref{6.3} we showed that the GSMF is clearly shaped by halo
mass, since $M^{\star}$ clearly increases, and $\alpha_1$ decreases
with $M_{\rm halo}$. The only result in agreement with ours is the
steep value of $\alpha_{2}$ observed by \cite{calvi+13} in their
ungrouped galaxy sample, similar to ours (cf.\ Table~\ref{tab:nuovo}).
It is important to note here that \cite{calvi+13} and
\cite{vulcani+13} find an excess of galaxies at $\log[M/(M_{\odot} \,
  h^{-2}_{70})] > 11.7$ in both cluster and group mass functions, with
the Schechter function not able to adequately describe this behaviour.

On the other hand, the results from \citet{vulcani+12} and
\citet{okane+24} seem to be compatible with ours, despite their use of
a different definition of environment. Besides our findings regarding
$M^{\star}$, as previously discussed, in Sects.~\ref{6.3} and
\ref{6.4} we also showed that at the highest halo masses, i.e.\ at
$\log[M_{\rm halo}/(M_{\odot} \, h^{-1}_{70})] \ga 13.5$, the GSMF is
best described by a single Schechter function, in which $\alpha_{1} =
\alpha_{2}$. In other words, at these halo masses, the physical
processes affecting galaxy formation and evolution seem to have a
similar impact at low and intermediate masses. Moreover, we confirm
that the most massive halos host the most massive galaxies.

Our clearest result is the evolution of $M^{\star}$, going from
ungrouped to group environments (Sect.~\ref{6.2}) as well as inside the
groups as a function of halo mass (Sect.~\ref{6.3}). We refer the reader to
\cite{robotham+06} and \cite{robotham+10b} for consistent results on
the variation of the luminosity function with group halo mass. The
same behaviour was already observed by \cite{baldry+06}. Starting from
a $z<0.1$ SDSS sample and averaging between a $4^{\rm th}$ and a
$5^{\rm th}$ nearest neighbour approach, they investigated the
dependence of the low-$z$ GSMF on projected density. They found that
$M^{\star}$ varied from $10^{10.6}$ to $10^{11}~M_{\odot}~h_{70}^{-2}$
with increasing environmental density, with the fractional
contribution of galaxies with $\log[M^{\star}/(M_{\odot} \,
  h_{70}^{-2})] \approx 11.5$ increasing by a factor of $\sim$$50$. On
the other hand, $\alpha$ did not vary strongly with environment, even
if they claim that a standard Schechter function does not fit their
entire mass range. When fitting with a double Schechter function with
fixed $\alpha_{2} = -1.5$, they found an overall steepening of
$\alpha_{1}$. This last result is also compatible with our findings:
In Sect.~\ref{6.3} we also showed that $\alpha_{1}$ steepens with
increasing $M_{\rm halo}$.

\section{Conclusions}\label{8}
We employed the equatorial GAMA~II dataset at $z \leq 0.213$ to
construct multiple GSMFs. For our analysis, we made use of various
GAMA data products: stellar masses (Sect.~\ref{2.1};
\citealp{taylor+11}), the G$^{3}$C for the identification of groups
(Sect.~\ref{2.2}; \citealp{robotham+11}), and the FC for the
identification of the filaments (Sect.~\ref{2.3};
\citealp{alpaslan+14}). Our magnitude-limited parent galaxy sample
(the selection of which is summarised in Table~\ref{tab:ss}) consisted
of $88\,093$ galaxies and $11\,725$ groups. These numbers were reduced
to $50\,089$ galaxies and $10\,429$ groups after applying our
redshift-dependent stellar mass limit (Sect.~\ref{smcl}).

Our global GSMF is well described by a double Schechter function with
the following parameters: $\log[M^{\star}/(M_{\odot} \, h_{70}^{-2})]
= 10.76 \pm 0.01$, $\Phi_1^{\star} = (3.72 \pm 0.09) \times
10^{-3}$~Mpc$^{-3} \, h_{70}^3$, $\alpha_{1} = - 0.86 \pm 0.03$,
$\Phi_2^{\star} = (0.13 \pm 0.05) \times 10^{-3}$~Mpc$^{-3} \,
h_{70}^3$, and $\alpha_{2} = - 1.72 \pm 0.06$. As shown in
Fig.~\ref{fig:comp1}, our global GSMF agrees well with the GSMFs of
prominent previous GAMA and SDSS studies, with the differences being
smaller than a factor of two over the mass range $10^8$ to $3 \times
10^11~M_{\odot}~h_{70}^{-2}$.

We investigated the variation of the low-redshift GSMF as a function
of four different environmental properties: orthogonal distance to the
nearest filament, $D_{\perp,\rm min}$ (in Sect.~\ref{6.1}), group
membership (Sect.~\ref{6.2}), group halo mass, $M_{\rm halo}$
(Sect.~\ref{6.3}), and the combination of group branch order, BO, and
group number of connecting links, $N_{\rm links}$ (Sect.~\ref{6.4}).
We noted first that the GSMF of filament galaxies shows a small change
in $M^{\star}$ as the size of the filaments is varied, whereas the
void GSMF does not change its shape at all. In other words, the
filament and void mass function do not depend strongly on how big or
small we assume the filaments to be (see Fig.~\ref{fig:group_uno}). We
did, however, detect a significant difference between the filament and
void GSMFs in each double Schechter function parameter. This
difference vanished, though, as soon as the grouped galaxies were
removed from filament samples, as shown in
Fig.~\ref{fig:group_due}. On the other hand, in Sect.~\ref{6.2} we
showed that the shape of the GSMF changes significantly between our
total group and ungrouped galaxy samples. We therefore concluded that
the mass function is not strongly affected by how close (or distant) a
galaxy is to a filament but rather by its membership of a group or
not. The apparent difference between the filament and void GSMFs is
thus entirely attributable to a much higher fraction of grouped
galaxies in filaments compared to voids.

We therefore studied the dependence of the GSMF on group halo mass
$M_{\rm halo}$, using both a dynamical and a luminosity-based halo
mass estimator. For the dynamical halo mass estimates, we found that
the characteristic mass of the GSMF, $M^\star$, increases with $M_{\rm
  halo}$, while the intermediate mass slope, $\alpha_1$, decreases
(cf.\ Table~\ref{tab:mdyn1} and Fig.~\ref{fig:dsmdyn}). Furthermore,
at the highest halo masses, the intermediate- and low-mass slopes of
the GSMF become very similar such that the GSMF is best described by a
single Schechter function in this halo mass regime. While the results
using the luminosity-based halo mass estimator are less conclusive,
they are nevertheless consistent with those derived from the dynamical
halo mass estimator.

Combining these results, we hence concluded that the GSMF primarily
depends on halo mass, while the larger-scale environment plays a
secondary role at best. While this finding is inconsistent with some
previous studies \citep{vulcani+13,calvi+13,guglielmo+18}, our
unambiguous evidence that the most massive halos host the most massive
galaxies in particular is clearly consistent with other works
\citep{baldry+06,vulcani+12,okane+24}.

Having said that, it would be incorrect to claim that the GSMF is
entirely oblivious to the existence of filaments. After all, it is in
the filament environment that the halo mass distribution is, to a
large extent, established, which in turn determines the distribution
of galaxy stellar masses. In this sense, the influence of filaments on
the GSMF is simultaneously fundamental and indirect.

This picture is further corroborated by our observation that the group
GSMF is almost entirely independent of the location within the
filamentary structure, as measured by the group branch order and the
group number of connecting links, except that the GSMF of highly
connected groups is best described by a single Schechter function.

Our overall conclusion is that all of our observations are fully
explained by the dependence of the GSMF on halo mass. The observed
differences between the GSMFs of void and filament galaxies, ungrouped
galaxies, grouped galaxies outside of filaments, and grouped galaxies
in filaments are all sourced by the different halo mass distributions
of these samples. In a future paper we plan to investigate the halo
mass dependence of the GSMF separately for star-forming and quiescent
galaxies.

\begin{acknowledgements}
GAMA is a joint European-Australasian project based around a
spectroscopic campaign using the Anglo-Australian Telescope. The GAMA
input catalogue is based on data taken from the SDSS and the UKIRT
Infrared Deep Sky Survey. Complementary imaging of the GAMA regions
was obtained by a number of independent survey programmes including
GALEX MIS, VST KiDS, VISTA VIKING, WISE, Herschel-ATLAS, GMRT and
ASKAP providing UV to radio coverage. GAMA is funded by the STFC (UK),
the ARC (Australia), the AAO, and the participating institutions. The
GAMA website is \url{https://www.gama-survey.org/}. AS is funded by,
and JL acknowledges support by the Deutsche Forschungsgemeinschaft
(DFG, German Research Foundation) under Germany's Excellence Strategy
-- EXC 2121 ``Quantum Universe'' -- 390833306.
\end{acknowledgements}

\bibliographystyle{aa}

\begin{thebibliography}{81}
\expandafter\ifx\csname natexlab\endcsname\relax\def\natexlab#1{#1}\fi

\bibitem[{{Alpaslan} {et~al.}(2015){Alpaslan}, {Driver}, {Robotham},
  {Obreschkow}, {Andrae}, {Cluver}, {Kelvin}, {Lange}, {Owers}, {Taylor},
  {Andrews}, {Bamford}, {Bland-Hawthorn}, {Brough}, {Brown}, {Colless},
  {Davies}, {Eardley}, {Grootes}, {Hopkins}, {Kennedy}, {Liske},
  {Lara-L{\'o}pez}, {L{\'o}pez-S{\'a}nchez}, {Loveday}, {Madore}, {Mahajan},
  {Meyer}, {Moffett}, {Norberg}, {Penny}, {Pimbblet}, {Popescu}, {Seibert}, \&
  {Tuffs}}]{alpaslan+15}
{Alpaslan}, M., {Driver}, S., {Robotham}, A. S.~G., {et~al.} 2015, \mnras, 451,
  3249

\bibitem[{{Alpaslan} {et~al.}(2014){Alpaslan}, {Robotham}, {Driver}, {Norberg},
  {Baldry}, {Bauer}, {Bland-Hawthorn}, {Brown}, {Cluver}, {Colless}, {Foster},
  {Hopkins}, {Van Kampen}, {Kelvin}, {Lara-Lopez}, {Liske}, {Lopez-Sanchez},
  {Loveday}, {McNaught-Roberts}, {Merson}, \& {Pimbblet}}]{alpaslan+14}
{Alpaslan}, M., {Robotham}, A. S.~G., {Driver}, S., {et~al.} 2014, \mnras, 438,
  177

\bibitem[{{Baldry} {et~al.}(2014){Baldry}, {Alpaslan}, {Bauer},
  {Bland-Hawthorn}, {Brough}, {Cluver}, {Croom}, {Davies}, {Driver},
  {Gunawardhana}, {Holwerda}, {Hopkins}, {Kelvin}, {Liske},
  {L{\'o}pez-S{\'a}nchez}, {Loveday}, {Norberg}, {Peacock}, {Robotham}, \&
  {Taylor}}]{baldry+14}
{Baldry}, I.~K., {Alpaslan}, M., {Bauer}, A.~E., {et~al.} 2014, \mnras, 441,
  2440

\bibitem[{{Baldry} {et~al.}(2006){Baldry}, {Balogh}, {Bower}, {Glazebrook},
  {Nichol}, {Bamford}, \& {Budavari}}]{baldry+06}
{Baldry}, I.~K., {Balogh}, M.~L., {Bower}, R.~G., {et~al.} 2006, \mnras, 373,
  469

\bibitem[{{Baldry} {et~al.}(2012){Baldry}, {Driver}, {Loveday}, {Taylor},
  {Kelvin}, {Liske}, {Norberg}, {Robotham}, {Brough}, {Hopkins}, {Bamford},
  {Peacock}, {Bland-Hawthorn}, {Conselice}, {Croom}, {Jones}, {Parkinson},
  {Popescu}, {Prescott}, {Sharp}, \& {Tuffs}}]{baldry+12}
{Baldry}, I.~K., {Driver}, S.~P., {Loveday}, J., {et~al.} 2012, \mnras, 421,
  621

\bibitem[{{Baldry} {et~al.}(2008){Baldry}, {Glazebrook}, \&
  {Driver}}]{baldry+08}
{Baldry}, I.~K., {Glazebrook}, K., \& {Driver}, S.~P. 2008, \mnras, 388, 945

\bibitem[{{Baldry} {et~al.}(2018){Baldry}, {Liske}, {Brown}, {Robotham},
  {Driver}, {Dunne}, {Alpaslan}, {Brough}, {Cluver}, {Eardley}, {Farrow},
  {Heymans}, {Hildebrandt}, {Hopkins}, {Kelvin}, {Loveday}, {Moffett},
  {Norberg}, {Owers}, {Taylor}, {Wright}, {Bamford}, {Bland-Hawthorn},
  {Bourne}, {Bremer}, {Colless}, {Conselice}, {Croom}, {Davies}, {Foster},
  {Grootes}, {Holwerda}, {Jones}, {Kafle}, {Kuijken}, {Lara-Lopez},
  {L{\'o}pez-S{\'a}nchez}, {Meyer}, {Phillipps}, {Sutherland}, {van Kampen}, \&
  {Wilkins}}]{baldry+18}
{Baldry}, I.~K., {Liske}, J., {Brown}, M.~J.~I., {et~al.} 2018, \mnras, 474,
  3875

\bibitem[{{Baldry} {et~al.}(2010){Baldry}, {Robotham}, {Hill}, {Driver},
  {Liske}, {Norberg}, {Bamford}, {Hopkins}, {Loveday}, {Peacock}, {Cameron},
  {Croom}, {Cross}, {Doyle}, {Dye}, {Frenk}, {Jones}, {van Kampen}, {Kelvin},
  {Nichol}, {Parkinson}, {Popescu}, {Prescott}, {Sharp}, {Sutherland},
  {Thomas}, \& {Tuffs}}]{baldry+10}
{Baldry}, I.~K., {Robotham}, A.~S.~G., {Hill}, D.~T., {et~al.} 2010, \mnras,
  404, 86

\bibitem[{{Balogh} {et~al.}(2001){Balogh}, {Christlein}, {Zabludoff}, \&
  {Zaritsky}}]{balogh+01}
{Balogh}, M.~L., {Christlein}, D., {Zabludoff}, A.~I., \& {Zaritsky}, D. 2001,
  \apj, 557, 117

\bibitem[{{Bell} \& {de Jong}(2001)}]{bell+01}
{Bell}, E.~F. \& {de Jong}, R.~S. 2001, \apj, 550, 212

\bibitem[{{Bell} {et~al.}(2003){Bell}, {McIntosh}, {Katz}, \&
  {Weinberg}}]{bell+03}
{Bell}, E.~F., {McIntosh}, D.~H., {Katz}, N., \& {Weinberg}, M.~D. 2003, \apjs,
  149, 289

\bibitem[{{Bellstedt} {et~al.}(2020){Bellstedt}, {Driver}, {Robotham},
  {Davies}, {Bogue}, {Cook}, {Hashemizadeh}, {Koushan}, {Taylor}, {Thorne},
  {Turner}, \& {Wright}}]{bellstedt+20}
{Bellstedt}, S., {Driver}, S.~P., {Robotham}, A. S.~G., {et~al.} 2020, \mnras,
  496, 3235

\bibitem[{{Blumenthal} {et~al.}(1984){Blumenthal}, {Faber}, {Primack}, \&
  {Rees}}]{blumenthal+84}
{Blumenthal}, G.~R., {Faber}, S.~M., {Primack}, J.~R., \& {Rees}, M.~J. 1984,
  \nat, 311, 517

\bibitem[{{Bond} {et~al.}(1996){Bond}, {Kofman}, \& {Pogosyan}}]{bond+96b}
{Bond}, J.~R., {Kofman}, L., \& {Pogosyan}, D. 1996, \nat, 380, 603

\bibitem[{{Bond} \& {Myers}(1996)}]{bond+96a}
{Bond}, J.~R. \& {Myers}, S.~T. 1996, \apjs, 103, 1

\bibitem[{{Calvi} {et~al.}(2013){Calvi}, {Poggianti}, {Vulcani}, \&
  {Fasano}}]{calvi+13}
{Calvi}, R., {Poggianti}, B.~M., {Vulcani}, B., \& {Fasano}, G. 2013, \mnras,
  432, 3141

\bibitem[{{Croton} {et~al.}(2006){Croton}, {Springel}, {White}, {De Lucia},
  {Frenk}, {Gao}, {Jenkins}, {Kauffmann}, {Navarro}, \& {Yoshida}}]{croton+06}
{Croton}, D.~J., {Springel}, V., {White}, S. D.~M., {et~al.} 2006, \mnras, 365,
  11

\bibitem[{{Davis} {et~al.}(1985){Davis}, {Efstathiou}, {Frenk}, \&
  {White}}]{davis+85}
{Davis}, M., {Efstathiou}, G., {Frenk}, C.~S., \& {White}, S.~D.~M. 1985, \apj,
  292, 371

\bibitem[{{Dekel} \& {Birnboim}(2006)}]{dekel+06}
{Dekel}, A. \& {Birnboim}, Y. 2006, \mnras, 368, 2

\bibitem[{{Dekel} \& {Silk}(1986)}]{dekel+86}
{Dekel}, A. \& {Silk}, J. 1986, \apj, 303, 39

\bibitem[{{Dickinson} {et~al.}(2003){Dickinson}, {Papovich}, {Ferguson}, \&
  {Budav{\'a}ri}}]{dickinson+03}
{Dickinson}, M., {Papovich}, C., {Ferguson}, H.~C., \& {Budav{\'a}ri}, T. 2003,
  \apj, 587, 25

\bibitem[{{Driver} {et~al.}(2022){Driver}, {Bellstedt}, {Robotham}, {Baldry},
  {Davies}, {Liske}, {Obreschkow}, {Taylor}, {Wright}, {Alpaslan}, {Bamford},
  {Bauer}, {Bland-Hawthorn}, {Bilicki}, {Bravo}, {Brough}, {Casura}, {Cluver},
  {Colless}, {Conselice}, {Croom}, {de Jong}, {D'Eugenio}, {De Propris},
  {Dogruel}, {Drinkwater}, {Dvornik}, {Farrow}, {Frenk}, {Giblin}, {Graham},
  {Grootes}, {Gunawardhana}, {Hashemizadeh}, {H{\"a}u{\ss}ler}, {Heymans},
  {Hildebrandt}, {Holwerda}, {Hopkins}, {Jarrett}, {Heath Jones}, {Kelvin},
  {Koushan}, {Kuijken}, {Lara-L{\'o}pez}, {Lange}, {L{\'o}pez-S{\'a}nchez},
  {Loveday}, {Mahajan}, {Meyer}, {Moffett}, {Napolitano}, {Norberg}, {Owers},
  {Radovich}, {Raouf}, {Peacock}, {Phillipps}, {Pimbblet}, {Popescu}, {Said},
  {Sansom}, {Seibert}, {Sutherland}, {Thorne}, {Tuffs}, {Turner}, {van der
  Wel}, {van Kampen}, \& {Wilkins}}]{driver+22}
{Driver}, S.~P., {Bellstedt}, S., {Robotham}, A. S.~G., {et~al.} 2022, \mnras,
  513, 439

\bibitem[{{Driver} {et~al.}(2011){Driver}, {Hill}, {Kelvin}, {Robotham},
  {Liske}, {Norberg}, {Baldry}, {Bamford}, {Hopkins}, {Loveday}, {Peacock},
  {Andrae}, {Bland-Hawthorn}, {Brough}, {Brown}, {Cameron}, {Ching}, {Colless},
  {Conselice}, {Croom}, {Cross}, {de Propris}, {Dye}, {Drinkwater}, {Ellis},
  {Graham}, {Grootes}, {Gunawardhana}, {Jones}, {van Kampen}, {Maraston},
  {Nichol}, {Parkinson}, {Phillipps}, {Pimbblet}, {Popescu}, {Prescott},
  {Roseboom}, {Sadler}, {Sansom}, {Sharp}, {Smith}, {Taylor}, {Thomas},
  {Tuffs}, {Wijesinghe}, {Dunne}, {Frenk}, {Jarvis}, {Madore}, {Meyer},
  {Seibert}, {Staveley-Smith}, {Sutherland}, \& {Warren}}]{driver+11}
{Driver}, S.~P., {Hill}, D.~T., {Kelvin}, L.~S., {et~al.} 2011, \mnras, 413,
  971

\bibitem[{{Driver} {et~al.}(2009){Driver}, {Norberg}, {Baldry}, {Bamford},
  {Hopkins}, {Liske}, {Loveday}, {Peacock}, {Hill}, {Kelvin}, {Robotham},
  {Cross}, {Parkinson}, {Prescott}, {Conselice}, {Dunne}, {Brough}, {Jones},
  {Sharp}, {van Kampen}, {Oliver}, {Roseboom}, {Bland-Hawthorn}, {Croom},
  {Ellis}, {Cameron}, {Cole}, {Frenk}, {Couch}, {Graham}, {Proctor}, {De
  Propris}, {Doyle}, {Edmondson}, {Nichol}, {Thomas}, {Eales}, {Jarvis},
  {Kuijken}, {Lahav}, {Madore}, {Seibert}, {Meyer}, {Staveley-Smith},
  {Phillipps}, {Popescu}, {Sansom}, {Sutherland}, {Tuffs}, \&
  {Warren}}]{driver+09}
{Driver}, S.~P., {Norberg}, P., {Baldry}, I.~K., {et~al.} 2009, Astronomy and
  Geophysics, 50, 5.12

\bibitem[{{Driver} {et~al.}(2016){Driver}, {Wright}, {Andrews}, {Davies},
  {Kafle}, {Lange}, {Moffett}, {Mannering}, {Robotham}, {Vinsen}, {Alpaslan},
  {Andrae}, {Baldry}, {Bauer}, {Bamford}, {Bland-Hawthorn}, {Bourne}, {Brough},
  {Brown}, {Cluver}, {Croom}, {Colless}, {Conselice}, {da Cunha}, {De Propris},
  {Drinkwater}, {Dunne}, {Eales}, {Edge}, {Frenk}, {Graham}, {Grootes},
  {Holwerda}, {Hopkins}, {Ibar}, {van Kampen}, {Kelvin}, {Jarrett}, {Jones},
  {Lara-Lopez}, {Liske}, {Lopez-Sanchez}, {Loveday}, {Maddox}, {Madore},
  {Mahajan}, {Meyer}, {Norberg}, {Penny}, {Phillipps}, {Popescu}, {Tuffs},
  {Peacock}, {Pimbblet}, {Prescott}, {Rowlands}, {Sansom}, {Seibert}, {Smith},
  {Sutherland}, {Taylor}, {Valiante}, {Vazquez-Mata}, {Wang}, {Wilkins}, \&
  {Williams}}]{driver+16}
{Driver}, S.~P., {Wright}, A.~H., {Andrews}, S.~K., {et~al.} 2016, \mnras, 455,
  3911

\bibitem[{{Efstathiou} {et~al.}(1988){Efstathiou}, {Frenk}, {White}, \&
  {Davis}}]{efstathiou+88}
{Efstathiou}, G., {Frenk}, C.~S., {White}, S. D.~M., \& {Davis}, M. 1988,
  \mnras, 235, 715

\bibitem[{{Fontana} {et~al.}(2006){Fontana}, {Salimbeni}, {Grazian},
  {Giallongo}, {Pentericci}, {Nonino}, {Fontanot}, {Menci}, {Monaco},
  {Cristiani}, {Vanzella}, {de Santis}, \& {Gallozzi}}]{fontana+06}
{Fontana}, A., {Salimbeni}, S., {Grazian}, A., {et~al.} 2006, \aap, 459, 745

\bibitem[{{Gallazzi} {et~al.}(2005){Gallazzi}, {Charlot}, {Brinchmann},
  {White}, \& {Tremonti}}]{gallazzi+05}
{Gallazzi}, A., {Charlot}, S., {Brinchmann}, J., {White}, S. D.~M., \&
  {Tremonti}, C.~A. 2005, \mnras, 362, 41

\bibitem[{{Guglielmo} {et~al.}(2018){Guglielmo}, {Poggianti}, {Vulcani},
  {Adami}, {Gastaldello}, {Ettori}, {Fotoupoulou}, {Koulouridis}, {Ramos Ceja},
  {Giles}, {McGee}, {Altieri}, {Baldry}, {Birkinshaw}, {Bolzonella},
  {Bongiorno}, {Brown}, {Chiappetti}, {Driver}, {Elyiv}, {Evrard}, {Garilli},
  {Grootes}, {Guennou}, {Hopkins}, {Horellou}, {Iovino}, {Lidman}, {Liske},
  {Maurogordato}, {Owers}, {Pacaud}, {Paltani}, {Pierre}, {Plionis}, {Ponman},
  {Robotham}, {Sadibekova}, {Scodeggio}, {Sereno}, {Smol{\v{c}}i{\'c}},
  {Tuffs}, {Valtchanov}, {Vignali}, \& {Willis}}]{guglielmo+18}
{Guglielmo}, V., {Poggianti}, B.~M., {Vulcani}, B., {et~al.} 2018, \aap, 620,
  A7

\bibitem[{{Han} {et~al.}(2015){Han}, {Eke}, {Frenk}, {Mandelbaum}, {Norberg},
  {Schneider}, {Peacock}, {Jing}, {Baldry}, {Bland-Hawthorn}, {Brough},
  {Brown}, {Liske}, {Loveday}, \& {Robotham}}]{han+15}
{Han}, J., {Eke}, V.~R., {Frenk}, C.~S., {et~al.} 2015, \mnras, 446, 1356

\bibitem[{{Hopkins} {et~al.}(2013){Hopkins}, {Cox}, {Hernquist}, {Narayanan},
  {Hayward}, \& {Murray}}]{hopkins+13}
{Hopkins}, P.~F., {Cox}, T.~J., {Hernquist}, L., {et~al.} 2013, \mnras, 430,
  1901

\bibitem[{{J{\~o}eveer} {et~al.}(1978){J{\~o}eveer}, {Einasto}, \&
  {Tago}}]{joeveer+78}
{J{\~o}eveer}, M., {Einasto}, J., \& {Tago}, E. 1978, \mnras, 185, 357

\bibitem[{{Jablonka} \& {Arimoto}(1992)}]{jablonka+92}
{Jablonka}, J. \& {Arimoto}, N. 1992, \aap, 255, 63

\bibitem[{{Kaiser}(1984)}]{kaiser+84}
{Kaiser}, N. 1984, \apjl, 284, L9

\bibitem[{{Kauffmann} {et~al.}(2003){Kauffmann}, {Heckman}, {White}, {Charlot},
  {Tremonti}, {Brinchmann}, {Bruzual}, {Peng}, {Seibert}, {Bernardi},
  {Blanton}, {Brinkmann}, {Castander}, {Cs{\'a}bai}, {Fukugita}, {Ivezic},
  {Munn}, {Nichol}, {Padmanabhan}, {Thakar}, {Weinberg}, \&
  {York}}]{kauffmann+03}
{Kauffmann}, G., {Heckman}, T.~M., {White}, S. D.~M., {et~al.} 2003, \mnras,
  341, 33

\bibitem[{{Kelvin} {et~al.}(2014){Kelvin}, {Driver}, {Robotham}, {Taylor},
  {Graham}, {Alpaslan}, {Baldry}, {Bamford}, {Bauer}, {Bland-Hawthorn},
  {Brown}, {Colless}, {Conselice}, {Holwerda}, {Hopkins}, {Lara-L{\'o}pez},
  {Liske}, {L{\'o}pez-S{\'a}nchez}, {Loveday}, {Norberg}, {Phillipps},
  {Popescu}, {Prescott}, {Sansom}, \& {Tuffs}}]{kelvin+14}
{Kelvin}, L.~S., {Driver}, S.~P., {Robotham}, A. S.~G., {et~al.} 2014, \mnras,
  444, 1647

\bibitem[{{Kere{\v{s}}} {et~al.}(2005){Kere{\v{s}}}, {Katz}, {Weinberg}, \&
  {Dav{\'e}}}]{keres+05}
{Kere{\v{s}}}, D., {Katz}, N., {Weinberg}, D.~H., \& {Dav{\'e}}, R. 2005,
  \mnras, 363, 2

\bibitem[{{Larson}(1974)}]{larson+74}
{Larson}, R.~B. 1974, \mnras, 169, 229

\bibitem[{{Larson} \& {Tinsley}(1978)}]{larson+78}
{Larson}, R.~B. \& {Tinsley}, B.~M. 1978, \apj, 219, 46

\bibitem[{{Liske} {et~al.}(2015){Liske}, {Baldry}, {Driver}, {Tuffs},
  {Alpaslan}, {Andrae}, {Brough}, {Cluver}, {Grootes}, {Gunawardhana},
  {Kelvin}, {Loveday}, {Robotham}, {Taylor}, {Bamford}, {Bland-Hawthorn},
  {Brown}, {Drinkwater}, {Hopkins}, {Meyer}, {Norberg}, {Peacock}, {Agius},
  {Andrews}, {Bauer}, {Ching}, {Colless}, {Conselice}, {Croom}, {Davies}, {De
  Propris}, {Dunne}, {Eardley}, {Ellis}, {Foster}, {Frenk}, {H{\"a}u{\ss}ler},
  {Holwerda}, {Howlett}, {Ibarra}, {Jarvis}, {Jones}, {Kafle}, {Lacey},
  {Lange}, {Lara-L{\'o}pez}, {L{\'o}pez-S{\'a}nchez}, {Maddox}, {Madore},
  {McNaught-Roberts}, {Moffett}, {Nichol}, {Owers}, {Palamara}, {Penny},
  {Phillipps}, {Pimbblet}, {Popescu}, {Prescott}, {Proctor}, {Sadler},
  {Sansom}, {Seibert}, {Sharp}, {Sutherland}, {V{\'a}zquez-Mata}, {van Kampen},
  {Wilkins}, {Williams}, \& {Wright}}]{liske+15}
{Liske}, J., {Baldry}, I.~K., {Driver}, S.~P., {et~al.} 2015, \mnras, 452, 2087

\bibitem[{{Marchesini} {et~al.}(2009){Marchesini}, {van Dokkum}, {F{\"o}rster
  Schreiber}, {Franx}, {Labb{\'e}}, \& {Wuyts}}]{marchesini+09}
{Marchesini}, D., {van Dokkum}, P.~G., {F{\"o}rster Schreiber}, N.~M., {et~al.}
  2009, \apj, 701, 1765

\bibitem[{{Moster} {et~al.}(2010){Moster}, {Somerville}, {Maulbetsch}, {van den
  Bosch}, {Macci{\`o}}, {Naab}, \& {Oser}}]{moster+10}
{Moster}, B.~P., {Somerville}, R.~S., {Maulbetsch}, C., {et~al.} 2010, \apj,
  710, 903

\bibitem[{{Muldrew} {et~al.}(2012){Muldrew}, {Croton}, {Skibba}, {Pearce},
  {Ann}, {Baldry}, {Brough}, {Choi}, {Conselice}, {Cowan}, {Gallazzi}, {Gray},
  {Gr{\"u}tzbauch}, {Li}, {Park}, {Pilipenko}, {Podgorzec}, {Robotham},
  {Wilman}, {Yang}, {Zhang}, \& {Zibetti}}]{muldrew+12}
{Muldrew}, S.~I., {Croton}, D.~J., {Skibba}, R.~A., {et~al.} 2012, \mnras, 419,
  2670

\bibitem[{{Obreschkow} {et~al.}(2018){Obreschkow}, {Murray}, {Robotham}, \&
  {Westmeier}}]{obreschkow+18}
{Obreschkow}, D., {Murray}, S.~G., {Robotham}, A.~S.~G., \& {Westmeier}, T.
  2018, \mnras, 474, 5500

\bibitem[{{O'Kane} {et~al.}(2024){O'Kane}, {Kuchner}, {Gray}, \&
  {Arag{\'o}n-Salamanca}}]{okane+24}
{O'Kane}, C.~J., {Kuchner}, U., {Gray}, M.~E., \& {Arag{\'o}n-Salamanca}, A.
  2024, \mnras

\bibitem[{{Oppenheimer} {et~al.}(2010){Oppenheimer}, {Dav{\'e}}, {Kere{\v{s}}},
  {Fardal}, {Katz}, {Kollmeier}, \& {Weinberg}}]{oppenheimer+10}
{Oppenheimer}, B.~D., {Dav{\'e}}, R., {Kere{\v{s}}}, D., {et~al.} 2010, \mnras,
  406, 2325

\bibitem[{{Panter} {et~al.}(2004){Panter}, {Heavens}, \& {Jimenez}}]{panter+04}
{Panter}, B., {Heavens}, A.~F., \& {Jimenez}, R. 2004, \mnras, 355, 764

\bibitem[{{P{\'e}rez-Gonz{\'a}lez} {et~al.}(2008){P{\'e}rez-Gonz{\'a}lez},
  {Rieke}, {Villar}, {Barro}, {Blaylock}, {Egami}, {Gallego}, {Gil de Paz},
  {Pascual}, {Zamorano}, \& {Donley}}]{perezgonzalez+08}
{P{\'e}rez-Gonz{\'a}lez}, P.~G., {Rieke}, G.~H., {Villar}, V., {et~al.} 2008,
  \apj, 675, 234

\bibitem[{{Pillepich} {et~al.}(2018){Pillepich}, {Springel}, {Nelson}, {Genel},
  {Naiman}, {Pakmor}, {Hernquist}, {Torrey}, {Vogelsberger}, {Weinberger}, \&
  {Marinacci}}]{pillepich+18}
{Pillepich}, A., {Springel}, V., {Nelson}, D., {et~al.} 2018, \mnras, 473, 4077

\bibitem[{{Pogosyan} {et~al.}(1996){Pogosyan}, {Bond}, {Kofman}, \&
  {Wadsley}}]{pogosyan+96}
{Pogosyan}, D., {Bond}, J.~R., {Kofman}, L., \& {Wadsley}, J. 1996, in American
  Astronomical Society Meeting Abstracts, Vol. 189, American Astronomical
  Society Meeting Abstracts, 13.03

\bibitem[{{Pozzetti} {et~al.}(2010){Pozzetti}, {Bolzonella}, {Zucca},
  {Zamorani}, {Lilly}, {Renzini}, {Moresco}, {Mignoli}, {Cassata}, {Tasca},
  {Lamareille}, {Maier}, {Meneux}, {Halliday}, {Oesch}, {Vergani}, {Caputi},
  {Kova{\v{c}}}, {Cimatti}, {Cucciati}, {Iovino}, {Peng}, {Carollo}, {Contini},
  {Kneib}, {Le F{\'e}vre}, {Mainieri}, {Scodeggio}, {Bardelli}, {Bongiorno},
  {Coppa}, {de la Torre}, {de Ravel}, {Franzetti}, {Garilli}, {Kampczyk},
  {Knobel}, {Le Borgne}, {Le Brun}, {Pell{\`o}}, {Perez Montero},
  {Ricciardelli}, {Silverman}, {Tanaka}, {Tresse}, {Abbas}, {Bottini}, {Cappi},
  {Guzzo}, {Koekemoer}, {Leauthaud}, {Maccagni}, {Marinoni}, {McCracken},
  {Memeo}, {Porciani}, {Scaramella}, {Scarlata}, \& {Scoville}}]{pozzetti+10}
{Pozzetti}, L., {Bolzonella}, M., {Zucca}, E., {et~al.} 2010, \aap, 523, A13

\bibitem[{{Press} \& {Schechter}(1974)}]{press+74}
{Press}, W.~H. \& {Schechter}, P. 1974, \apj, 187, 425

\bibitem[{{Quadri} {et~al.}(2012){Quadri}, {Williams}, {Franx}, \&
  {Hildebrandt}}]{quadri+12}
{Quadri}, R.~F., {Williams}, R.~J., {Franx}, M., \& {Hildebrandt}, H. 2012,
  \apj, 744, 88

\bibitem[{{Robotham} {et~al.}(2010{\natexlab{a}}){Robotham}, {Driver},
  {Norberg}, {Baldry}, {Bamford}, {Hopkins}, {Liske}, {Loveday}, {Peacock},
  {Cameron}, {Croom}, {Doyle}, {Frenk}, {Hill}, {Jones}, {van Kampen},
  {Kelvin}, {Kuijken}, {Nichol}, {Parkinson}, {Popescu}, {Prescott}, {Sharp},
  {Sutherland}, {Thomas}, \& {Tuffs}}]{robotham+10a}
{Robotham}, A., {Driver}, S.~P., {Norberg}, P., {et~al.} 2010{\natexlab{a}},
  \pasa, 27, 76

\bibitem[{{Robotham} {et~al.}(2010{\natexlab{b}}){Robotham}, {Phillipps}, \&
  {de Propris}}]{robotham+10b}
{Robotham}, A., {Phillipps}, S., \& {de Propris}, R. 2010{\natexlab{b}},
  \mnras, 403, 1812

\bibitem[{{Robotham} {et~al.}(2006){Robotham}, {Wallace}, {Phillipps}, \& {De
  Propris}}]{robotham+06}
{Robotham}, A., {Wallace}, C., {Phillipps}, S., \& {De Propris}, R. 2006, \apj,
  652, 1077

\bibitem[{{Robotham} {et~al.}(2020){Robotham}, {Bellstedt}, {Lagos}, {Thorne},
  {Davies}, {Driver}, \& {Bravo}}]{robotham+20}
{Robotham}, A.~S.~G., {Bellstedt}, S., {Lagos}, C. d.~P., {et~al.} 2020,
  \mnras, 495, 905

\bibitem[{{Robotham} {et~al.}(2018){Robotham}, {Davies}, {Driver}, {Koushan},
  {Taranu}, {Casura}, \& {Liske}}]{robotham+18}
{Robotham}, A.~S.~G., {Davies}, L.~J.~M., {Driver}, S.~P., {et~al.} 2018,
  \mnras, 476, 3137

\bibitem[{{Robotham} {et~al.}(2014){Robotham}, {Driver}, {Davies}, {Hopkins},
  {Baldry}, {Agius}, {Bauer}, {Bland-Hawthorn}, {Brough}, {Brown}, {Cluver},
  {De Propris}, {Drinkwater}, {Holwerda}, {Kelvin}, {Lara-Lopez}, {Liske},
  {L{\'o}pez-S{\'a}nchez}, {Loveday}, {Mahajan}, {McNaught-Roberts}, {Moffett},
  {Norberg}, {Obreschkow}, {Owers}, {Penny}, {Pimbblet}, {Prescott}, {Taylor},
  {van Kampen}, \& {Wilkins}}]{robotham+14}
{Robotham}, A.~S.~G., {Driver}, S.~P., {Davies}, L.~J.~M., {et~al.} 2014,
  \mnras, 444, 3986

\bibitem[{{Robotham} {et~al.}(2011){Robotham}, {Norberg}, {Driver}, {Baldry},
  {Bamford}, {Hopkins}, {Liske}, {Loveday}, {Merson}, {Peacock}, {Brough},
  {Cameron}, {Conselice}, {Croom}, {Frenk}, {Gunawardhana}, {Hill}, {Jones},
  {Kelvin}, {Kuijken}, {Nichol}, {Parkinson}, {Pimbblet}, {Phillipps},
  {Popescu}, {Prescott}, {Sharp}, {Sutherland}, {Taylor}, {Thomas}, {Tuffs},
  {van Kampen}, \& {Wijesinghe}}]{robotham+11}
{Robotham}, A.~S.~G., {Norberg}, P., {Driver}, S.~P., {et~al.} 2011, \mnras,
  416, 2640

\bibitem[{{Sandage} {et~al.}(1979){Sandage}, {Tammann}, \&
  {Yahil}}]{sandage+79}
{Sandage}, A., {Tammann}, G.~A., \& {Yahil}, A. 1979, \apj, 232, 352

\bibitem[{{Saunders} {et~al.}(2004){Saunders}, {Bridges}, {Gillingham},
  {Haynes}, {Smith}, {Whittard}, {Churilov}, {Lankshear}, {Croom}, {Jones}, \&
  {Boshuizen}}]{saunders+04}
{Saunders}, W., {Bridges}, T., {Gillingham}, P., {et~al.} 2004, in Society of
  Photo-Optical Instrumentation Engineers (SPIE) Conference Series, Vol. 5492,
  Ground-based Instrumentation for Astronomy, ed. A.~F.~M. {Moorwood} \&
  M.~{Iye}, 389--400

\bibitem[{{Scharr{\'e}} {et~al.}(2024){Scharr{\'e}}, {Sorini}, \&
  {Dav{\'e}}}]{scharre+24}
{Scharr{\'e}}, L., {Sorini}, D., \& {Dav{\'e}}, R. 2024, \mnras, 534, 361

\bibitem[{{Schechter}(1976)}]{schechter+76}
{Schechter}, P. 1976, \apj, 203, 297

\bibitem[{{Schmidt}(1968)}]{schmidt+68}
{Schmidt}, M. 1968, \apj, 151, 393

\bibitem[{{Sharp} {et~al.}(2006){Sharp}, {Saunders}, {Smith}, {Churilov},
  {Correll}, {Dawson}, {Farrel}, {Frost}, {Haynes}, {Heald}, {Lankshear},
  {Mayfield}, {Waller}, \& {Whittard}}]{sharp+06}
{Sharp}, R., {Saunders}, W., {Smith}, G., {et~al.} 2006, in Society of
  Photo-Optical Instrumentation Engineers (SPIE) Conference Series, Vol. 6269,
  Ground-based and Airborne Instrumentation for Astronomy, ed. I.~S. {McLean}
  \& M.~{Iye}, 62690G

\bibitem[{{Smith} {et~al.}(2004){Smith}, {Saunders}, {Bridges}, {Churilov},
  {Lankshear}, {Dawson}, {Correll}, {Waller}, {Haynes}, \& {Frost}}]{smith+04}
{Smith}, G.~A., {Saunders}, W., {Bridges}, T., {et~al.} 2004, in Society of
  Photo-Optical Instrumentation Engineers (SPIE) Conference Series, Vol. 5492,
  Ground-based Instrumentation for Astronomy, ed. A.~F.~M. {Moorwood} \&
  M.~{Iye}, 410--420

\bibitem[{{Taylor} {et~al.}(2011){Taylor}, {Hopkins}, {Baldry}, {Brown},
  {Driver}, {Kelvin}, {Hill}, {Robotham}, {Bland-Hawthorn}, {Jones}, {Sharp},
  {Thomas}, {Liske}, {Loveday}, {Norberg}, {Peacock}, {Bamford}, {Brough},
  {Colless}, {Cameron}, {Conselice}, {Croom}, {Frenk}, {Gunawardhana},
  {Kuijken}, {Nichol}, {Parkinson}, {Phillipps}, {Pimbblet}, {Popescu},
  {Prescott}, {Sutherland}, {Tuffs}, {van Kampen}, \& {Wijesinghe}}]{taylor+11}
{Taylor}, E.~N., {Hopkins}, A.~M., {Baldry}, I.~K., {et~al.} 2011, \mnras, 418,
  1587

\bibitem[{{Tomczak} {et~al.}(2014){Tomczak}, {Quadri}, {Tran}, {Labb{\'e}},
  {Straatman}, {Papovich}, {Glazebrook}, {Allen}, {Brammer}, {Kacprzak},
  {Kawinwanichakij}, {Kelson}, {McCarthy}, {Mehrtens}, {Monson}, {Persson},
  {Spitler}, {Tilvi}, \& {van Dokkum}}]{tomczak+14}
{Tomczak}, A.~R., {Quadri}, R.~F., {Tran}, K.-V.~H., {et~al.} 2014, \apj, 783,
  85

\bibitem[{{Tonry} {et~al.}(2000){Tonry}, {Blakeslee}, {Ajhar}, \&
  {Dressler}}]{tonry+00}
{Tonry}, J.~L., {Blakeslee}, J.~P., {Ajhar}, E.~A., \& {Dressler}, A. 2000,
  \apj, 530, 625

\bibitem[{{V{\'a}zquez-Mata} {et~al.}(2020){V{\'a}zquez-Mata}, {Loveday},
  {Riggs}, {Baldry}, {Davies}, {Robotham}, {Holwerda}, {Brown}, {Cluver},
  {Wang}, {Alpaslan}, {Bland-Hawthorn}, {Brough}, {Driver}, {Hopkins},
  {Taylor}, \& {Wright}}]{vazquez+20}
{V{\'a}zquez-Mata}, J.~A., {Loveday}, J., {Riggs}, S.~D., {et~al.} 2020,
  \mnras, 499, 631

\bibitem[{{Viola} {et~al.}(2015){Viola}, {Cacciato}, {Brouwer}, {Kuijken},
  {Hoekstra}, {Norberg}, {Robotham}, {van Uitert}, {Alpaslan}, {Baldry},
  {Choi}, {de Jong}, {Driver}, {Erben}, {Grado}, {Graham}, {Heymans},
  {Hildebrandt}, {Hopkins}, {Irisarri}, {Joachimi}, {Loveday}, {Miller},
  {Nakajima}, {Schneider}, {Sif{\'o}n}, \& {Verdoes Kleijn}}]{viola+15}
{Viola}, M., {Cacciato}, M., {Brouwer}, M., {et~al.} 2015, \mnras, 452, 3529

\bibitem[{{Vulcani} {et~al.}(2012){Vulcani}, {Poggianti}, {Fasano}, {Desai},
  {Dressler}, {Oemler}, {Calvi}, {D'Onofrio}, \& {Moretti}}]{vulcani+12}
{Vulcani}, B., {Poggianti}, B.~M., {Fasano}, G., {et~al.} 2012, \mnras, 420,
  1481

\bibitem[{{Vulcani} {et~al.}(2013){Vulcani}, {Poggianti}, {Oemler}, {Dressler},
  {Arag{\'o}n-Salamanca}, {De Lucia}, {Moretti}, {Gladders}, {Abramson}, \&
  {Halliday}}]{vulcani+13}
{Vulcani}, B., {Poggianti}, B.~M., {Oemler}, A., {et~al.} 2013, \aap, 550, A58

\bibitem[{{Weigel} {et~al.}(2016){Weigel}, {Schawinski}, \&
  {Bruderer}}]{weigel+16}
{Weigel}, A.~K., {Schawinski}, K., \& {Bruderer}, C. 2016, \mnras, 459, 2150

\bibitem[{{White} \& {Frenk}(1991)}]{white+91}
{White}, S. D.~M. \& {Frenk}, C.~S. 1991, \apj, 379, 52

\bibitem[{{White} \& {Rees}(1978)}]{white+78}
{White}, S.~D.~M. \& {Rees}, M.~J. 1978, \mnras, 183, 341

\bibitem[{{Wright} {et~al.}(2016){Wright}, {Robotham}, {Bourne}, {Driver},
  {Dunne}, {Maddox}, {Alpaslan}, {Andrews}, {Bauer}, {Bland-Hawthorn},
  {Brough}, {Brown}, {Clarke}, {Cluver}, {Davies}, {Grootes}, {Holwerda},
  {Hopkins}, {Jarrett}, {Kafle}, {Lange}, {Liske}, {Loveday}, {Moffett},
  {Norberg}, {Popescu}, {Smith}, {Taylor}, {Tuffs}, {Wang}, \&
  {Wilkins}}]{wright+16}
{Wright}, A.~H., {Robotham}, A.~S.~G., {Bourne}, N., {et~al.} 2016, \mnras,
  460, 765

\bibitem[{{Wright} {et~al.}(2017){Wright}, {Robotham}, {Driver}, {Alpaslan},
  {Andrews}, {Baldry}, {Bland-Hawthorn}, {Brough}, {Brown}, {Colless}, {da
  Cunha}, {Davies}, {Graham}, {Holwerda}, {Hopkins}, {Kafle}, {Kelvin},
  {Loveday}, {Maddox}, {Meyer}, {Moffett}, {Norberg}, {Phillipps}, {Rowlands},
  {Taylor}, {Wang}, \& {Wilkins}}]{wright+17}
{Wright}, A.~H., {Robotham}, A.~S.~G., {Driver}, S.~P., {et~al.} 2017, \mnras,
  470, 283

\bibitem[{{Yang} {et~al.}(2009){Yang}, {Mo}, \& {van den Bosch}}]{yang+09}
{Yang}, X., {Mo}, H.~J., \& {van den Bosch}, F.~C. 2009, \apj, 695, 900

\bibitem[{{Zel'dovich}(1970)}]{zeldovich+70}
{Zel'dovich}, Y.~B. 1970, \aap, 5, 84

\end{thebibliography}

\end{document}